\documentclass[aps,prl,twocolumn,showpacs,superscriptaddress,groupedaddress]{revtex4}  
\usepackage{graphicx}  
\usepackage{dcolumn}   
\usepackage{bm}        
\usepackage{amssymb}   
\usepackage[final]{epsfig}
\usepackage{amssymb}
\usepackage{latexsym}
\usepackage{wrapfig}
\usepackage{amsmath}
\usepackage{booktabs}
\usepackage{threeparttable}
\usepackage{xcolor}
\usepackage{amsthm}
\usepackage{lipsum}
\usepackage{hyperref}
\usepackage{enumitem}
\newcommand{\bfa}{{\bf a}}

\newcommand{\bfc}{{\bf c}}

\newcommand{\bfe}{{\bf e}}

\newcommand{\bfx}{{\bf x}}
\newcommand{\bfy}{{\bf y}}

\newcommand{\bfz}{{\bf z}}

\newcommand{\bfC}{{\bf C}}

\newcommand{\bfI}{{\bf I}}

\newcommand{\bfR}{{\bf R}}

\newcommand{\beq}{\begin{equation}}
\newcommand{\eeq}{\end{equation}}
\newcommand{\beqs}{\begin{eqnarray}}
\newcommand{\eeqs}{\end{eqnarray}}

\begin{document}



\title{Helical Miura Origami}
\author{Fan Feng}
\author{Paul Plucinsky}%
\author{Richard D. James}
\email{james@umn.edu}
\affiliation{Aerospace Engineering and Mechanics, University of Minnesota, Minneapolis, MN 55455, USA}%
\date{\today}

\begin{abstract}
We characterize the phase-space of all Helical Miura Origami.  These structures are obtained by taking a partially folded Miura parallelogram as the unit cell, applying a generic helical or rod group to the cell, and characterizing all the parameters that lead to a globally compatible origami structure.  When such compatibility is achieved, the result is cylindrical-type origami that can be manufactured from a suitably designed flat tessellation and  ``rolled-up" by a rigidly foldable motion into a cylinder. We find that the closed Helical Miura Origami are generically rigid to deformations that preserve cylindrical symmetry, but multistable.  We are inspired by the ways atomic structures deform \cite{feng2019phase} to develop two broad strategies for reconfigurability: motion by slip, which involves relaxing the closure condition; and motion by phase transformation, which exploits multistability.  Taken together, these results provide a comprehensive description of the phase-space of cylindrical origami, as well as quantitative design guidance for their use as actuators or metamaterials that  exploit twist, axial extension, radial expansion, and symmetry.  
\end{abstract}

\maketitle

Origami is the ancient Japanese art of paper folding. In recent years, this art form has been appreciated not only for its aesthetics \footnote{See Robert Lang's  \href{https://langorigami.com/}{https://langorigami.com/}
	and Tomohiro Tachi's  \href{http://www.flickr.com/photos/tactom/}{http://www.flickr.com/photos/tactom/} for aesthetically pleasing examples of origami.}, but also for its potential functionality \cite{peraza2014origami}\textemdash including in space technologies \cite{wilson2013origami, schenk2013inflatable}, transforming architectures \cite{reis2015transforming, tachi2011designing}, multistability and topological properties \cite{waitukaitis2015origami, waitukaitis2016origami, chen2016topological}, biological structures \cite{Faber1386, DNAorigami, han2011dna}, deployable antennas \cite{Liu_helical_antenna_2014, yao2014novel}, metamaterials \cite{schenk2013geometry, yasuda2015reentrant, silverberg2014using}, and mechanical properties \cite{wei2013geometric, ftp_PNAS_2015, dudte2016programming}.  Origami design utilizes the shape change induced by piecewise affine isometric deformations (i.e., folding along creases)\textemdash from, say, an easy-to-manufacture flat reference sheet with a pre-designed folding crease pattern\textemdash to achieve a desired configuration in 3-D space. We call such designs {\it rigidly foldable} if each panel can rotate along the folding crease lines and remain rigid (without stretch or flexure) during the folding process.  The classical Miura origami pattern \cite{miura_93} is the simplest example of this type, and its generalizations lead to the study of systems of equations that are highly nonlinear and geometrically constrained.  As a result, characterizing global properties of broad classes of origami structures\textemdash such as whether they are rigid, multistable or rigidly foldable\textemdash is a challenge that has attracted significant research interest. One way to study this problem is by using iterative algorithms that enforce a certain topology and foldability \cite{paul_miura, bowers2015lang, dudte2016programming,lh_RFF_2018}.  Another approach is to focus on patterns consistent with a certain symmetry.  

In this work, we follow the symmetry approach to characterize, in a quite general way, Helical Miura Origami (HMO).  These are cylindrical type origami obtained by repeated application of a \textit{helical or rod group} to a partially folded unit cell, which we call a \textit{Miura parallelogram}.  In this procedure, the parameters are kept completely general and on full display, and we are able to address the global problem of closing the cylinder by a straightforward numerical algorithm.  In group theory language ``closing the cylinder'' is ensuring the group is discrete.  As a result, we can completely characterize the phase-space of all HMO, i.e., all cylindrical origami consistent with helical or rod symmetry and the Miura parallelogram as the unit cell.  By exhaustive numerical treatment, we find that HMO are generically rigid to deformations that preserve cylindrical symmetry, but multistable.  This rigidity is not all that surprising; the well-known cylindrical origami are either rigid (for example the Yoshimura pattern \cite{wang2011folding, bos_incompressibility_2016}, Kresling pattern \cite{jianguo2015bistable}) or they lose the cylindrical symmetry while folding \cite{tachi2009generalization}.  Nevertheless, we show that reconfigurability can be achieved.  Inspired by atomistic theory, we discuss two strategies for doing so: one involving \textit{motion by slip} and the other involving \textit{phase transformation}.

\begin{figure}
	\includegraphics[width=\columnwidth]{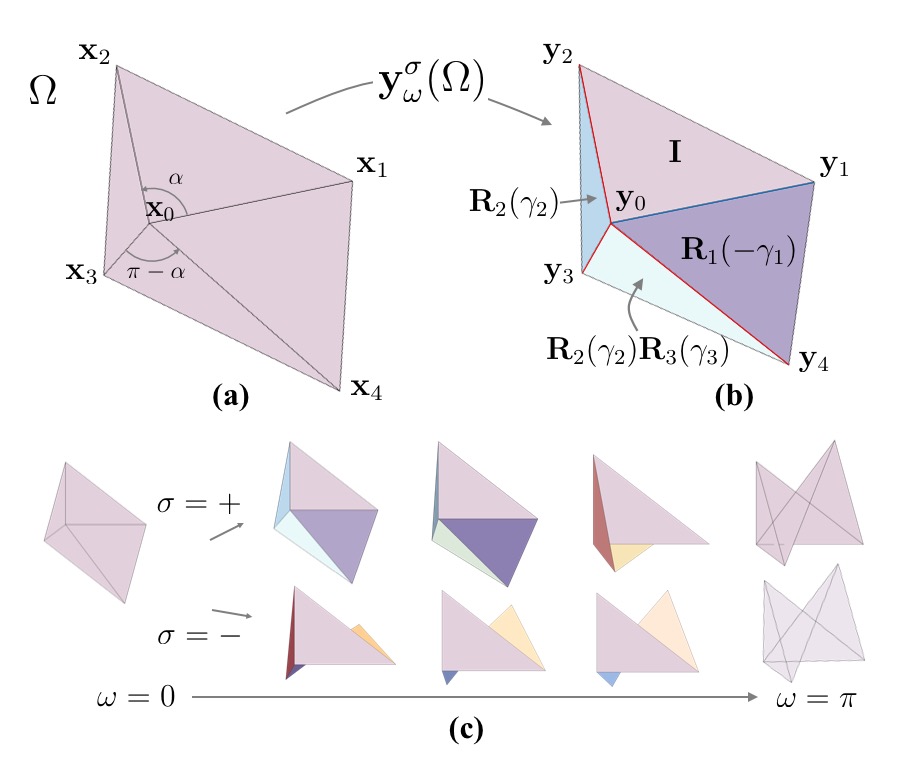}
	\caption{The folding kinematics of Miura parallelograms.  (a) The crease pattern prior to folding.  (b) The kinematics are described by a deformation $\mathbf{y}^{\sigma}_\omega(\Omega)$ whose gradients are as depicted, i.e., rotations of the panels along the creases that are  subject to the compatibility condition (\ref{compat}).   (c) There are exactly two continuous one parameter families which take the Miura parallelogram from flat ($\omega = 0$) to folded flat $(\omega = \pm \pi)$ as rigidly foldable origami. The two branches correspond to distinct mountain-valley assignments, as indicated with $\sigma = \pm$. }
	\label{fig:unit_cell}
\end{figure}

\textbf{Characterization of a Miura parallelogram.} We begin by analyzing the kinematics of a parallelogram with a four-fold vertex satisfying the Kawasaki's condition:\;``opposite sector angles sum to $\pi$".  This defines a Miura parallelogram unit cell (Fig.\;\ref{fig:unit_cell}(a)) with $\angle \bfx_1 \bfx_0 \bfx_2 + \angle \bfx_4 \bfx_0 \bfx_3 = \pi$, where $0 < \angle \bfx_i \bfx_j \bfx_k < \pi $ is the angle between $\bfx_k - \bfx_j$ and $\bfx_i - \bfx_j$ \footnote{Kawasaki condition is necessary and sufficient for the flat and rigid foldability of a four-fold origami.}.  To characterize an isometric (piecewise rigid) folding of this crease pattern, we fix one of the panels by setting $\mathbf{y}(\bfx_{i}) = \mathbf{x}_{i},\ i = 0, 1, 2$,  without loss of generality.  In our notation
$\bfy(\bfx)$ represents the deformation from the flat state in Lagrangian form (see the Supplement).  The kinematics of this pattern (i.e., its deformation gradients) are then described by a composition of $3 \times 3$ rotation matrices  $\mathbf{R}_i(\gamma_i)$ whose axes are tangent to the crease pattern $\mathbf{x}_i - \mathbf{x}_0$ in flat state \footnote{Specifically, $\mathbf{R}_i(\gamma_i)$ is the unique the right-hand rotation of angle $\gamma_i$ that satisfies $\mathbf{R}_i(\gamma_i)(\mathbf{x}_i - \mathbf{x}_0) = \mathbf{x}_i - \mathbf{x}_0$}. Specifically, the necessary and sufficient condition for isometric origami is
\begin{equation}\label{compat}
\mathbf{R}_1(\gamma_1)  \bfR(\gamma_2) \bfR_3(\gamma_3)\bfR_4(\gamma_4) = \mathbf{I}.
\end{equation}
Note, the solutions of this equation describe a folding where panels deform as depicted in Fig.\;\ref{fig:unit_cell}(a-b).  Further, a positive folding angle describes a \textit{valley} and a negative a \textit{mountain} here (red and blue, respectively, in Fig.\;\ref{fig:unit_cell}(b)). 

By solving (\ref{compat}), we derive the full kinematics of the Miura parallelogram.  Generically, the solutions are described by a continuous one-parameter family for which  the four folding angles $(\gamma_1, \gamma_2, \gamma_3, \gamma_4)$ are given by the following expression:
\begin{equation}
\begin{aligned}\label{kin1}
&\gamma_1=-\sigma \bar{\gamma}_3^{\sigma} (\omega),~ \gamma_2 = \sigma \omega,~ \gamma_3=\bar{\gamma}_3^{\sigma} (\omega),~ \gamma_4 = \omega, \\
&\bar{\gamma}_3^{\sigma}(\omega) = \text{sign}\Big((c_{\alpha} - \sigma c_{\beta}) \omega \Big) \arccos\Big(\tfrac{(\sigma 1 - c_{\alpha} c_{\beta})c_{\omega} + s_{\alpha} s_{\beta}}{(\sigma 1 - c_{\alpha} c_{\beta}) + s_{\alpha} s_{\beta} c_{\omega} } \Big), \\
&\sigma \in \mathcal{A} = \begin{cases}
\emptyset & \text{ if } \alpha = \beta = \tfrac{\pi}{2} \\
\{ - \} & \text{ if }  \alpha = \beta \neq \tfrac{\pi}{2} \\
\{ + \} & \text{ if } \alpha = \pi - \beta \neq \tfrac{\pi}{2} \\
\{\pm \} & \text{ otherwise.}
\end{cases}
\end{aligned}
\end{equation}
Here, $\sigma \in \mathcal{A}$ denotes one of the (at most) two branches of solutions corresponding to different mountain-valley crease assignments, the folding angles are parameterized by $-\pi \leq \omega \leq \pi$, and we employ the shorthand notation $c_{\theta} = \cos(\theta)$, $s_{\theta} = \sin(\theta)$, $\alpha = \angle \bfx_1 \bfx_0 \bfx_2$ and $\beta = \angle \bfx_2 \bfx_0 \bfx_3$.   In addition to this generic family, there is a degenerate family of solutions for certain Miura parallelograms; specifically, those characterized by a solution branch $\sigma = +$ or $-$ that does not belong to $\mathcal{A}$.  These cases describe folding-in-half along a single crease:
\begin{align}
\begin{cases}\label{kin2}
\gamma_{1,3} = 0, ~ \gamma_{2,4} = \omega  & \text{ if } \sigma = +, \alpha = \beta \\
\gamma_{1,3} = \omega, ~  \gamma_{2,4} = 0 & \text{ if } \sigma = -, \alpha  = \pi- \beta 
\end{cases}
\end{align}
for the folding parameter $-\pi \leq \omega \leq \pi$.
We provide a brief derivation of these results in the Supplement, and this viewpoint of the kinematics of origami is further developed in \cite{paul_miura}. 

The folding angle parameterizations (\ref{kin1}) and (\ref{kin2}) are sufficient and (with a minor caveat \footnote{Technically, for the Miura parallelogram with crease sector angles as in (\ref{kin2}), we can ``fold-in-half" again. However, these additional kinematically admissible foldings are not relevant to the construction of a HMO.}) necessary for solving (\ref{compat}) and, as such, completely characterize the folding of a Miura parallelogram.  Importantly, the parameterizations highlight two universal features of kinematics: There are two branches of solutions  $\sigma =  \pm$ corresponding to the different mountain-valley crease assignments, and each branch is described by a single folding parameter $\omega$.  Accordingly, the explicit folding deformation  is a continuous piecewise rigid deformation $\mathbf{y}_{\omega}^{\sigma} \colon \Omega \rightarrow \mathbb{R}^3$ with deformation gradients as shown in Fig.\;\ref{fig:unit_cell}(a) for $\gamma_{1,2,3,4} \equiv \gamma_{1,2,3,4}(\omega, \sigma)$ satisfying one of the parameterizations in (\ref{kin1}-\ref{kin2}).  This furnishes the deformed unit cell $\bfy^{\sigma}_{\omega}(\Omega)$ with corner positions $\bfy_i = \bfy^{\sigma}_{\omega}(\bfx_i), i=1,2,3,4$ after folding \footnote{Explicitly, $\mathbf{y}_{1,2} = \mathbf{x}_{1,2}$, $\mathbf{y}_3 = \mathbf{y}_0 +  \mathbf{R}_2(\gamma_2) (\mathbf{x}_3 - \mathbf{x}_0)$ and $\mathbf{y}_4 = \mathbf{y}_0 + \mathbf{R}_1(-\gamma_1)(\mathbf{x}_4 - \mathbf{x}_0)$ under one of the parameterizations in (\ref{kin1}-\ref{kin2}).}.  These observations are known in a different way in the literature on origami, but it will be important for our purposes to write the deformation explicitly.

To construct the HMO, we will apply rotations and translations to $\bfy^{\sigma}_{\omega}(\Omega)$  that map one side of the deformed unit cell to its opposite side.  Thus, as a final point of characterization for the unit cell, we require the parallelogram condition to hold (i.e.,  $|{\bf x}_4-{\bf x}_3|= |{\bf x}_1-{\bf x}_2| $ and $|{\bf x}_1-{\bf x}_4| = |{\bf x}_2-{\bf x}_3|$).   This condition, combined with Kawasaki's condition, constrains the four creases $\mathbf{x}_i - \mathbf{x}_0$.  We parameterize these two conditions up to a trivial rescaling, rotation and translation as follows:  We assume $|\mathbf{x}_2 - \mathbf{x}_1| =  |\mathbf{x}_4- \mathbf{x}_3| = 1$ without loss of generality \footnote{Any Miura parallelogram $\Omega$ can be rescaled by a $\lambda > 0$ such that the four corners of $\lambda \Omega$ satisfy $1 =|\mathbf{x}_2 - \mathbf{x}_1| = |\mathbf{x}_4- \mathbf{x}_3|$}, and we introduce the angle between $\mathbf{x}_2 - \mathbf{x}_1$ and $\mathbf{x}_4 - \mathbf{x}_1$ as $\eta$ (where $0 < \eta < \pi$), and the length $|\mathbf{x}_4 - \mathbf{x}_1| = |\mathbf{x}_3 - \mathbf{x}_2|= l >0$.  This completely parameterizes the boundary of the parallelogram (Fig.\;\ref{fig:reference}(a)).  Additionally, we show in the Supplement that the creases satisfy Kawasaki's condition if and only if the vertex $\mathbf{x}_0$ lies on one of two curves in the interior of the parallelogram pictured in red in Fig.\;\ref{fig:reference}  These curves are parameterized as follows.
\begin{enumerate}
	\item[(i).]  Case  $l > 1$:  The two curves are given by $\bfx_0(\lambda)=\bfx_1 + \lambda (\bfx_2 - \bfx_1) + f^{\pm}(\lambda) (\bfx_4 - \bfx_1)$, where $0 < \lambda < 1$ and the two functions $f^{\pm}$ satisfy
	\beq
	f^{\pm}(\lambda) = \frac{1}{2} \pm \sqrt{\frac{1}{l^2}\big(\lambda - \frac{1}{2}\big)^2 + \frac{1}{l^2} \frac{l^2-1}{4}}.
	\eeq
	\item[(ii).] Case $l < 1$: The two curves are given by $\bfx_0(\lambda)=\bfx_1 + \lambda (\bfx_4 - \bfx_1) + g^{\pm}(\lambda) (\bfx_2 - \bfx_1)$, where $0 < \lambda < 1$ and the two functions $g^{\pm}$ satisfy
	\beq
	g^{\pm}(\lambda) = \frac{1}{2} \pm \sqrt{l^2\big(\lambda - \frac{1}{2}\big)^2 + \frac{1-l^2}{4}}.
	\eeq
	\item[(iii).] Case $l = 1$: The two curves are given by  $\mathbf{x}_0(\lambda) = \lambda \mathbf{x}_2 + (1-\lambda) \mathbf{x}_4$ and $\mathbf{x}_0(\lambda) = \lambda \mathbf{x}_1 + (1-\lambda) \mathbf{x}_3$ for  $0 < \lambda <1$.
\end{enumerate}
As there is an underlying reflection symmetry to this geometry of these curves, we are free to restrict our attention to either the $+$ or $-$ case in (i-iii) without loss of generality.  This fully defines the crease pattern.

\begin{figure}[t!]
	\includegraphics[width=.9\columnwidth]{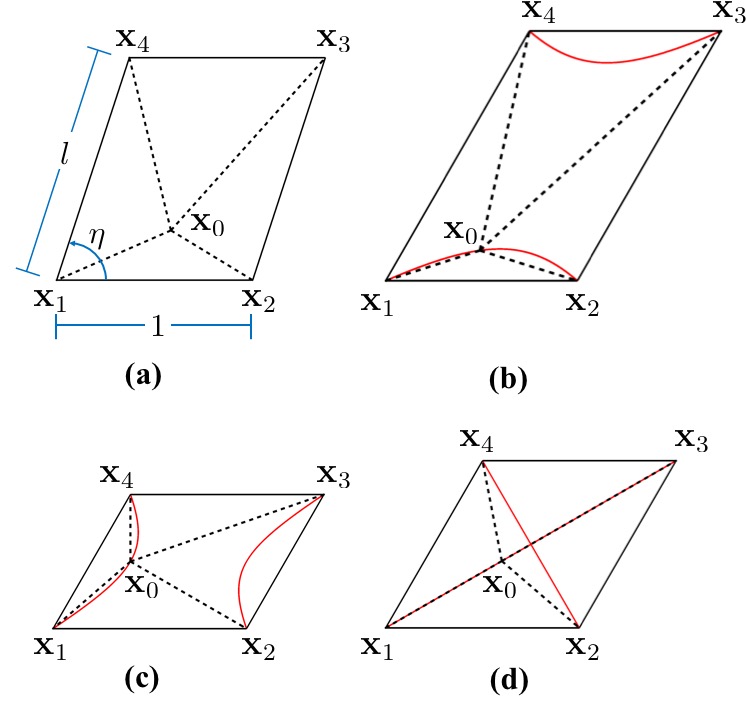}
	\caption{The reference geometry of a Miura parallelogram. (a) The boundary of the parallelogram is parameterized by $\eta$ and $l$ (up to a uniform rescaling).  (b-d) The admissible curves on which the interior vertex satisfies Kawasaki's condition are shown in red.  There are three cases, each parameterized by a $\lambda$ taking values in the interval $(0,1)$.}
	\label{fig:reference}
\end{figure}

To summarize, the parameters $(\eta, l, \lambda,\omega, \sigma)$ given above completely characterize all possible Miura parallelograms\textemdash up to trivial rescaling, translation, overall rigid rotation, and reflection\textemdash and all possible ways of folding origami using these parallelograms.  

\textbf{HMO are objective structures.}  We now define precisely what it means for a structure to be HMO, and we discuss the implications of this definition as it relates to characterizing \textit{all} such structures.  The line of thinking here is based on a systematic and complete characterization of helical and rod symmetry that we developed for an analogous problem: describing all possible phases in nanotubes \cite{feng2019phase}.  To avoid being  redundant, we simply borrow (and state without proof) many ideas from this work that are used in the constructions here.   

Briefly, we define an HMO as any compatible origami structure obtained by a suitable group action $\mathcal{G}$ (see below) on the partially folded Miura parallelogram $\mathbf{y}_{\omega}^{\sigma}(\Omega)$. The groups we consider are discrete, Abelian (i.e., the elements commute), contain only isometries, and have an \textit{orbit} $\{ g(\mathbf{x}) \colon g \in \mathcal{G}\}$ for each point $\mathbf{x} \in \mathbb{R}^3$ that gives a collection of points that all lie on a cylinder.  (The cylinders can be different for different choices of $\bfx$.)  This means that every Miura parallelogram in the structure  ``sees the same environment" \cite{james_JMPS_06}, which is the natural generalization of periodicity to cylindrical origami. 

\begin{figure*}
	\includegraphics[width= \textwidth]{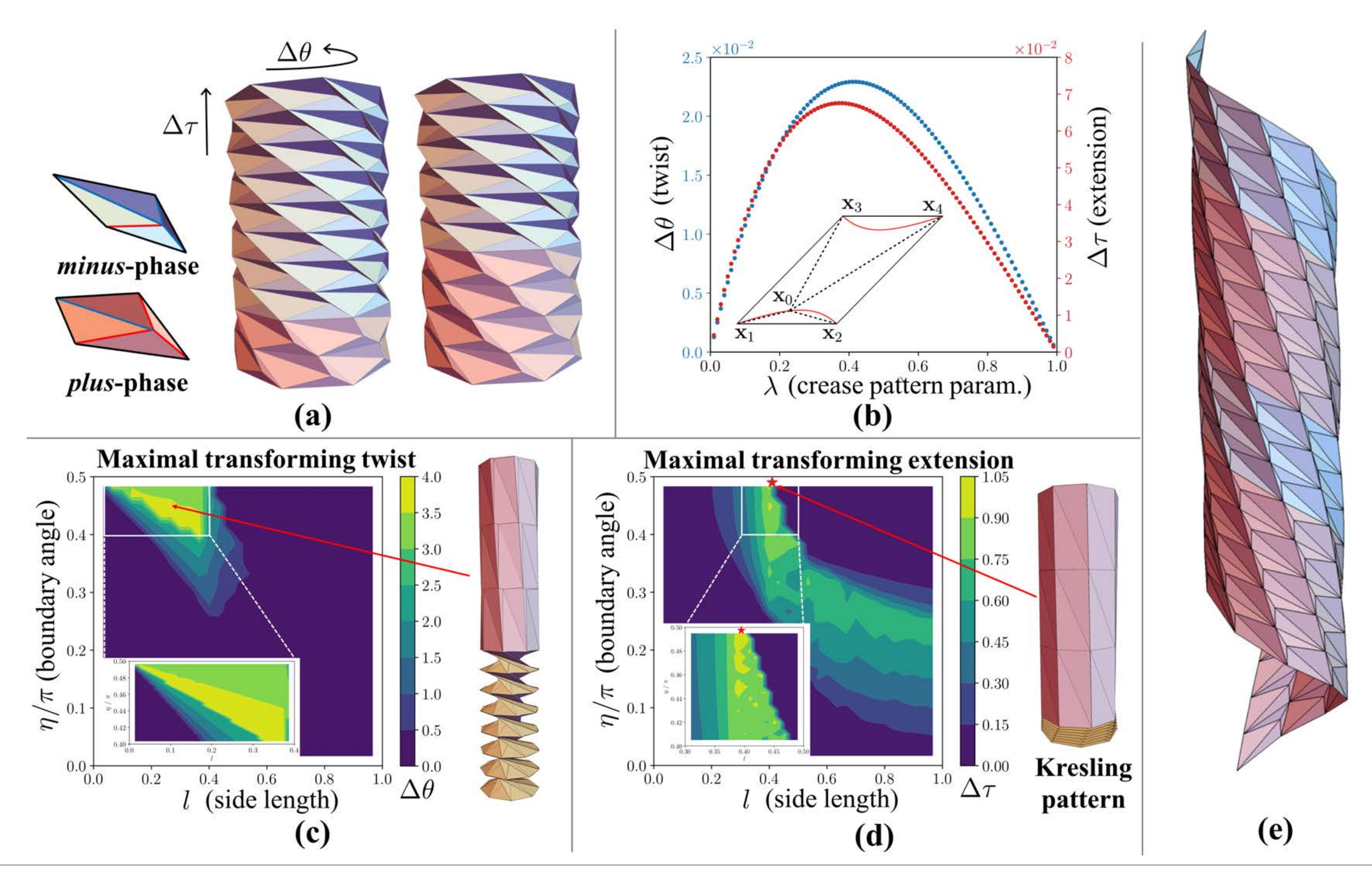}
	\caption{Phase transforming HMO. (a) Overall twist and extension obtained by transformation via a compatible horizontal interface. (b) . The twist $\Delta \theta$ and extension $\Delta \tau$ with $(l, \eta) = (1.5, \pi/4)$ for $0 < \lambda<1$.  (c-d) The maximal twist and extension as a function of the reference parameters $(l,\eta, \lambda)$. Each point in the phase space indicates the twist and extension after maximizing with respect to $\lambda$.  For a more detailed diagram that highlights the $\lambda$ dependence, see the supplemental.  (e) An example of cylindrical origami with helical interfaces.}
	\label{fig:PhaseTran}
\end{figure*}

An \textit{isometry} is simply a map $g = (\mathbf{R}| \mathbf{c})$ defined by $g(\mathbf{x}) = \mathbf{R} \mathbf{x} + \mathbf{c}$, where $\mathbf{R}$ is a $3 \times 3$ orthogonal matrix and $\mathbf{c} \in \mathbb{R}^3$.  Below, we use O(3) to denote the $3 \times 3$ orthogonal matrices, and SO(3) to denote rotations (i.e., the subset of O(3) with determinant $+1$).  One can multiply isometries 
$g_1 = (\mathbf{R}_1| \mathbf{c}_1)$ and $g_2 = (\mathbf{R}_2| \mathbf{c}_2)$ using the standard rule $g_1 g_2 = (\bfR_1 \bfR_2 | \bfc_1 + \bfR_1 \bfc_2)$.  Under this rule, the collection of all isometries is a group, and it has many subgroups.  Thus, one might worry that the aforementioned\textemdash and rather general\textemdash family of groups lacks meaningful structure.  Strikingly though (and this is made precise in \cite{feng2019phase}), discrete and Abelian isometry groups subject to the stated cylinder condition are quite restrictive.  They must be described as the product of powers of two generators on the set of pairs of integers $\mathbb{Z}^2$, i.e.,  
\begin{align}\label{gGroup}
\mathcal{G} = \{ g_1^{p} g_2^q \colon (p, q) \in \mathbb{Z}^2 \},
\end{align} 
in which the {\it generators} of the group $g_1$ and $g_2$ are two screw isometries
\begin{align}\label{param1}
g_i = (\bfR_{\theta_i} | \tau_i \bfe  + (\bfI - \bfR_{\theta_i})\bfz),\;\;~  i = 1,2,
\end{align}
with parameters $\mathbf{R}_{\theta_i} \in$ SO(3), $-\pi < \theta_{i} \leq \pi$, $\tau_{i} \in \mathbb{R}$, $\mathbf{e} \in \mathbb{R}^3$, $|\mathbf{e}| = 1$, and $\mathbf{z} \in \mathbb{R}^3$, $\mathbf{z} \cdot \mathbf{e} =0$ characterizing the rotation, rotation angle, translation, rotation axis and origin of the isometry, respectively.  These parameters are subject to a \textit{discreteness condition} 
\begin{align}
\begin{cases}\label{param2}
p^{\star} \tau_1 + q^{\star} \tau_2 = 0  \\
p^{\star} \theta_1 + q^{\star} \theta_2 = 2 \pi 
\end{cases}
\end{align}
for some pair of integers $ (p^{\star}, q^{\star}) \in \mathbb{Z}^2$.   Technically, we should also enforce $\tau_1^2 + \tau_2^2 > 0$, as the violation of this condition results in a flattened ring rather than a cylinder.  However, we avoid this restriction since the flattened ring is of technological interest: the folded flat portion of the Kresling pattern in Fig.\;\ref{fig:PhaseTran}(d) is one example.

Finally, we should point out the groups $\mathcal{G}$ in (\ref{gGroup}) are not uniquely described by a single parameterization satisfying (\ref{param1}-\ref{param2}). This should not be unexpected.  In periodic structures, there are many equivalent choices of lattice vectors which generate the same lattice.  In fact, the degeneracy here\textemdash much like the 2-D lattice\textemdash is fully characterized by a $\boldsymbol{\mu} \in GL(\mathbb{Z}^2)$. Here, $GL(\mathbb{Z}^2)$ is the set of $2 \times 2$ matrices with integer entries and determinant $\pm 1$.  That is, for any group $\mathcal{G}$ satisfying (\ref{gGroup}-\ref{param2}), we can replace the parameters by a linear transformation $(\theta_1,\theta_2) \mapsto \boldsymbol{\mu}(\theta_1, \theta_2)$, $(\tau_1, \tau_2) \mapsto \boldsymbol{\mu}(\tau_1, \tau_2)$, and $(p^{\star}, q^{\star}) \mapsto \boldsymbol{\mu}^{-T} (p^{\star}, q^{\star})$ and generate the same structure (i.e., $\mathcal{G} = \mathcal{G}_{\boldsymbol{\mu}}$) if $\boldsymbol{\mu} \in GL(\mathbb{Z}^2)$.  

Thus, the aim in what follows is to characterize the sets of parameters $(\eta,l, \lambda, \omega, \sigma, \theta_{1,2}, \tau_{1,2}, \mathbf{e}, \mathbf{z}, p^{\star}, q^{\star})$ (up to this trivial degeneracy $\boldsymbol{\mu} \in GL(\mathbb{Z}^2))$ that lead to a fully compatible cylindrical origami structure.  This then captures the phase-space of \textit{all} HMO.

\begin{figure*}
	\includegraphics[width=\textwidth]{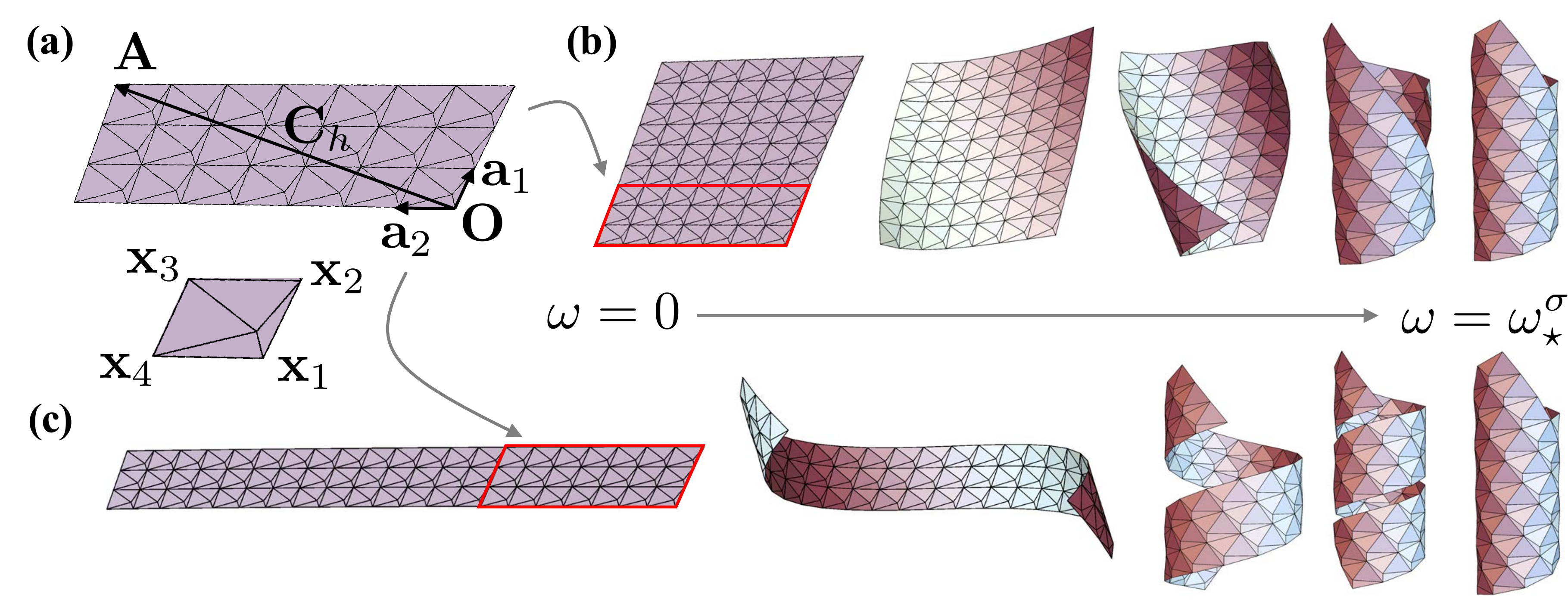}
	\caption{Constructing an HMO from a flat tessellation.  (a) Each HMO is characterized by a  chiral vector $\mathbf{C}_h = \protect\overrightarrow{\bf OA} = p^\star \bfa_1 + q^\star \bfa_2$, which is a linear combination of the side length vectors $\mathbf{a}_{1} = \mathbf{x}_2 - \mathbf{x}_1$ and $\mathbf{a}_2 = \mathbf{x}_4 - \mathbf{x}_1$. This describes a $(p^{\star}, q^{\star})$ building block for the structure.  (b-c) In extending this building block to a tessellated strip in the ``vertical" (b) or ``horizontal" (c) direction, an HMO structure is achieved by a rigidly foldable motion that connects the points $\mathbf{O}$ and $\mathbf{A}$ at  $\omega = \omega_{\star}^{\sigma}$ (i.e., the angle(s) solving (\ref{finalpq})).}
	\label{fig:chirality}
\end{figure*}

\textbf{Design equations for HMO.}  These are obtained systematically by satisfying all compatibility conditions, i.e., the conditions under which the folded tiles of the structure fit together
perfectly without gaps.  For developing these ideas, we find it convenient to introduce the tessellation $\mathcal{T}\Omega = \{ g(\Omega) \colon g \in \mathcal{T} \}$ for the translation group $\mathcal{T} = \{ t_1^{p} t_2^q \colon (p,q) \in \mathbb{Z}^2 \}$ such that $t_1= (\mathbf{I}|\mathbf{x}_1 - \mathbf{x}_4)$ and $t_2 = (\mathbf{I}|\mathbf{x}_2 - \mathbf{x}_1)$.   Since 
\begin{equation}
\begin{aligned}\label{tesselate}
t_1(\mathbf{x}_4) = \mathbf{x}_1, \quad t_2(\mathbf{x}_3) = \mathbf{x}_2, \\
t_2(\mathbf{x}_1) = \mathbf{x}_2, \quad t_2(\mathbf{x}_4) = \mathbf{x}_3,
\end{aligned}
\end{equation} 
this gives a tessellated plane in $\mathbb{R}^3$ of Miura parallelograms prior to folding.  Suitably defined strips of this tessellation will be used to construct the HMO from this easy-to-manufacture flat state (e.g., Fig\;\ref{fig:chirality}).

  We begin with \textit{local compatibility}: Consider a partially folded Miura parallelogram $\mathbf{y}_{\omega}^{\sigma}(\Omega)$ with it corners denoted as $\mathbf{y}_{i}$, $i =1,2,3,4$ (Fig.\;\ref{fig:unit_cell}(a)), consider a group $\mathcal{G}$ satisfying (\ref{gGroup}-\ref{param2}), and consider the structure $\mathcal{G}\mathbf{y}_{\omega}^{\sigma}(\Omega) = \{ g_1^{p}g_2^{q}(\mathbf{y}_{\omega}^{\sigma}(\Omega)) \colon (p,q) \in \mathbb{Z}^2\}$.  The nearest neighbors to $\mathbf{y}_{\omega}^{\sigma}(\Omega)$ on the structure are, therefore, obtained by the application of group elements to this domain.  Without loss of generality \footnote{As this assumption is equivalent to alleviating the degeneracy discussed above by fixing a $\boldsymbol{\mu} \in GL(\mathbb{Z}^2)$.}, we assume the neighbor to the``left" of the unit cell is $g_1(\mathbf{y}_{\omega}^{\sigma}(\Omega))$ and the neighbor ``above" the unit cell is $g_2(\mathbf{y}_{\omega}^{\sigma}(\Omega))$.    Then, one condition of compatibility is that the unit cell is connected to its neighbors; particularly, to its neighbor on the left along the line $\ell_{23} =  \{ \delta \mathbf{y}_2 + (1-\delta) \mathbf{y}_3 \colon 0 \leq \delta \leq 1 \}$ and to its neighbor up above along the line $\ell_{12} = \{ \delta \mathbf{y}_1 + (1-\delta) \mathbf{y}_2 \colon 0 \leq \delta \leq 1 \}$.  This gives four restrictions on the group elements:
\begin{equation}
\begin{aligned}\label{localCompat}
g_1(\mathbf{y}_4) = \mathbf{y}_1,\quad g_1(\mathbf{y}_3) = \mathbf{y}_2, \\
g_2(\mathbf{y}_1) = \mathbf{y}_2,\quad  g_2(\mathbf{y}_4) = \mathbf{y}_3,
\end{aligned} 
\end{equation}
which we term local compatibility. 

The reason for the terminology is that (\ref{localCompat}) is a discrete and symmetry-related version of the local curl-free and jump \textit{compatibility} conditions that indicate whether a prescribed deformation gradient can describe a continuous deformation on a simply connected domain.  Indeed, $g_1$ and $g_2$ are commutative  (i.e., $g_1 g_2 = g_2 g_1$) under the multiplication rule $g_1 g_2 (\bfx) = g_1(g_2(\bfx))$.  This means that $g_1$ and $g_2$ satisfy the loop condition $g_1g_2(\bfy_i) = g_2g_1(\bfy_i)$. As a result, the four nearest neighbor Miura parallelograms $\bfy_{\omega}^{\sigma}(\Omega)$, $g_1(\bfy_{\omega}^{\sigma}(\Omega))$, $g_2(\bfy_{\omega}^{\sigma}(\Omega))$ and $g_1g_2(\bfy_{\omega}^{\sigma}(\Omega)) = g_2g_1(\bfy_{\omega}^{\sigma}(\Omega))$ fit together automatically whenever (\ref{localCompat}) holds.  Combining (\ref{tesselate}) and (\ref{localCompat}), it then follows that the induced deformation 
given by 
\begin{align}\label{flatToCyl}
\mathbf{y}(t_1^{p} t_2^{q}(\mathbf{x})) = g_1^p g_2^{q}(\mathbf{y}_{\omega}^{\sigma}(\mathbf{x})), ~ \forall ~ \mathbf{x} \in \overline{\Omega}, ~ (p, q) \in \mathbb{Z}^2
\end{align} 
is a continuous isometric origami deformation that maps the tessellated plane to the origami structure.  This is the key advantage of bringing out the group structure: simply solve the four
equations (\ref{localCompat}), and the entire structure fits together perfectly without  gaps (\ref{flatToCyl}).


In the Supplement, we solve the conditions of local compatibility  (\ref{localCompat}) explicitly. To explain the parameterization obtained, we assume the Miura parallelogram is partially folded (i.e., $-\pi < \omega < \pi$ and $\omega \neq 0$), and we let the side length vectors be given by $\mathbf{u}_a = \mathbf{y}_3 - \mathbf{y}_4$, $\mathbf{u}_b = \mathbf{y}_2 - \mathbf{y}_1$, $\mathbf{v}_a = \mathbf{y}_1 -\mathbf{y}_4$ and $\mathbf{v}_b = \mathbf{y}_2 - \mathbf{y}_3$.   We can then always define the right-hand orthonormal frame $\{\mathbf{f}_1, \mathbf{f}_2, \mathbf{f}_3 \}$,
\begin{align}
\mathbf{f}_1 = \frac{\mathbf{u}_a + \mathbf{u}_b}{|\mathbf{u}_a + \mathbf{u}_b|}, ~ \mathbf{f}_2 = \frac{\mathbf{u}_a \times \mathbf{u}_b}{|\mathbf{u}_a \times \mathbf{u}_b|}, ~ \mathbf{f}_3 = \frac{\mathbf{u}_a - \mathbf{u}_b}{|\mathbf{u}_a - \mathbf{u}_b|}
\end{align}
(since $\omega \neq  0, \pm \pi$ implies $\mathbf{u}_{a} \times \mathbf{u}_b \neq 0$).  The necessary and sufficient conditions for local compatibility in this setting are thus
\begin{equation}
\begin{aligned}\label{paramLocal}
&\mathbf{e} \equiv \mathbf{e}^{\sigma}(\omega, \varphi) = c_{\varphi} \mathbf{f}_1  + s_{\varphi} \mathbf{f}_2, \\
&\theta_1 \equiv \theta_1^{\sigma}(\omega, \varphi) = \text{sign}( \mathbf{e} \cdot (\mathbf{u}_a \times \mathbf{u}_b)) \arccos\Big( \frac{\mathbf{u}_a \cdot \mathbf{P}_{\mathbf{e}} \mathbf{u}_b}{|\mathbf{P}_{\mathbf{e}} \mathbf{u}_a|^2} \Big), \\
&\theta_2 \equiv \theta_2^{\sigma}(\omega, \varphi) = \text{sign}( \mathbf{e} \cdot (\mathbf{v}_a \times \mathbf{v}_b)) \arccos\Big( \frac{\mathbf{v}_a \cdot \mathbf{P}_{\mathbf{e}} \mathbf{v}_b}{|\mathbf{P}_{\mathbf{e}} \mathbf{v}_a|^2} \Big), \\
&\tau_1 \equiv \tau_1^{\sigma}(\omega, \varphi) = \mathbf{e} \cdot \mathbf{v}_a, \quad \tau_2 \equiv \tau_2^{\sigma}(\omega, \varphi) = \mathbf{e} \cdot \mathbf{u}_a,  \\
&\mathbf{z} \equiv \mathbf{z}^{\sigma}(\omega, \varphi)  = (\mathbf{I} - \mathbf{R}_{\theta_1} + \mathbf{e} \otimes \mathbf{e})^{-1} \mathbf{P}_{\mathbf{e}} (\mathbf{y}_2 - \mathbf{R}_{\theta_1} \mathbf{y}_3),
\end{aligned}
\end{equation}
where $\mathbf{P}_{\mathbf{e}} = \mathbf{I} - \mathbf{e} \otimes \mathbf{e}$ denotes the linear transformation that projects vectors onto the plane with normal $\mathbf{e}$ and  the angle $-\pi/2 < \varphi \leq \pi/2$ (describing the axis $\mathbf{e}$) is a free parameter \footnote{Two points: 1).\;Here, we define the sign function as $\text{sign}(t) = +1$ if $t \geq 0$ and $-1$ if  $t < 0$.  Note $\text{sign}(0) = 1$. 2).\;Technically, $\varphi \in (-\pi, \pi]$ is the necessary condition, not $\varphi \in (-\pi/2,\pi/2]$.  However, there is an additional degeneracy related to the axis: if $\mathbf{e} \mapsto - \mathbf{e}$ (i.e., $\varphi \mapsto \varphi + \pi$), then $(\theta_{1,2}, \tau_{1,2}, \mathbf{z}) \mapsto (-\theta_{1,2}, -\tau_{1,2}, \mathbf{z})$ due to the formulas (\ref{paramLocal}), but this does not change the generators $g_{1}$ and $g_2$.}.  For completeness, note that the fully folded cases $\omega = \pm \pi$ and fully unfolded case $\omega=0$ are of course included (see the Supplement).  For definiteness, we will focus on the solutions governed by (\ref{paramLocal}).   Examples of the exceptional cases are  fully degenerate cylinders ($\omega = 0$) or flattened ring structures $(\omega = \pi)$. 

Importantly, the corners $\mathbf{y}_i$, $i =1,2,3,4$ (and thereby $\mathbf{u}_{a,b}, \mathbf{v}_{a,b},$ and $\mathbf{f}_{1,2,3}$ above) depend only on the folding parameter $\omega$ and the mountain-valley assignment $\sigma$.  
Consequently,  after satisfying local compatibility, the kinematic freedom in (\ref{paramLocal})
is  
$-\pi < \omega < \pi,\ \omega \ne 0,\ -\pi/2 < \varphi \le \pi/2$ and  $\sigma = \pm$.  We utilize this freedom to solve the discreteness condition (\ref{param2}).  This, in turn, is equivalent to \textit{closing the cylinder}; see Fig.\;\ref{fig:chirality}.

Indeed, given integers $(p^{\star}, q^{\star})$, not both zero,  we observe that the discreteness condition (i.e., $p^{\star} \tau_1 + q^{\star} \tau_2 = 0$)  uniquely determines  the angle $\varphi$ under the parameterization (\ref{paramLocal}).  The explicit form is
\begin{align}\label{varphiDef}
\varphi \equiv \varphi_{\star}^{\sigma}(\omega) = \arctan \Big(\frac{-\mathbf{f}_1\cdot (p^{\star} \mathbf{v}_a + q^{\star} \mathbf{u}_a)}{\mathbf{f}_2 \cdot (p^{\star} \mathbf{v}_a + q^{\star} \mathbf{u}_a)}\Big).
\end{align}
(Note, $\varphi \equiv \varphi_{\star}^{\sigma}(\omega) = \pi/2$ if $\mathbf{f}_2 \cdot (p^{\star} \mathbf{v}_a + q^{\star} \mathbf{u}_a)=0$, and $p^{\star} \mathbf{v}_a + q^{\star} \mathbf{u}_a$ is never parallel to $\mathbf{f}_3$ for $-\pi < \omega < \pi, \omega \ne 0$.  So the parameterization is always well-defined.)  We then substitute (\ref{varphiDef})  into (\ref{paramLocal}), to get the final form of the discreteness condition, 
\begin{align}\label{finalpq}
p^{\star} \theta^{\sigma}_1(\omega, \varphi_{\star}^{\sigma}(\omega)) + q^{\star} \theta_2^{\sigma}(\omega, \varphi_{\star}^{\sigma}(\omega)) = 2\pi
\end{align}
which is to be solved for $\omega$.  This we evaluate numerically by cycling through the folding parameter $\omega \ne 0, -\pi < \omega < \pi$.  The solutions then correspond to parameters that give a HMO structure.  Specifically, consider the \textit{chiral vector} $\bfC_h = p^{\star} \mathbf{a}_1 + q^{\star} \mathbf{a}_2$ (see Fig.\;\ref{fig:chirality}(a)), i.e.,  the widely used descriptor of chirality in carbon nanotubes \cite{dresselhaus1995physics, james_icm}.  Upon substituting $\varphi \equiv \varphi_{\star}^{\sigma}(\omega)$ into the group parameters (\ref{paramLocal}), and using these to generate an origami structure (\ref{flatToCyl}) from the flat tessellation, we make the striking observation related to $\mathbf{C}_h$: as $\omega$ monotonically increases (or decreases) from zero, the structure is simply ``rolling up" as rigidly foldable origami, with the line traced by $\mathbf{C}_h$ deforming effectively as a singly curved arc.  Further, the points of $\omega$ at which $\mathbf{O}$ and $\mathbf{A}$ connect perfectly during this rolling up process are exactly the points $\omega = \omega_{\star}^{\sigma}$ such that (\ref{finalpq}) holds.  Finally, because of the underlying symmetry of the group $\mathcal{G}$, the boundaries of a suitable $\mathbf{C}_h$ tessellated strip (Fig.\;\ref{fig:chirality}(b-c)) connect perfectly if and only if (\ref{finalpq}) holds.  Thus, (\ref{finalpq}) is the necessary and sufficient condition on the kinematic parameters for closing the cylinder and generating a HMO with $\mathbf{C}_h$ chirality.

\begin{figure*}
	\includegraphics[width=\textwidth]{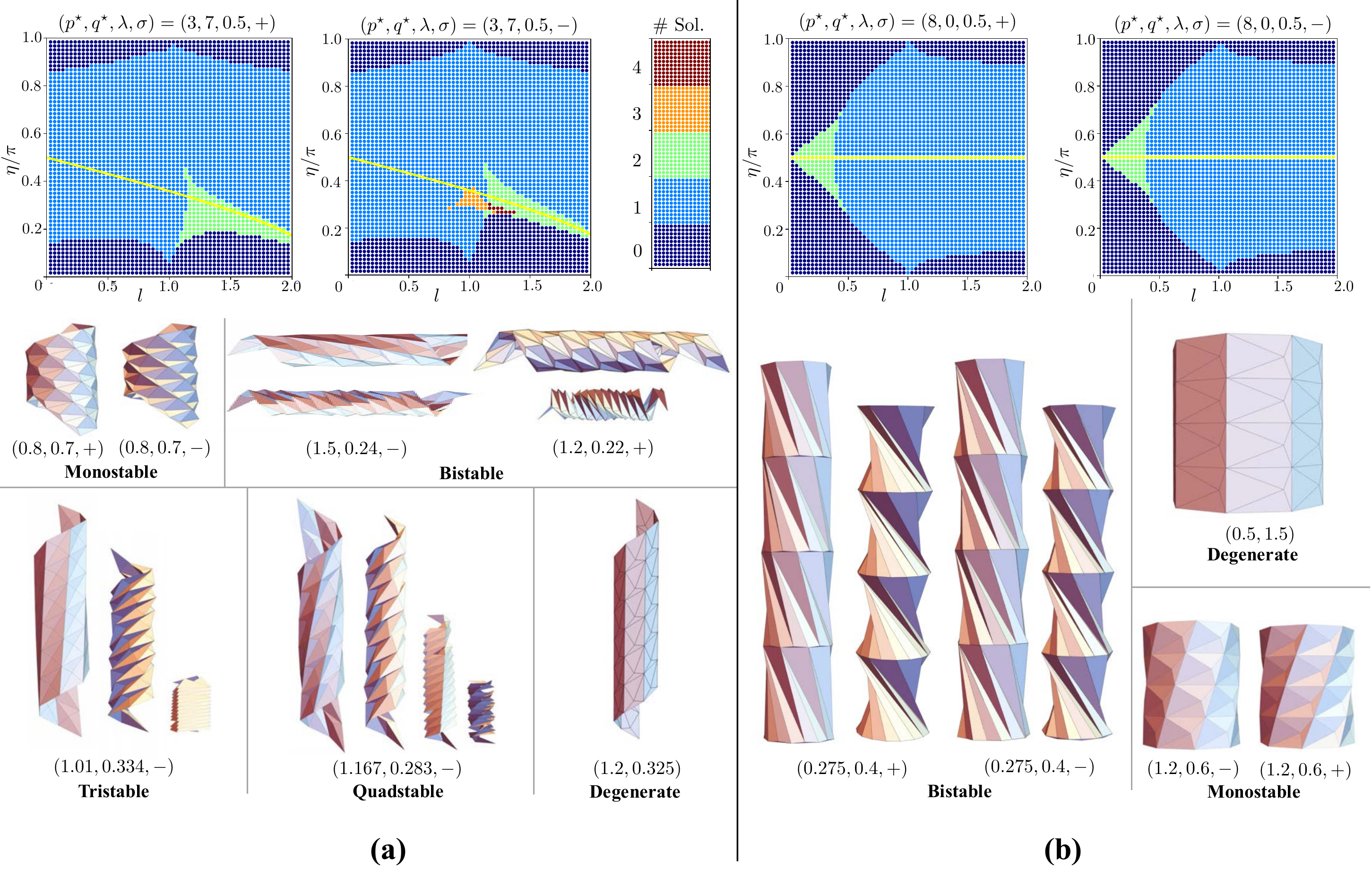}
	\caption{Phase diagrams of HMO structures parameterized by the three parameters $(l, \eta, \lambda)$ of the Miura parallelogram unit cell, as well as the mountain-valley assignment $\sigma \in \{ \pm \}$. The slices of the phase diagrams displayed are taken at $\lambda = 1/2$.  (a) The discreteness is given by 
$(p^{\star}, q^{\star}) = (3,7)$ in this case, and the coloring scheme is in accordance to the number of HMO solutions: 0 (purple), 1 (light blue), 2 (green), 3 (orange) and  4 (red). The curve in yellow indicates a degenerate HMO that is achieved for a fully unfolded Miura parallelogram.  As such, it is independent of $\lambda$ since the creases are not being utilized.  Further, it always corresponds to a vertical interface.  The examples below the diagrams give HMO in the different regimes of stability and are parameterized by $(l, \eta, \sigma)$, as shown.  (b) An example of ring-like structures given by fixing the discreteness $(p^{\star} ,q^{\star}) = (8,0)$.  The characterization is similar.}
	\label{fig:AllCases}
\end{figure*}

\textbf{The phase-space of HMO.}  The design equations above lead to a comprehensive and explicit recipe to  determine all HMO: 
\begin{enumerate}
\item Fix the reference geometry of the Miura parallelogram and chirality by assigning $(\eta, l, \lambda)$ as $0 < \eta < \pi$, $l > 0$ and $0 < \lambda < 1$, and by assigning a non-zero pair of integers $(p^{\star}, q^{\star})$. 
\item Assign the group parameters by the design equations in (\ref{paramLocal}-\ref{varphiDef}).
\item Cycle through the folding parameter $-\pi <  \omega < \pi$, $\omega \neq 0$, and solution branch $\sigma = \pm$ to numerically compute all solutions $\omega = \omega_{\star}^{\sigma}$ to the final design equation (\ref{finalpq}).  This closes the cylinder and generates a HMO with the parameters $\eta, l, \lambda, p^{\star}, q^{\star}$ and group parameters (\ref{paramLocal}-\ref{varphiDef}) for $\omega = \omega_{\star}^{\sigma}$.
\item Cycle through the reference geometry $(\eta, l ,\lambda)$ and chirality $(p^{\star}, q^{\star})$ in Step 1 and repeat Steps 2 and 3 for each case to determine all HMO structures.  
\end{enumerate}

A natural design tool for HMO solutions is to fix the {\it discreteness} $(p^{\star}, q^{\star})$ and cycle through reference parameters $(\eta, l , \lambda)$ to create a three dimensional \textit{phase diagram}.  In Fig.\;\ref{fig:AllCases} below, we present two examples for illustrative purposes.  These describe two-dimensional slices of such phase diagrams at $\lambda = 1/2$ \footnote{Additional slices of the phase diagrams at $\lambda = 0.2,0.8$ are provided in the Supplement for these cases.}.  In the diagrams, the coloring scheme is in accordance to the number of HMO configurations (solutions to (\ref{finalpq})) for a fixed mountain-valley assignment ($\sigma = +$ for the left diagram and $-$ for the right diagram; Fig.\;\ref{fig:AllCases}(a) and (b), respectively). We also highlight examples of HMO in each of the respective regions.  

The first phase diagram  (Fig.\;\ref{fig:AllCases}(a)) is a generic helical case in which  the discreteness is $(p^{\star}, q^{\star})= (3,7)$.  Notice the reference parameters $(l,\eta, \lambda)$ typically furnish a single HMO solution for a fixed mountain-valley assignment (light blue).  However, there are regions with no solutions (purple), and regions of multistability.  The $\sigma = -$ case is particularly interesting, as it has regions of bistability (green), tristability (orange) and quadstability (red).  This is quite striking: by carefully designing reference parameters in the multistable regimes, the HMO achieved by such design can transform from one stable state to another by stress-induce twist and contraction (or expansion). As evidenced by the examples, the induced deformation for such transformation can be quite dramatic; for instance, the displayed tristable HMO, when deformed from its most unfolded state (denoted by folding parameter $\omega_{1}$) to its most folded state (denoted $\omega_3$), contracts by a factor $r_{\omega_3}/r_{\omega_1} \approx 0.7$ along its radius and by a factor $L_{\omega_3}/L_{\omega_1} \approx 0.2$ along its axial length.  It also experiences $\approx 470$ degrees of twist under this transformation.

The second phase diagram  (Fig.\;\ref{fig:AllCases}(b)) is for discreteness $(p^{\star}, q^{\star}) = (8,0)$. This characterizes ring-type HMO described by a closed ring of $8$ Miura parallelograms repeated along the axis $\mathbf{e}$ in a periodic fashion. We again see that the reference parameters $(l,\eta, \lambda)$ typically furnish a single HMO solution for a given mountain-valley assignment (light blue), but there are also regions of bistability (green).  Interestingly, transformation between the two stable states in the bistable regime of parameters induces axial contraction (expansion) and twist, but no change in the radius.  This means that each ring layer can be transformed independently to form a structure which is a mixture of the two different HMO states. This stands in stark contrast to the generic case $p^{\star},q^{\star} \neq 0$, where the entire structure must fully participate in the transformation from one state to the other \footnote{We will make this notion precise below with the forthcoming discussion on \textit{phase transformations}.}.

Importantly, our exhaustive numerical treatment beyond these examples suggests there are no parameters for which HMO structures exhibit rigidly-foldable motions that preserve helical symmetry.  However, large regions of multistability appear to be ubiquitous.  

As a final comment before shifting viewpoints, we note that one drawback of this design procedure is it does not take into account self-intersection: it is actually possible for the Miura parallelogram $\mathbf{y}_\omega^{\sigma}(\Omega)$ and one of its neighbors $g_{1,2}(\mathbf{y}_{\omega}^{\sigma}(\Omega))$ to overlap in an unphysical way at large values of $|\omega|$ and still solve the condition (\ref{finalpq}).   One such example of self-intersection is the fourth and most folded configuration (Fig.\;\ref{fig:AllCases}(a), the quadstable case).  We did not exclude self-intersection in the phase diagram, as it is far too numerically laborious to do so while simultaneously exploring large regions of the configuration space.  So this procedure does, in some cases, overestimate the number of stable HMO states for given set of reference parameters.  Nevertheless, these self-intersecting configurations may be relevant\textemdash in the sense that, mechanistically, they suggest the possible existence of a stressed but stable mechanical equilibrium described by a tubular structure with the panels in direct contact.

\begin{figure*}
	\includegraphics[width=\textwidth]{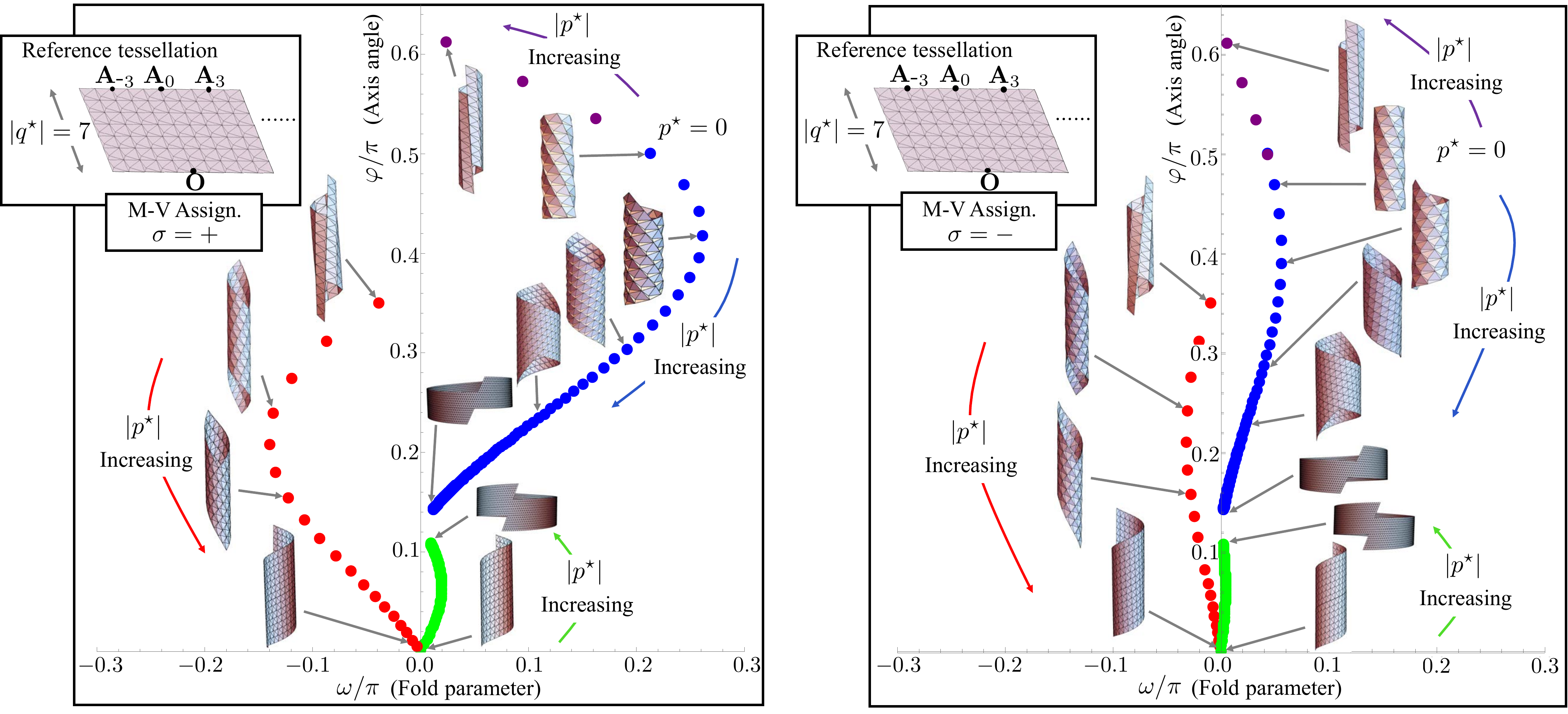}
	\caption{A diagram describing all the ways the reference tessellation, i.e., an infinite strip with width $|q^{\star}| = 7$ Miura parallelogram in the top left corner, can be rolled up to form a HMO.  These solve (\ref{varphiDef}) and (\ref{finalpq}) for $p^{\star} = 0,\ldots, 150$ and $q^{\star} =  7$ in blue, $p^\star = 4, \ldots, 22$ and  $q^{\star} =-7$ in red, and $p^{\star} = 23, \ldots, 150$ and $q^{\star} =  -7$ in green.  The purple solutions depicted are obtained  differently for aesthetic reasons: First, we solve (\ref{varphiDef}) and (\ref{finalpq}) for $p^{\star} = 1,2,3$ and $q^{\star} =  -7$.  This yields the axis orientation $\varphi = \varphi_{\star}^{\sigma}(\omega_{\star}^{\sigma}) < 0$.  We then notice that, by replacing the axis orientation with $\varphi =  \varphi_{\star}^{\sigma}(\omega_{\star}^{\sigma}) + \pi$ and the discreetness $(p^{\star}, q^{\star})$ with its minus, we achieve exactly the same structure.  For all these solutions (blue/purple/red/green), there are also trivial solutions of mirrored chirality which correspond to $(p^{\star}, q^{\star}, \omega, \varphi) \mapsto - (p^{\star}, q^{\star}, \omega, \varphi)$.   We did not bother to plot these, as they do not provide any additional information.  Finally, we did not calculate the solutions beyond $|p^{\star}| = 150$ since the accumulation points along the blue and green trajectories as $|p^{\star}| \rightarrow \infty$ are evident. }
	\label{fig:Rigidity1}
\end{figure*}

Now, an alternative way to view this phase-space is to fix a tessellated strip and classify all the HMO that can be obtained from this strip by the ``rolling up" process (Fig.\;\ref{fig:chirality}).  For example, consider the tessellation in the top-left corner of Fig.\;\ref{fig:Rigidity1}. This has a width of $|q^{\star}| = 7$ Miura parallelograms that are repeated along the length the strip.  The boundaries of this tessellation can fit together to form a HMO in different ways; particularly, in all the ways $(p^{\star}, \pm 7)$ for $p^{\star} \in \mathbb{Z}$ that solve the discreteness conditions in (\ref{varphiDef}) and (\ref{finalpq}).  This corresponds to different points $\mathbf{A}_i$ on the boundary that connect to $\mathbf{O}$ on the opposite boundary. 

To clarify this viewpoint, we have completely evaluated the phase-space\textemdash in this particular sense\textemdash for the tessellation shown in Fig.\;\ref{fig:chirality}.  Strikingly, this tessellation admits a HMO solution for \textit{all} $(p^{\star}, \pm 7)$ and $\sigma \in \{ \pm \}$, meaning that it can be isometrically rolled up to form a HMO of arbitrary chirality for either choice of mountain-valley assignment.  This is highlighted graphically in Fig. \ref{fig:Rigidity1}.  Starting at $p^{\star} = 0$, the solutions for increasing integer values of $|p^{\star}|$ are plotted in the $(\omega, \varphi)$ phase-space. Physically, this integer increase describes a shift in the structure by one Miura parallelogram along the helical interface corresponding to the boundary of the tessellation\textemdash a process completely encapsulated by a change in both the folding parameter  $\omega$ of the Miura parallelogram and axis orientation $\varphi$.    We find it instructive to elaborate on this diagram in detail, as it is a natural lead-in to a mechanism for reconfigurability in HMO structures.

Briefly: For the trajectory describing increasing $|p^{\star}|$ in blue, the shift simultaneously involves widening the radius and contracting along the axis. Interestingly, the helical interface traced out by this shift is gradually flattening out as $|p^{\star}| \rightarrow \infty$.  However, for reasons of discreteness, it cannot  go completely flat since a horizontal interface can only exist for $p^{\star} =0$ \footnote{This is proved in the supplemental.}.  Instead, we see the emergence of an accumulation point (limit as $|p^{\star}| \rightarrow \infty$) at roughly $ \pi^{-1} (\omega, \varphi) \approx (0, 0.13)$.  In contrast, the trajectory in purple is also for increasing integer values of $|p^{\star}|$, starting from zero, but describes a shift along the helical interface in the opposite sense, which takes the interface ever more vertical.  This involves extension along the axis and contraction of the radius; a process that, evidently, cannot continue indefinitely.  Instead, there is a transition in the chirality from $|p^{\star}| = 3$ to $|p^{\star}| = 4$ (purple to red) that is achieved by flipping each mountain and valley of the Miura parallelogram, as indicated by the sign change in $\omega$.  After the transition, the trajectory for increasing $|p^{\star}|$ in red describes a shifting helical interface that goes evermore horizontal again, resulting in expansion of the radius and contraction along the axis.  This also does not continue indefinitely, as there is a final transition between $|p^{\star}| = 22$ and $|p^{\star}| = 23$ (red to green) which again flips each mountain and valley of the Miura parallelogram.  Along the green trajectory, we can take $|p^{\star}| \rightarrow \infty$.  The shifting helical interface is flattening out giving, seemingly, the same accumulation point in the phase-space  $ \pi^{-1} (\omega, \varphi) \approx (0, 0.13)$ as the blue trajectory, but corresponding to solutions of the opposite chirality.

Importantly, these \textit{transitions}\textemdash whereby $\omega$ changes signs and induces a flip in the mountain valley assignment\textemdash occur exactly when one of the helical interfaces is nearly vertical.   Our musings in this direction, guided by further numerical evidence, suggest this is an observation generic to all HMO structures.  Recall that vertical interfaces correspond to the fully degenerate lines in the phase diagrams (Fig.\;\ref{fig:AllCases}), and describe HMO solutions for  which the Miura parallelogram is completely unfolded (i.e., $\omega = 0$).  We observe that configurations``above" the line and ``below" the line correspond to a change in the sign of $\omega$, and this is apparently what is happening in the transition of $|p^{\star}| = 3$ to $4$ and $|p^{\star}| = 22$ to $23$ in Fig\;\ref{fig:Rigidity1}.  As a final comment related to the diagrams, it is tantalizing to think that the solutions for the two mountain-valley assignments can be directly mapped onto each other by a linear transformation in the $(\omega,\varphi)$ phase space, as it very much looks like the $\sigma = -$ solutions are related to the $+$ solutions by a contraction of the folding angle $\omega$.   Alas, we have checked this carefully, and it is not the case.

In summary, we have presented a  general framework to investigate the phase-space of HMO structures.   Our numerical efforts in this direction suggest that rigidly foldable motions that preserve helical symmetry are \textit{impossible} in such structures.  Nevertheless, as the examples in Fig.\ref{fig:AllCases} and \ref{fig:Rigidity1} highlight, multistability for a fixed discreteness $(p^{\star}, q^{\star})$ is an ubiquitous feature, and a rich variety of configurations can generically be achieved by rolling up a reference tessellation in different ways.  In what follows, we exploit these two features to discuss approaches for making these structures  \textit{reconfigurable}.

%

\textbf{Motion by slip.}  One such means of reconfigurability is suggested by the diagram in Fig.\;\ref{fig:Rigidity1}: A variety of HMO are achieved from the same underlying tessellation by solving the discreteness conditions  (\ref{varphiDef}) and (\ref{finalpq}) for different discrete values of $p^{\star} \in \mathbb{Z}$ (with $|q^{\star}| = const.$ a fixed integer describing the number of unit cells along the width of the tessellated strip of interest).   However, notice that the parameterizations also make sense when treating $p^{\star}$ as a continuous parameter and solving these equations. We simply ``connect the dots" along a continuous curve in the $(\omega, \varphi)$ phase-space. While this continuation does not give a HMO for non-integer values of $p^{\star}$, it \textit{does} describe a  rigidly foldable isometric origami motion of this tessellation. Precisely, consider any continuous curve $(\omega_{p^\star}^{\sigma}, \varphi_{\star}^{\sigma}(\omega_{p^\star}^{\sigma}))$ that solves (\ref{varphiDef}-\ref{finalpq}) for $p^{\star}$ in some connected interval of $\mathbb{R}$.  Substituting this curve into the group parameters (\ref{paramLocal}) and then all the parameters into (\ref{flatToCyl}), we observe that the deformation \textit{must} describe rigidly foldable origami as a function of $p^{\star}$ due to the underlying continuity and distance preserving nature of all of these maps.  In fact, there is a quite simple physical interpretation; this is nothing but \textit{motion by slip} along the helical interface that connects the boundaries of the tessellation.

\begin{figure}
	\includegraphics[width=\columnwidth]{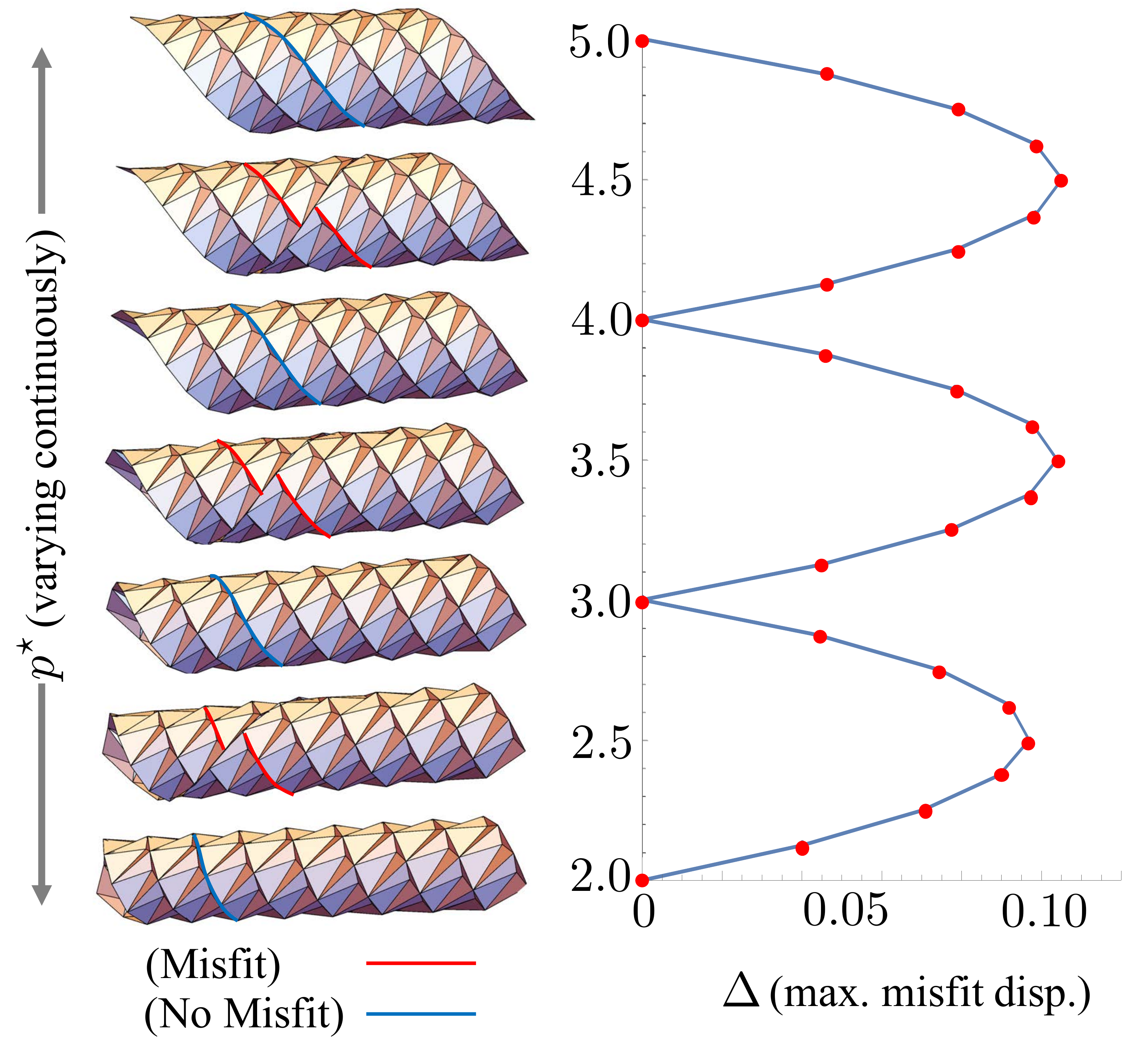}
	\caption{By treating $p^{\star}$ as a continuous parameter in the design equations, a rigidly foldable motion corresponding to slip along a helical interface is achieved (left).  This induces a misfit displacement in the radial direction at the boundaries of the tessellation for non-discrete values of $p^{\star}$ (right).}
	\label{fig:SlipMotion}
\end{figure}

An example to this effect is provided in Fig.\;\ref{fig:SlipMotion}.  For the same underlying tessellation, we vary $p^{\star}$ continuously from $2$ to $5$ to generate  a continuous curve of solutions to  (\ref{varphiDef}) and (\ref{finalpq}) in the $(\omega,\varphi)$ phase-space.  The origami structures (obtained by substituting solutions on this curve into (\ref{paramLocal}) and (\ref{flatToCyl})) are displayed at integers and half-integers.  Notice at the half-integers $p^{\star} = 2.5,3.5,4.5$, the \textit{misfit} (in red) is along a single helical interface and exactly halfway between two HMO structures $p^{\star} \pm 1/2$ (in blue).  This clearly indicates motion by slip along the helical interface.  

We should point out that this slip motion is by no means special to the particular example shown but rather generic to these origami structures.  Thus, it would seem a natural means of reconfigurability in engineering design: For example, one could design a slider mechanism that attaches to the two boundaries of the underlying tessellation. For the design, we envision that, once the two sides are connected to form a HMO, this slider would allow for easy motion along the helical interface but would otherwise act as a linear spring  for distortion in the radial direction and be (ideally) rigid for distortion normal to the radial and helical tangent directions.  In this sense then, the square of the max misfit displacement (e.g.,  $\Delta^2$ for $\Delta$ in the example Fig.\;\ref{fig:SlipMotion}) would provide a reasonable proxy to the energy barrier to motion up to, say, a constant depending only  on the design of the slider.  Since the motion is otherwise rigidly foldable origami\textemdash and, particularly, involves no change in the mountain-valley assignment in most instances\textemdash it is reasonable that all other sources of energy in the system (e.g., a ``bending energy" of the folds or friction in the hinges) could be made negligible by comparison.   As a result, we expect such designs to achieve equilibrium states at exactly each discrete value of $p^{\star}$  that admits a HMO along a continuous path in the $(\omega, \varphi)$ phase-space.  We also expect a modest energy barrier for transitioning between these discrete states.  Thus, reconfigurability here would presumably involve an actuation or loading that exceeds the energy barrier; thereby, allowing the structure to ``jump" from one HMO to its neighbor.  

\textbf{Motion by phase transformation.} Multistabilty is ubiquitous in mechanical systems \cite{Kebadze20042801, bistable_helices_00}, and one can often leverage this to obtain overall motion by transforming the system from one stable state to another.  In HMO, we have a  generically multistable mechanical system for a fixed discreteness due to the underlying constraints imposed by cylindrical origami; there is typically a \textit{plus}-phase and \textit{minus}-phase  (sometimes multiple such states) corresponding to the two different mountain-valley assignments for the Miura parallelogram (Fig.\;\ref{fig:AllCases}).  We exploit this feature to study \textit{coexistence of phases}, i.e., whether the two phases can exist as mixtures that result in cylindrical origami, with the potential to produce overall motion. The line of thinking  here is inspired by geometric compatibility in martensitic phase transformations \cite{ball_fine_1987, song_enhanced_2013, bhattacharya_microstructure_2003} and its analog for discrete helical structures \cite{feng2019phase}.

To address coexistence of phases, the naive idea is to transform one of the phases generating a HMO to another via the propagation of geometrically compatible interfaces, but there is admittedly some subtlety.  We begin by considering two locally compatible origami structures generated by the same underlying tessellation: $\check{\mathcal{G}} \mathbf{y}_{\check{\omega}}^{\check{\sigma}}(\Omega) = \{ \check{g}_1^p \check{g}_2^q( \mathbf{y}_{\check{\omega}}^{\check{\sigma}}(\Omega)) \colon (p,q) \in \mathbb{Z}^2 \}$ and $\hat{\mathcal{G}} \mathbf{y}_{\hat{\omega}}^{\hat{\sigma}}(\Omega) = \{\hat{g}_1^p \hat{g}_2^q( \mathbf{y}_{\hat{\omega}}^{\hat{\sigma}}(\Omega)) \colon (p,q) \in \mathbb{Z}^2\}$, where $\check{g}_i$ (respectively, $\hat{g}_i$) have group parameters as in (\ref{paramLocal}) with $(\omega, \varphi, \sigma) = (\check{\omega}, \check{\varphi}, \check{\sigma})$ (respectively, $(\omega, \varphi, \sigma) = (\hat{\omega}, \hat{\varphi}, \hat{\sigma})$) on the domain given above.  Note that we are not enforcing the discreteness condition (\ref{varphiDef}-\ref{finalpq}), as this will be relaxed since the origami here can involve more than one phase.   In fact, by arguing rigorously in the Supplement, we  show that necessary and sufficient conditions for a closed cylindrical origami of these two phases are
\begin{equation}
\begin{aligned}\label{twoPhase}
&\tau_{1}^{\check{\sigma}}(\check{\omega}, \check{\varphi}) = \tau_{1}^{\hat{\sigma}}(\hat{\omega}, \hat{\varphi}),  \\
&\theta_{1}^{\check{\sigma}}(\check{\omega}, \check{\varphi}) = \theta_{1}^{\hat{\sigma}}(\hat{\omega}, \hat{\varphi}),  \\
&p^{\star} \tau_1^{\check{\sigma}}(\check{\omega}, \check{\varphi}) +  \tilde{q} \check{\tau}_2^{\check{\sigma}}(\check{\omega}, \check{\varphi}) + (q^{\star}-\tilde{q})  \tau_{2}^{\hat{\sigma}}(\hat{\omega}, \hat{\varphi}) = 0, \\
&p^{\star} \theta_1^{\check{\sigma}}(\check{\omega}, \check{\varphi}) + \tilde{q} \theta_2^{\check{\sigma}}(\check{\omega}, \check{\varphi}) + (q^{\star}-\tilde{q})  \theta_{2}^{\hat{\sigma}}(\hat{\omega}, \hat{\varphi}) = 2\pi,
\end{aligned}
\end{equation}
for some $p^{\star},\tilde{q},q^{\star} \in \mathbb{Z}$ with $|\tilde{q}| \leq |q^{\star}|$ and $q^{\star} \tilde{q} \geq 0$.  (Trivially, we can also exchange the roles of $(\cdot)_2$ and $(\cdot)_1$ above.)  We focus on the system in (\ref{twoPhase}) without loss of generality. 

This system of equations, when solved, admits three types of cylindrical origami depending on the values of the various parameters: HMO (i.e., single-phased cylindrical origami), those with \textit{horizontal interfaces}, and those with \textit{helical interfaces}.  The former is obvious; if we set $(\check{\omega}, \check{\varphi}, \check{\sigma})= (\hat{\omega}, \hat{\varphi}, \hat{\sigma})$, then the system in (\ref{twoPhase}) degenerates to the original discreteness condition (\ref{param2}), which is solved via the procedure in (\ref{varphiDef}-\ref{finalpq}).  Alternatively, horizontal interfaces correspond to $q^{\star} = \tilde{q} = 0$ and $(\check{\omega}, \check{\varphi}, \check{\sigma}) \neq (\hat{\omega}, \hat{\varphi}, \hat{\sigma})$. Finally, helical interfaces correspond to everything else, i.e., $p^{\star}, \tilde{p} \neq 0$ and $(\check{\omega}, \check{\varphi}, \check{\sigma}) \neq (\hat{\omega}, \hat{\varphi}, \hat{\sigma})$.  In particular, the latter two formulas for helical interfaces describe a $\rho$-\textit{averaged} discreteness condition given the former two. That is, these formulas can be written as $p^{\star} \langle \tau_1 \rangle_{\rho} + q^{\star} \langle \tau_2 \rangle_{\rho} = 0$   and $p^{\star} \langle \theta_1 \rangle_{\rho} + q^{\star} \langle \theta_2 \rangle_{\rho} = 2\pi$ with $\rho = \tilde{q}/q^{\star}$ being the density of $\check{(\cdot)}$-phase.  In this sense, we will show that $|\tilde{q}|$ and $|q^{\star}| - |\tilde{q}|$ are the number of rows of the $\check{(\cdot)}$-phase and $\hat{(\cdot)}$-phase, respectively, for this type of cylindrical origami. 

Focusing first on simpler case of a horizontal interface (i.e., $q^{\star} = \tilde{q} = 0$), we see that (\ref{twoPhase}) reduces to 
\begin{equation}
\begin{aligned}\label{simplePhase}
&\tau_{1}^{\check{\sigma}}(\check{\omega}, \check{\varphi}) = \tau_{1}^{\hat{\sigma}}(\hat{\omega}, \hat{\varphi}) = 0, \\
&p^{\star} \theta_{1}^{\check{\sigma}}(\check{\omega}, \check{\varphi}) = p^{\star} \theta_{1}^{\hat{\sigma}}(\hat{\omega}, \hat{\varphi}) = 2\pi.
\end{aligned}
\end{equation}
As a consequence, the complete characterization of the solutions here is rather trivial. Specifically, \textit{any} set of parameters which generates a ring-type HMO (i.e., solves (\ref{varphiDef}-\ref{finalpq}) with $q^{\star} = 0$) can evidently be connected to \textit{any other} along a horizontal interface, and these are the only types of solutions in this case.  This is quite striking: As HMO solutions in this case typically come in pairs (a \textit{plus} and \textit{minus} phase; sometimes two each), it is a generic fact that such structures can form mixtures of the two (or four) phases along horizontal interfaces.  

One example, involving a change in the mountain-valley assignment for the Miura parallelograms, is provided in Fig\;\ref{fig:PhaseTran}(a).  Notice that, when a ring is transformed, it produces an overall twist and extension of the structure: The parameters $\theta_1$ and  $\tau_1$ for the generators $\check{g}_1$ and $\hat{g}_1$ are identical (as they solve (\ref{simplePhase})), but their analogs $\theta_2$ and $\tau_2$ in $\check{g}_2$ and $\hat{g}_2$ need not be.  The overall twist and extension is simply the manifestation of this difference. That is, whenever a single ring  is transformed, the magnitudes of overall twist and extension are  $\Delta \theta = | (\check{\theta}_2 - \hat{\theta}_2)|$ and $\Delta \tau = |\check{\tau}_2 - \hat{\tau}_2|$, respectively \footnote{Here, we define $\check{\theta}_{2} = \theta_2^{\check{\sigma}}(\hat{\omega}, \check{\varphi})$ and the other parameters likewise.}. 

As these quantities are often figures of merit in design, it seems quite natural to address \textit{what ring-structures ($q^{\star} = 0$) give the maximal overall twist or maximal overall extension when a layer is transformed}.  This can be done systematically.  We first fix the boundary of of Miura parallelogram.  Then, as a function of the crease pattern parameter $0< \lambda <1$ defining the interior vertex, we compute the parameters that give a multistable HMO and find the maximum difference $\Delta \tau$ and $\Delta \theta$ for this $\lambda$-dependence (Fig.\;\ref{fig:PhaseTran}(b)).  Finally, we cycle through the boundary parameters $(l,\eta)$ and repeat.  The results of this procedure for $p^{\star} = 8$ are highlighted graphically in Fig.\;\ref{fig:PhaseTran}(c-d) and are quite illuminating for design.  For ring-type HMO with horizontal interfaces, we  conclude:
\begin{enumerate}
\item {\it The transforming twist is largest in the triangular region depicted in Fig. \ref{fig:PhaseTran}(c).  These involve a twist with angle $\approx \pi$ per transforming layer.}
\item {\it The maximal extension is unambiguously  achieved as $(l, \eta, \lambda) \rightarrow (0.414, \pi/2, 0)$, which exactly corresponds to a special Kresling pattern.}
\end{enumerate}
A transformation inducing the maximal twist (i.e., on in the triangular region) and the special Kresling pattern are also provided in the figures.  The latter has the feature that the Miura parallelogram\textemdash actually, the limit $\lambda \rightarrow 0$ of Miura parallelograms\textemdash generates a HMO in the completely unfolded and the fully folded-in-half  states. Hence, the extension achieved is as large as can ever be expected, as it is the full height of the layer itself. The former is one of many examples for which a nearly degenerate HMO can transform along the same mountain-valley assignment, inducing a twist-per-layer that is essentially half the circumference of the cylinder. 

We end the discussion of compatible interfaces by briefly introducing an example corresponding to the helical case (Fig.\;\ref{fig:PhaseTran}(e)).  This is obtained by solving the system of equations in (\ref{twoPhase}) with $\tilde{q} = 3$, $q^{\star} = 8$ and $p^{\star} = 8$.  We do not present here a general strategy for solving this class of problems.  Instead, the solution is achieved by an iterative procedure involving slight perturbations of both \textit{plus}-phase and \textit{minus}-phase HMO with this discreteness.   Observe that the perturbed phases are such that a three layered \textit{plus}-phase can connect to a five layered $\textit{minus}$-phase in such a way to be perfectly compatible  along two infinitely long helical interfaces.  Additional perturbations are required to propagate the geometrically compatible helical interfaces (i.e., solve the system with different $\tilde{q}$) and induce an overall motion (twist and extension).  Such change in solutions is significant, as it avoids the issue of rigidity we observed with helical interfaces of two phases in atomic structures \cite{feng2019phase}.

\textbf{Discussion.}  In this work, we have presented a thorough\textemdash if not complete\textemdash characterization of the phase-space of Helical Miura Origami.  The results are mostly explicit and quantitative, but otherwise involve only the simplest of numerical implementation.  As such, this characterization should make for an efficient design tool for the myriad of applications seeking multi-functional and tunable structures:  The \textit{motion by slip}, which induces expansion (or contraction) radially in the structure, would seem a natural mechanism in the design of medical stents.  The quadstable ring structures\textemdash really their ``fold-in-half" analogs\textemdash are already being explored as a concept for deployable space structures \cite{ftp_PNAS_2015}.  And the \textit{helical symmetry}, which is on full display in this methodology, has the potential to be exploited for the design of novel electro-magnetic antennas \footnote{Presumably also acoustic metamaterials$\ldots$}; particularly, since discrete symmetries can interact with Maxwell's equations to produce highly directionalized electro-magnetic profiles \cite{jfj_actacryst_16}.  In fact, the helical interfaces are an interesting example in this setting, as they involve both a short-wavelength helical symmetry (the unit cell) and a long-wavelength symmetry (e.g., a shift and potential twist of the eight layers along the axis in Fig\;\ref{fig:PhaseTran}(e)).

On the theoretical front, the abstraction underlying all the results here is: \textit{discrete and Abelian groups of isometries interact naturally with origami unit cells to produce complex\textemdash hopefully interesting\textemdash structures}.  While we have chosen to focus our attention on one type of unit cell (the Miura parallelogram), the approach applies much more broadly.  For instance, Kawasaki's condition can certainly be relaxed \cite{paul_miura} and still yield an origami unit cell.  In addition, advances in 3-D/4-D printing and manufacturing make it sensible to also consider: 1) origami that is absent a flat configuration, and 2) simple building blocks of \textit{nonisometric origami}  \cite{plucinsky_actuation_2018, plucinsky2018patterning} made of active materials.  In all such cases, as long as the unit cell has four and only four corners, the design equations for the group parameters  (i.e., the equations (\ref{paramLocal}-\ref{finalpq})) directly apply to construct any Helical Origami.  Thus, the group theoretic approach to origami design would seem to have a broad and remarkable scope.

\textbf{Acknowledgment.}  The authors gratefully acknowledge the support of the Air
Force Office of Scientific Research through the MURI grant no. FA9550-16-1-0566.

\bibliographystyle{ieeetr}
\bibliography{origami} 

\begin{thebibliography}{10}

\bibitem{feng2019phase}
F.~Feng, P.~Plucinsky, and R.~D. James, ``Phase transformations and
  compatibility in helical structures,'' {\em Journal of the Mechanics and
  Physics of Solids}, 2019.

\bibitem{peraza2014origami}
E.~A. Peraza-Hernandez, D.~J. Hartl, R.~J. Malak~Jr, and D.~C. Lagoudas,
  ``Origami-inspired active structures: a synthesis and review,'' {\em Smart
  Materials and Structures}, vol.~23, no.~9, p.~094001, 2014.

\bibitem{wilson2013origami}
L.~Wilson, S.~Pellegrino, and R.~Danner, ``Origami sunshield concepts for space
  telescopes,'' in {\em 54th AIAA/ASME/ASCE/AHS/ASC Structures, Structural
  Dynamics, and Materials Conference}, p.~1594, 2013.

\bibitem{schenk2013inflatable}
M.~Schenk, S.~Kerr, A.~Smyth, and S.~Guest, ``Inflatable cylinders for
  deployable space structures,'' in {\em Proceedings of the First Conference
  Transformables, Sept}, pp.~18--20, 2013.

\bibitem{reis2015transforming}
P.~M. Reis, F.~L. Jim{\'e}nez, and J.~Marthelot, ``Transforming architectures
  inspired by origami,'' {\em Proceedings of the National Academy of Sciences},
  vol.~112, no.~40, pp.~12234--12235, 2015.

\bibitem{tachi2011designing}
T.~Tachi and G.~Epps, ``Designing one-dof mechanisms for architecture by
  rationalizing curved folding,'' in {\em International Symposium on
  Algorithmic Design for Architecture and Urban Design (ALGODE-AIJ). Tokyo},
  2011.

\bibitem{waitukaitis2015origami}
S.~Waitukaitis, R.~Menaut, B.~G.-g. Chen, and M.~van Hecke, ``Origami
  multistability: From single vertices to metasheets,'' {\em Physical review
  letters}, vol.~114, no.~5, p.~055503, 2015.

\bibitem{waitukaitis2016origami}
S.~Waitukaitis and M.~van Hecke, ``Origami building blocks: Generic and special
  four-vertices,'' {\em Physical Review E}, vol.~93, no.~2, p.~023003, 2016.

\bibitem{chen2016topological}
B.~G.-g. Chen, B.~Liu, A.~A. Evans, J.~Paulose, I.~Cohen, V.~Vitelli, and
  C.~Santangelo, ``Topological mechanics of origami and kirigami,'' {\em
  Physical review letters}, vol.~116, no.~13, p.~135501, 2016.

\bibitem{Faber1386}
J.~A. Faber, A.~F. Arrieta, and A.~R. Studart, ``Bioinspired spring origami,''
  {\em Science}, vol.~359, no.~6382, pp.~1386--1391, 2018.

\bibitem{DNAorigami}
P.~W.~K. Rothemund, ``Folding dna to create nanoscale shapes and patterns,''
  {\em Nature}, vol.~440, pp.~297 EP --, 03 2006.

\bibitem{han2011dna}
D.~Han, S.~Pal, J.~Nangreave, Z.~Deng, Y.~Liu, and H.~Yan, ``Dna origami with
  complex curvatures in three-dimensional space,'' {\em Science}, vol.~332,
  no.~6027, pp.~342--346, 2011.

\bibitem{Liu_helical_antenna_2014}
X.~Liu, S.~Yao, S.~V. Georgakopoulos, B.~S. Cook, and M.~M. Tentzeris,
  ``Reconfigurable helical antenna based on an origami structure for wireless
  communication system,'' in {\em 2014 IEEE MTT-S International Microwave
  Symposium (IMS2014)}, pp.~1--4, June 2014.

\bibitem{yao2014novel}
S.~Yao, X.~Liu, S.~V. Georgakopoulos, and M.~M. Tentzeris, ``A novel
  reconfigurable origami spring antenna,'' in {\em Antennas and Propagation
  Society International Symposium (APSURSI), 2014 IEEE}, pp.~374--375, IEEE,
  2014.

\bibitem{schenk2013geometry}
M.~Schenk and S.~D. Guest, ``Geometry of miura-folded metamaterials,'' {\em
  Proceedings of the National Academy of Sciences}, vol.~110, no.~9,
  pp.~3276--3281, 2013.

\bibitem{yasuda2015reentrant}
H.~Yasuda and J.~Yang, ``Reentrant origami-based metamaterials with negative
  poisson's ratio and bistability,'' {\em Physical review letters}, vol.~114,
  no.~18, p.~185502, 2015.

\bibitem{silverberg2014using}
J.~L. Silverberg, A.~A. Evans, L.~McLeod, R.~C. Hayward, T.~Hull, C.~D.
  Santangelo, and I.~Cohen, ``Using origami design principles to fold
  reprogrammable mechanical metamaterials,'' {\em science}, vol.~345, no.~6197,
  pp.~647--650, 2014.

\bibitem{wei2013geometric}
Z.~Y. Wei, Z.~V. Guo, L.~Dudte, H.~Y. Liang, and L.~Mahadevan, ``Geometric
  mechanics of periodic pleated origami,'' {\em Physical review letters},
  vol.~110, no.~21, p.~215501, 2013.

\bibitem{ftp_PNAS_2015}
E.~T. Filipov, T.~Tachi, and G.~H. Paulino, ``Origami tubes assembled into
  stiff, yet reconfigurable structures and metamaterials,'' {\em Proceedings of
  the National Academy of Sciences}, vol.~112, no.~40, pp.~12321--12326, 2015.

\bibitem{dudte2016programming}
L.~H. Dudte, E.~Vouga, T.~Tachi, and L.~Mahadevan, ``Programming curvature
  using origami tessellations,'' {\em Nature materials}, vol.~15, no.~5,
  p.~583, 2016.

\bibitem{miura_93}
K.~Miura, ``Concepts of deployable space structures,'' {\em International
  Journal of Space Structures}, vol.~8, no.~1-2, pp.~3--16, 1993.

\bibitem{paul_miura}
P.~Plucinsky, F.~Feng, and R.~D. James, ``The design and deformations of
  generalized miura origami,'' {\em preprint}.

\bibitem{bowers2015lang}
J.~C. Bowers and I.~Streinu, ``Lang'€™s universal molecule algorithm,'' {\em
  Annals of Mathematics and Artificial Intelligence}, vol.~74, no.~3-4,
  pp.~371--400, 2015.

\bibitem{lh_RFF_2018}
R.~J. Lang and L.~Howell, ``Rigidly foldable quadrilateral meshes from angle
  arrays,'' {\em Journal of Mechanisms and Robotics}, vol.~10, no.~2,
  p.~021004, 2018.

\bibitem{wang2011folding}
K.~Wang and Y.~Chen, ``Folding a patterned cylinder by rigid origami,'' {\em
  Origami}, vol.~5, pp.~265--276, 2011.

\bibitem{bos_incompressibility_2016}
F.~Bös, M.~Wardetzky, E.~Vouga, and O.~Gottesman, ``On the {Incompressibility}
  of {Cylindrical} {Origami} {Patterns},'' {\em Journal of Mechanical Design},
  vol.~139, pp.~021404--021404--9, Dec. 2016.

\bibitem{jianguo2015bistable}
C.~Jianguo, D.~Xiaowei, Z.~Ya, F.~Jian, and T.~Yongming, ``Bistable behavior of
  the cylindrical origami structure with kresling pattern,'' {\em Journal of
  Mechanical Design}, vol.~137, no.~6, p.~061406, 2015.

\bibitem{tachi2009generalization}
T.~Tachi, ``Generalization of rigid foldable quadrilateral mesh origami,'' in
  {\em Symposium of the International Association for Shell and Spatial
  Structures (50th. 2009. Valencia). Evolution and Trends in Design, Analysis
  and Construction of Shell and Spatial Structures: Proceedings}, Editorial
  Universitat Polit{\`e}cnica de Val{\`e}ncia, 2009.

\bibitem{james_JMPS_06}
R.~D. James, ``Objective structures,'' {\em Journal of the Mechanics and
  Physics of Solids}, vol.~54, no.~11, pp.~2354--2390, 2006.

\bibitem{dresselhaus1995physics}
M.~Dresselhaus, G.~Dresselhaus, and R.~Saito, ``Physics of carbon nanotubes,''
  {\em Carbon}, vol.~33, no.~7, pp.~883--891, 1995.

\bibitem{james_icm}
R.~D. James, ``Symmetry, invariance and the structure of matter,'' in {\em
  Proceedings of the International Congress of Mathematicians 2018 (ICM 2018)},
  World Scientific, 2019.

\bibitem{Kebadze20042801}
E.~Kebadze, S.~Guest, and S.~Pellegrino, ``Stable prestressed shell
  structures,'' {\em International Journal of Solids and Structures}, vol.~41,
  no.~11–12, pp.~2801 -- 2820, 2004.

\bibitem{bistable_helices_00}
R.~E. Goldstein, A.~Goriely, G.~Huber, and C.~W. Wolgemuth, ``Bistable
  helices,'' {\em Physical review letters}, vol.~84, no.~7, p.~1631, 2000.

\bibitem{ball_fine_1987}
J.~M. Ball and R.~D. James, ``Fine phase mixtures as minimizers of energy,''
  {\em Archive for Rational Mechanics and Analysis}, vol.~100, no.~1,
  pp.~13--52, 1987.

\bibitem{song_enhanced_2013}
Y.~Song, X.~Chen, V.~Dabade, T.~W. Shield, and R.~D. James, ``Enhanced
  reversibility and unusual microstructure of a phase-transforming material,''
  {\em Nature}, vol.~502, pp.~85--88, Oct. 2013.

\bibitem{bhattacharya_microstructure_2003}
K.~Bhattacharya, {\em Microstructure of martensite : why it forms and how it
  gives rise to the shape-memory effect}.
\newblock Oxford: Oxford University Press, 2003.

\bibitem{jfj_actacryst_16}
D.~J{\"u}stel, G.~Friesecke, and R.~D. James, ``Bragg--von laue diffraction
  generalized to twisted x-rays,'' {\em Acta Crystallographica Section A:
  Foundations and Advances}, vol.~72, no.~2, pp.~190--196, 2016.

\bibitem{plucinsky_actuation_2018}
P.~Plucinsky, M.~Lemm, and K.~Bhattacharya, ``Actuation of {Thin} {Nematic}
  {Elastomer} {Sheets} with {Controlled} {Heterogeneity},'' {\em Archive for
  Rational Mechanics and Analysis}, vol.~227, pp.~149--214, Jan. 2018.

\bibitem{plucinsky2018patterning}
P.~Plucinsky, B.~A. Kowalski, T.~J. White, and K.~Bhattacharya, ``Patterning
  nonisometric origami in nematic elastomer sheets,'' {\em Soft matter},
  vol.~14, no.~16, pp.~3127--3134, 2018.

\end{thebibliography}


\begin{thebibliography}{1}

\bibitem{paul_miura}
P.~Plucinsky, F.~Feng, and R.~D. James, ``The design and deformations of
  generalized miura origami,'' {\em preprint}.

\bibitem{Note1}
There is a degeneracy in rotations $\protect \mathbf {R}_i(\pi ) = \protect
  \mathbf {R}_i(-\pi )$. This means that when $\protect \qopname \relax
  o{cos}(\gamma _3) = -1$ and $\protect \qopname \relax o{sin}(\gamma _3) = 0$,
  we are free to choose $\gamma _3 = \pi $ or $-\pi $. Nevertheless, physically
  a rotation by $\pi $ folds a region on top of another and a rotation by $-\pi
  $ folds that same region underneath the other. In another sense, it is
  well-know that the four-fold setting corresponds to three mountains and a
  valley or three valleys and a mountain (Maekawa's theorem). We will preserve
  this formalism in the case that a folding angle attains the magnitude of $\pi
  $ to overcome this degeneracy.

\bibitem{Note2}
One could, of course, also deduce this result by direct calculation without
  resorting to Maekawa's theorem.

\bibitem{Note3}
Technically, one should also include reflections in the definition of an
  isometry, but the reflections are not relevant to this work. Hence, their
  exclusion here.

\bibitem{dayal2010objective}
K.~Dayal, R.~Elliott, and R.~D. James, ``Objective formulas,'' {\em preprint},
  2010.

\bibitem{jianguo2015bistable}
C.~Jianguo, D.~Xiaowei, Z.~Ya, F.~Jian, and T.~Yongming, ``Bistable behavior of
  the cylindrical origami structure with kresling pattern,'' {\em Journal of
  Mechanical Design}, vol.~137, no.~6, p.~061406, 2015.

\end{thebibliography}

%
%
%
%
%
%
%
%
%
%

\end{document}


\preprint{APS/123-QED}

\title{Helical Miura Origami: Supplement} 

\author{Fan Feng}
\author{Paul Plucinsky}%
\author{Richard D. James}
\email{james@umn.edu}
\affiliation{Aerospace Engineering and Mechanics, University of Minnesota, Minneapolis, MN 55455, USA}%
\date{\today}

\pacs{Valid PACS appear here}
\keywords{}
\maketitle


\subsection{On the parameterization of a Miura parallelogram.}
Let $\{ \mathbf{e}_1, \mathbf{e}_2, \mathbf{e}_3\}$ denote the standard basis on $\mathbb{R}^3$, and let $\bfx_1, \bfx_2, \bfx_3, \bfx_4 \in \mathbb{R}^3$ denote the corner points of the Miura parallelogram, which in the flat state is represented by a domain $\Omega \subset \mathbb{R}^3$ (a two-dimensional domain embedded into $\mathbb{R}^3$).   To parameterize these points, we choose $\mathbf{x}_1 = \mathbf{0}$,  $\bfx_2 = \bfe_1$, $\bfx_4 =l( \cos\eta\, \bfe_1 + \sin \eta\, \bfe_2)$, and $\bfx_3 = \bfx_2 + \bfx_4 - \bfx_1$ for $\eta =\angle \bfx_2\bfx_1\bfx_4 \in (0,\pi)$ and $l>0$.   This is a general parameterization of a parallelogram up to translation, rotation and uniform rescaling.  The four creases of the Miura parallelogram merge to a point on the interior, which we denote with $\bfx_0$.  This has the form
\beq
\bfx_0 = \bfx_1 + \lambda_1(\bfx_2 - \bfx_1)+ \lambda_2(\bfx_4 - \bfx_1), \lambda_1, \lambda_2 \in (0,1),
\eeq
where $\lambda_1$ and $\lambda_2$ are chosen so that the sector angles satisfy the well-know Kawasaki's condition for flat foldability, i.e., that opposite angles sum to $\pi$. 

To characterized $\lambda_{1}$ and $\lambda_2$, note that Kawasaki's condition $\angle \bfx_1 \bfx_0 \bfx_4 + \angle \bfx_2 \bfx_0 \bfx_3 = \pi$ is equivalent to 
\beq\label{eq:cos}
\frac{(\bfx_4 - \bfx_0)\cdot (\bfx_1 - \bfx_0)}{|\bfx_4 - \bfx_0||\bfx_1 - \bfx_0|} + \frac{(\bfx_3 - \bfx_0)\cdot(\bfx_2 - \bfx_0)}{|\bfx_3 - \bfx_0||\bfx_2 - \bfx_0|}=0
\eeq 
by the law of cosines.  A necessary condition is thus
\begin{equation}
\begin{aligned}\label{eq:preLongcos}
&((\bfx_4 - \bfx_0)\cdot (\bfx_1 - \bfx_0)|\bfx_3 - \bfx_0||\bfx_2 - \bfx_0|)^2  \\
&\quad = ((\bfx_3 - \bfx_0)\cdot(\bfx_2 - \bfx_0)|\bfx_4 - \bfx_0||\bfx_1 - \bfx_0|)^2=0.
\end{aligned}
\end{equation}
With our parameterizations $\mathbf{x}_i \equiv \mathbf{x}_i(\eta,l,\lambda_1, \lambda_2)$ for $i = 0,1,2,3,4$, the latter can be expressed as 
	\begin{equation}
	\begin{aligned} \label{eq:longcos}
	&p_1p_2 = 0, \\
	&p_1 = \Big(2\lambda_1- 1 + \frac{2(\lambda_1 - 1) \lambda_1 (2\lambda_2 - 1)l \cos(\eta)}{(\lambda_1 - 1) \lambda_1 + (\lambda_2 - 1) \lambda_2 l^2}\Big),\\
	&p_2 = ({\lambda_1}-1) {\lambda_1} -({\lambda_2}-1) {\lambda_2}l^2,
	\end{aligned}
	\end{equation}
	after some algebra that uses the fact that $(\lambda_1-1)\lambda_1 + (\lambda_2 - 1)\lambda_2 \neq 0$ and $\sin^2(\eta) l^2 \neq 0$.  From this point, we can actually argue that the necessary and sufficient condition for (\ref{eq:cos}) is, in fact, only $p_2 = 0$.  
	
  Indeed, suppose this is not necessary.  Then there is a solution to (\ref{eq:cos}) with $p_2 \neq 0$. Evidently then $p_1 = 0$ is necessary by (\ref{eq:longcos}).  But this is solved if and only if $\lambda_1 = \lambda_2 = 1/2$ with $l = 1$ or 
\begin{align}
\cos(\eta) = \frac{(1-2\lambda_1)(({\lambda_1}-1) {\lambda_1} +({\lambda_2}-1) {\lambda_2} {l}^2)}{2 ({\lambda_1}-1) {\lambda_1} (2 {\lambda_2}-1) {l}}.
\end{align}
The former gives $p_2 = 0$,  so we need only focus on the latter.  In this case, we observe that $(\bfx_4 - \bfx_0)\cdot(\bfx_1 - \bfx_0)$ in (\ref{eq:cos}) simplifies to 
   \beq \label{eq:sign1}
 \frac{(\lambda_1-1){\lambda_1} -({\lambda_2}-1) {\lambda_2} {l}^2}{2({\lambda_1}-1)}
\eeq
by substituting the formula for $\cos(\eta)$.  Likewise, $(\bfx_3 -\bfx_0) \cdot (\bfx_2 - \bfx_0)$ in (\ref{eq:cos}) simplifies to
\beq \label{eq:sign2}
\frac{({\lambda_2}-1) {\lambda_2} {l}^2-({\lambda_1}-1) {\lambda_1}}{2 {\lambda_1}}
\eeq
by the same substitution.  But again, these observations lead to the conclusion that
\beq \label{eq:sol}
p_2 = ({\lambda_2}-1) {\lambda_2} {l}^2-({\lambda_1}-1) {\lambda_1} =0
\eeq
since, if not, (\ref{eq:sign1}) and (\ref{eq:sign2}) have the same sign which violates (\ref{eq:cos}).  Thus, $p_2 = 0$ is necessary.

For the sufficiency of $p_2 = 0$, we observe that (\ref{eq:preLongcos}) holds since $p_2 = 0$ solves (\ref{eq:longcos}).  This means that either (\ref{eq:cos}) is solved or $(\mathbf{x}_4 - \mathbf{x}_0) \cdot (\mathbf{x}_1 - \mathbf{x}_0)$ is non-zero and has the same sign as $(\mathbf{x}_3 - \mathbf{x}_0) \cdot (\mathbf{x}_2 - \mathbf{x}_0)$.   We show that the latter is false, and, as such, the former is true.  Indeed, since $\mathbf{x}_i \equiv \mathbf{x}_i(\eta,l,\lambda_1, \lambda_2)$ for $i = 0,1,2,3,4$, we find that  
\begin{equation}
\begin{aligned}
&(\bfx_4 - \bfx_0) \cdot  (\bfx_1 - \bfx_0) =  \\
&\quad \lambda_1^2 +{\lambda_1} (2 {\lambda_2}-1) {l} \cos (\eta )+({\lambda_2}-1) {\lambda_2} {l}^2, \label{eq:eq1}\\
&(\bfx_3 - \bfx_0)\cdot (\bfx_2 - \bfx_0) =\\
&\quad ({\lambda_1}-1)^2 +({\lambda_1}-1) (2 {\lambda_2}-1)  {l} \cos (\eta )+({\lambda_2}-1) {\lambda_2} {l}^2.
\end{aligned}
\end{equation}
By making the substitution $(\lambda_2 - 1) \lambda_2 l^2 = (\lambda_1-1) \lambda_1$ (which is $p_2 = 0$), the two equations in (\ref{eq:eq1}) can be written as 
\begin{equation}
\begin{aligned}
&(\bfx_4 - \bfx_0) \cdot  (\bfx_1 - \bfx_0) = \\
&\quad  \lambda_1 ((2\lambda_1 -1) + l(2\lambda_2 -1) \cos(\eta)),  \\
&(\bfx_3 - \bfx_0)\cdot (\bfx_2 - \bfx_0) =  \\
&\quad (\lambda_1-1) ((2\lambda_1 -1) + l(2\lambda_2 -1) \cos(\eta)). 
\end{aligned}
\end{equation}
Thus clearly, $(\bfx_4 - \bfx_0) \cdot  (\bfx_1 - \bfx_0)$ and $(\bfx_3 - \bfx_0)\cdot (\bfx_2 - \bfx_0) $ have opposite signs or they are both zero.  This is all that is required to show that (\ref{eq:sol}) implies (\ref{eq:cos}).

In summary, $p_2 = 0$ is necessary and sufficient for a solution to (\ref{eq:cos}), which is the condition that the crease pattern satisfies Kawaski's conidition.   This furnishes a $(\lambda_1, \lambda_2)$:
If $l>1$, the solutions of (\ref{eq:sol}) are
\beq \label{eq:lambda1}
\lambda_2 = \frac{1}{2} \pm \sqrt{\frac{1}{l^2}(\lambda_1 - \frac{1}{2})^2 + \frac{1-1^2/l^2}{4}}.
\eeq
If $l<1$, the solutions are
\beq \label{eq:lambda2}
\lambda_1 = \frac{1}{2} \pm \sqrt{l^2(\lambda_2 - \frac{1}{2})^2 + \frac{1-l^2}{4}}.
\eeq
Finally, the case $l = 1$ is solved by
\begin{align} \label{eq:lambda12}
	\lambda_2 = 
	\lambda_1~\text{or}~ 
	1- \lambda_1.
\end{align} 
This is the complete parameterization for (\ref{eq:sol}) and gives the stated results (i-iii) in the main text.

\subsection{Kinematics of a four-fold origami satisfying Kawasaki's condition}

We define the four tangent and normal vectors to the crease pattern of the Miura parallelogram by 
\begin{align}
\mathbf{t}_i = \frac{\mathbf{x}_i - \mathbf{x}_0}{|\mathbf{x}_i - \mathbf{x}_0|}, \quad \mathbf{n}_i = -(\mathbf{t}_i \cdot \mathbf{e}_2) \mathbf{e}_1 + (\mathbf{t}_i \cdot \mathbf{e}_1) \mathbf{e}_2
\end{align}
for $i = 1,2,3,4$.  In addition, we denote two of the sector angles with $\alpha = \angle \bfx_1\bfx_0\bfx_2 = \arccos(\mathbf{t}_2 \cdot \mathbf{t}_1)$ and $\beta = \angle \mathbf{x}_2 \mathbf{x}_0 \mathbf{x}_3 = \arccos(\mathbf{t}_3 \cdot \mathbf{t}_2)$.  Since the interior vertex of the Miura parallelogram satisfies Kawasaki's condition, the other two are prescribed (i.e., $\angle \bfx_3\bfx_0\bfx_4 = \arccos(\mathbf{t}_4 \cdot \mathbf{t}_3) = \pi - \alpha$ and $\angle \bfx_4\bfx_0\bfx_1 = \arccos(\mathbf{t}_1 \cdot \mathbf{t}_4) = \pi - \beta$).   Finally, we introduce the right-hand rotations 
\begin{equation}
\begin{aligned}
\mathbf{R}_i(\gamma_i) &= \mathbf{t}_i \otimes \mathbf{t}_i + \cos(\gamma_i)(\mathbf{n}_i \otimes \mathbf{n}_i + \mathbf{e}_3 \otimes \mathbf{e}_3)  \\
&\qquad \sin(\gamma_i) (\mathbf{e}_3 \otimes \mathbf{n}_i - \mathbf{n}_i \otimes  \mathbf{e}_3),
\end{aligned}
\end{equation}
for $i = 1,2,3,4$, and we introduce the sets 
\begin{equation}
\begin{aligned}
&\mathcal{S}_{\alpha_{1}} = \big\{ \mathbf{x} \in \text{Conv }\{\mathbf{x}_{4}, \mathbf{x}_0, \mathbf{x}_{1}\} \big\} \\
&\mathcal{S}_{\alpha_{i}}  =  \big\{ \mathbf{x} \in \text{Conv }\{\mathbf{x}_{i}, \mathbf{x}_0, \mathbf{x}_{i-1}\} \big\}, \;\; i = 2,3,4.
\end{aligned}
\end{equation}
Observe that the rotation satisfy $\mathbf{R}_i(\gamma_i) \mathbf{t}_i = \mathbf{t}_i$ for all angles $\gamma_i$.  In addition,  the sets $\mathcal{S}_{\alpha_i}$ describe the convex hull of  two corner points and the interior vertex. As such,  each gives a triangular region with sector angle $\alpha_i$, and their union gives the domain of the entire Miura parallelogram, i.e., $\cup_{i = 1,2,3,4} \mathcal{S}_{\alpha_i} = \Omega$.

With the notation now set, we note that, up to rotation and translation, a general isometric origami deformation of the Miura parallelogram is given by $\mathbf{y} \colon \Omega \rightarrow \mathbb{R}^3$ such that 
\begin{align}\label{eq:yDefSup}
\mathbf{y}(\mathbf{x}) = \begin{cases}
\mathbf{x}, & \text{ if } \mathbf{x} \in \mathcal{S}_{\alpha_2} \\
\mathbf{R}_2(\gamma_2) (\mathbf{x} - \mathbf{x}_0)  + \mathbf{x}_0& \text{ if } \mathbf{x} \in \mathcal{S}_{\alpha_3} \\ 
\mathbf{R}_2(\gamma_2) \mathbf{R}_3(\gamma_3) (\mathbf{x} - \mathbf{x}_0)  + \mathbf{x}_0 & \text{ if } \mathbf{x} \in \mathcal{S}_{\alpha_4} \\
\mathbf{R}_1(-\gamma_1)(\mathbf{x} - \mathbf{x}_0) + \mathbf{x}_0 & \text{ if } \mathbf{x} \in \mathcal{S}_{\alpha_1},
\end{cases}
\end{align}
and subject to compatibility conditions that the jump in the deformation gradient along each of the four tangents $\mathbf{t}_i$ is zero, i.e.,
\begin{equation}
\begin{aligned}
\begin{cases}
(\mathbf{I} - \mathbf{R}_1(-\gamma_1)) \mathbf{t}_1 = \mathbf{0} \\ 
(\mathbf{R}_2(\gamma_2) - \mathbf{I})\mathbf{t}_2 = \mathbf{0} \\
(\mathbf{R}_2(\gamma_2)\mathbf{R}_3(\gamma_3)  - \mathbf{R}_2(\gamma_2))\mathbf{t}_3 = \mathbf{0} \\ 
(\mathbf{R}_1(-\gamma_1) - \mathbf{R}_2(\gamma_2) \mathbf{R}_3(\gamma_3)) \mathbf{t}_4 = \mathbf{0}.
\end{cases}
\end{aligned}
\end{equation}
These compatibility conditions are necessary and sufficient for continuity of the deformation $\mathbf{y}$, and thus, ensure that $\mathbf{y}$ is a continuous isometric origami deformation of the Miura parallelogram.  The angles are restricted to the interval $-\pi \le \gamma_1, \gamma_2, \gamma_3 \le \pi$.  This natural physical restriction
avoids paper passing through itself but allows the structure to be folded flat ($\gamma_1,\gamma_2,\gamma_3= \pm \pi$). 

As it happens, the first three compatibility conditions are solved trivially since $\mathbf{R}_i(\gamma_i) \mathbf{t}_i = \mathbf{t}_i$.  The fourth can be rewritten as 
\begin{align}
\mathbf{R}_1(\gamma_1) \mathbf{R}_2(\gamma_2) \mathbf{R}_3(\gamma_3) \mathbf{t}_4  =\mathbf{t}_4. 
\end{align}
In turn, this demands that a product of rotations $\mathbf{R} = \mathbf{R}_1(\gamma_1) \mathbf{R}_2(\gamma_2) \mathbf{R}_3(\gamma_3)$ satisfy $\mathbf{R} \mathbf{t}_4 = \mathbf{t}_4$.  Since this product is clearly also a rotation, it must be that $\mathbf{R} = \mathbf{R}_4(-\gamma_4)$ for some $\gamma_4 \in [-\pi, \pi]$.  Thus, we can write the final compatibility condition\textemdash really, the only one that is not solved trivially\textemdash as 
\begin{equation}
\begin{aligned}\label{eq:compatSup}
&\gamma_1, \gamma_2, \gamma_3, \gamma_4 \in [-\pi, \pi], \;\;  \text{ s.t. }\\
& \qquad  \mathbf{R}_1(\gamma_1) \mathbf{R}_2(\gamma_2) \mathbf{R}_3(\gamma_3) \mathbf{R}_4(\gamma_4) = \mathbf{I}.
\end{aligned}
\end{equation}
This is the condition of compatibility stated in the main text.  

We now introduce a theorem on the kinematics of a  Miura four-fold intersection, which simply characterizes all solutions to (\ref{eq:compatSup}) when the tangents (and therefore, the rotations) are related by Kawasaki's condition.  A generalized version of this theorem, in which Kawasaki's condition is relaxed, is introduced in \cite{paul_miura}.  For the result at hand, we find it convenient to employ the shorthand notation $c_{\theta} = \cos(\theta)$ and $s_{\theta} = \sin(\theta)$.

\begin{theorem}
	\label{thm:kinematics_proof}
	For a four-fold vertex satisfying Kawasaki's condition, the compatibility condition (\ref{eq:compatSup}) holds if and only if: 
	\begin{itemize}[leftmargin = *]
	\item The folding angle belong to the four-fold families
	\begin{equation}
	\begin{aligned}
	\label{kin1Sup}
&\gamma_1=-\sigma \bar{\gamma}_3^{\sigma} (\omega),~ \gamma_2 = \sigma \omega,~ \gamma_3=\bar{\gamma}_3^{\sigma} (\omega),~ \gamma_4 = \omega, \\
&\bar{\gamma}_3^{\sigma}(\omega) = \text{sign}\Big((c_{\alpha} - \sigma c_{\beta}) \omega \Big) \arccos\Big(\tfrac{(\sigma 1 - c_{\alpha} c_{\beta})c_{\omega} + s_{\alpha} s_{\beta}}{(\sigma 1 - c_{\alpha} c_{\beta}) + s_{\alpha} s_{\beta} c_{\omega} } \Big), \\
&\sigma \in \mathcal{A} = \begin{cases}
\emptyset & \text{ if } \alpha = \beta = \tfrac{\pi}{2} \\
\{ - \} & \text{ if }  \alpha = \beta \neq \tfrac{\pi}{2} \\
\{ + \} & \text{ if } \alpha = \pi - \beta \neq \tfrac{\pi}{2} \\
\{\pm \} & \text{ otherwise} 
\end{cases} \quad \text{ for } \omega \in [-\pi, \pi].
\end{aligned}
	\end{equation}
	\item The folding angle belong to the fold-in-half families
	\begin{equation}
	\begin{aligned}
&\begin{cases}\label{kin2Sup}
\gamma_{1,3} = 0, ~ \gamma_{2,4} = \omega \in [-\pi, \pi] & \text{ if } \sigma = +, \alpha = \beta \\
\gamma_{1,3} = \omega \in [-\pi, \pi],~ \gamma_{2,4} = 0 & \text{ if } \sigma = -, \alpha  = \pi- \beta;
\end{cases} \\
&\begin{cases}
\gamma_{1} = -\gamma_3 = \omega \in [-\pi,\pi],~ \gamma_{2,4} = \pm \pi & \text{ if } \sigma = +, \alpha = \beta \\
\gamma_{1,3} = \pm \pi,~ \gamma_{2} = -\gamma_4 = \omega \in [-\pi, \pi] & \text{ if } \sigma = -, \alpha  = \pi- \beta. 
\end{cases}
\end{aligned}
\end{equation}
	\end{itemize}
\end{theorem}

\begin{proof}
A necessary condition for (\ref{eq:compatSup}) is 
\begin{equation}
\begin{aligned}\label{eq:firstNec}
\mathbf{t}_3 \cdot \Big(\mathbf{R}_3(\gamma_3)  \mathbf{R}_4(\gamma_4) - \mathbf{R}_2^T(\gamma_2) \mathbf{R}_1^T(\gamma_1)\Big) \mathbf{t}_1 = 0 
\end{aligned}
\end{equation}
By a direct calculation and since $\mathbf{R}_i(\theta) \mathbf{t}_i = \mathbf{t}_i$, we observe that 
\begin{equation}
\begin{aligned}\label{eq:derive1}
&\mathbf{t}_3 \cdot \Big(\mathbf{R}_3(\gamma_3)  \mathbf{R}_4(\gamma_4) - \mathbf{R}_2^T(\gamma_2) \mathbf{R}_1^T(\gamma_1)\Big) \mathbf{t}_1 \\
&\qquad = \mathbf{t}_3 \cdot \Big( \mathbf{R}_4(\gamma_4) - \mathbf{R}_2^T(\gamma_2)\Big) \mathbf{t}_1 = \big(c_{\gamma_2} -c_{\gamma_4}\big) s_{\alpha} s_{\beta}.
\end{aligned}
\end{equation}  
Consequently, (\ref{eq:firstNec}) is solved if and only if $\gamma_2 = \pm \gamma_4$ since $\alpha, \beta \in (0,\pi)$.  Thus, from herein, we set $\gamma_4 = \omega$ and $\gamma_2  = \sigma \omega$ for $\sigma \in \{ \pm \}$ and $\omega \in [-\pi, \pi]$, as this necessary.

Another necessary condition is simply
\begin{equation}
\begin{aligned}\label{eq:secondNec}
\Big(\mathbf{R}_3(\gamma_3)  \mathbf{R}_4(\omega) - \mathbf{R}_2^T(\sigma \omega) \mathbf{R}_1^T(\gamma_1)\Big) \mathbf{t}_1 = \mathbf{0}. 
\end{aligned}
\end{equation}
The left-hand side can be rewritten in a form that is more reveling.   For instance, as we have already solved for the $\mathbf{t}_3$ component of this equation (by (\ref{eq:firstNec}) and (\ref{eq:derive1})), we can project the equation onto the plane normal to $\mathbf{t}_3$ and use the identity $\mathbf{R}_1(\theta) \mathbf{t}_1 = \mathbf{t}_1$.  These two observations give the identities 
\begin{equation}
\begin{aligned}\label{eq:derive2}
 &\Big(\mathbf{R}_3(\gamma_3)  \mathbf{R}_4(\omega) - \mathbf{R}_2^T(\sigma \omega) \mathbf{R}_1^T(\gamma_1)\Big) \mathbf{t}_1  \\
&\qquad = \mathbf{P}_3 \Big(\mathbf{R}_3(\gamma_3) \mathbf{R}_4(\omega) - \mathbf{R}_2^T(\sigma \omega)\mathbf{R}_1^T(\gamma_1) \Big) \mathbf{t}_1 \\
&\qquad = \mathbf{R}_3(\gamma_3) \mathbf{P}_3\mathbf{R}_4(\omega)\mathbf{t}_1 - \mathbf{P}_3 \mathbf{R}_2^T(\sigma \omega) \mathbf{t}_1,
\end{aligned}
\end{equation}
where $\mathbf{P}_3 = \mathbf{I} - \mathbf{t}_3 \otimes \mathbf{t}_3$.  Note that $\mathbf{P}_3$ and $\mathbf{R}_3(\gamma_3)$ commute, which gives the last identity.  

Now, since trivially $|\mathbf{R}_4(\omega) \mathbf{t}_1| = |\mathbf{R}_2^T(\sigma \omega) \mathbf{t}_1| = 1$ and since  $\mathbf{t}_3 \cdot \mathbf{R}_4(\omega) \mathbf{t}_1 = \mathbf{t}_3 \cdot \mathbf{R}_2^T(\sigma \omega) \mathbf{t}_1$ (by (\ref{eq:firstNec}) and (\ref{eq:derive1})), we have $|\mathbf{P}_3 \mathbf{R}_4(\omega)\mathbf{t}_1| = | \mathbf{P}_3 \mathbf{R}_2^T(\sigma \omega) \mathbf{t}_1|$.  Combining this with (\ref{eq:derive2}), we observe that the needed identity in (\ref{eq:secondNec}) is solved in exactly one of two ways:
\begin{enumerate}
\item If $\mathbf{P}_3 \mathbf{R}_4(\omega)\mathbf{t}_1 = 0$, then (\ref{eq:secondNec}) holds trivially for all $\gamma_{1,3} \in  [-\pi,\pi]$.  
\item Otherwise, $\mathbf{P}_3\mathbf{R}_4(\omega) \mathbf{t}_1 \neq 0$.  Consequently, this case is solved by letting $\gamma_3 \in [-\pi, \pi]$ be the angle that takes the vector $\mathbf{P}_3 \mathbf{R}_4(\omega)\mathbf{t}_1$ to the vector of equal (and non-zero) magnitude $\mathbf{P}_3 \mathbf{R}_2^T(\sigma \omega) \mathbf{t}_1$.
\end{enumerate}

(\textit{Solutions in Case 1.}) For this case, we characterize the $(\omega, \alpha, \beta)$ such that $\mathbf{P}_3 \mathbf{R}_4(\omega) \mathbf{t}_1 = 0$.  Clearly, we must have $\mathbf{R}_4(\omega) \mathbf{t}_1  \in \{ \pm \mathbf{t}_3\}$.  Further, since $\mathbf{t}_1, \mathbf{t}_3$ and $\mathbf{t}_4$ are all in the same plane and $\mathbf{t}_1 \cdot \mathbf{t}_4 = \cos(\pi - \beta) \neq 0$, we must also have $\omega \in \{ 0, \pm \pi\}$.  Consequently, the solution for $\omega = 0$ is evidently $\mathbf{t}_1 = - \mathbf{t}_3$ since $\mathbf{t}_1$ cannot equal $\mathbf{t}_3$.  In other words, one solution is 
\begin{align}\label{eq:case11}
\omega = 0 \quad \text{ and } \quad \alpha = \pi - \beta.
\end{align}
For the other case, $\omega \in \{ \pm \pi \}$, we evidently must have $\mathbf{R}_4(\pm \pi) \mathbf{t}_1 =  \mathbf{t}_3$.  This is solved if and only if $\mathbf{t}_4 \cdot \mathbf{t}_1 = \mathbf{t}_4 \cdot \mathbf{t}_3$.  In other words, the other solution in this case is 
\begin{align}\label{eq:case12}
\omega \in \{ \pm \pi \} \quad \text{ and } \quad \alpha = \beta.
\end{align}
The results (\ref{eq:case11}) and (\ref{eq:case12}) provide a complete parameterization of $\mathbf{P}_3 \mathbf{R}_4(\omega) \mathbf{t}_1 = 0$.

(\textit{Solutions in Case 2.})  We now assume $\mathbf{P}_3 \mathbf{R}_4(\omega)\mathbf{t}_1 \neq 0$; equivalently, $(\omega, \alpha, \beta,\sigma)$ do not belong to the parameterizations in (\ref{eq:case11}) and (\ref{eq:case12}).  It follows that $\gamma_3 \in [-\pi, \pi]$ obeys the parameterization
\begin{equation}
\begin{aligned}\label{eq:cosSinGamma3}
&\cos(\gamma_3) = \frac{\mathbf{R}_4(\omega)\mathbf{t}_1 \cdot \mathbf{P}_3 \mathbf{R}^T_2(\sigma \omega) \mathbf{t}_1}{|\mathbf{P}_3\mathbf{R}_4(\omega) \mathbf{t}_1|^2}, \\
&\sin(\gamma_3) =\frac{\mathbf{t}_3 \cdot ( \mathbf{R}_4(\omega) \mathbf{t}_1 \times \mathbf{R}^T_2(\sigma \omega) \mathbf{t}_1)}{|\mathbf{P}_3\mathbf{R}_4(\omega) \mathbf{t}_1|^2}.
\end{aligned}
\end{equation}
using the geometric definitions of the dot product and cross product, as well as several identities that we have either justified previously or can be easily shown: $|\mathbf{P}_3 \mathbf{R}_4(\omega)\mathbf{t}_1| = | \mathbf{P}_3 \mathbf{R}_2^T(\sigma \omega) \mathbf{t}_1|$, $\mathbf{P}_3^T \mathbf{P}_3 = \mathbf{P}_3$, and $\mathbf{t}_3 \cdot ( \mathbf{R}_4(\omega) \mathbf{t}_1 \times \mathbf{R}_2(\sigma \omega) \mathbf{t}_1) = \mathbf{t}_3 \cdot (\mathbf{P}_3 \mathbf{R}_4(\omega) \mathbf{t}_1 \times \mathbf{P}_3 \mathbf{R}_3(\sigma \omega) \mathbf{t}_1)$ since $\mathbf{t}_3 \cdot \mathbf{R}_4(\omega) \mathbf{t}_1 = \mathbf{t}_3 \cdot \mathbf{R}_2^T(\sigma \omega) \mathbf{t}_1$.

Now, we should point out that it is always possible to find an angle $\gamma_3 \in [-\pi, \pi]$ which obeys the relation (\ref{eq:cosSinGamma3}) when $\mathbf{P}_3 \mathbf{R}_4(\omega)\mathbf{t}_1 \neq 0$.  In fact, this relation uniquely determines the angle  (up to a minor caveat \footnote{There is a degeneracy in rotations $\mathbf{R}_i(\pi) = \mathbf{R}_i(-\pi)$.  This means that when $\cos(\gamma_3) = -1$ and $\sin(\gamma_3) = 0$, we are free to choose $\gamma_3 = \pi$ or $-\pi$.  Nevertheless, physically a rotation by $\pi$ folds a region on top of another and a rotation by $-\pi$ folds that same region underneath the other.  In another sense, it is well-know that the four-fold setting corresponds to three mountains and a valley or three valleys and a mountain (Maekawa's theorem).  We will preserve this formalism in the case that a folding angle attains the magnitude of  $\pi$ to overcome this degeneracy.} that we will disregard in what follows, as it is of no physical consequence).  

To describe the explicit function for this relation, we first note that Kawasaki's condition furnishes the identity 
\begin{equation}
\begin{aligned}\label{eq:simplifyCos}
&\frac{\mathbf{R}_4(\omega)\mathbf{t}_1 \cdot \mathbf{P}_3 \mathbf{R}^T_2(\sigma \omega) \mathbf{t}_1}{|\mathbf{P}_3\mathbf{R}_4(\omega) \mathbf{t}_1|^2} = \frac{(\sigma 1 - c_{\alpha} c_{\beta})c_{\omega} + s_{\alpha} s_{\beta}}{(\sigma 1 - c_{\alpha} c_{\beta}) + s_{\alpha} s_{\beta} c_{\omega} } 
\end{aligned}
\end{equation}
by explicit verification.  There are evidently two special cases to (\ref{eq:cosSinGamma3}) made apparent by this formula, 
\begin{equation}
\begin{aligned}
\begin{cases}\label{eq:specialAgain}
\gamma_3 = 0 \;\; \forall \;\; \omega \in (-\pi, \pi) & \text{ if } \sigma = +, \alpha = \beta \\
\gamma_3 \in \{ \pm \pi \}  \;\; \forall \;\; \omega \in [-\pi, \pi]\setminus\{0\}  & \text{ if } \sigma = -, \alpha = \pi -\beta.
\end{cases}
\end{aligned}
\end{equation}  
(Remember, $\mathbf{P}_3 \mathbf{R}_4(\omega) \mathbf{t}_4 \neq 0$ for this characterization; hence, the restrictions on the domain for $\omega$.)  However, if $(\omega, \sigma, \alpha, \beta)$ do not belong to one of these two sets, then it is straightforward to observe that $\cos(\gamma_3) = 1$ if and only if $\omega = 0$ and $\cos(\gamma_3) = -1$ if and only if $\omega \in \{ \pm \pi\}$.  This means that $\sin(\gamma_3)$, as given in (\ref{eq:cosSinGamma3}), does not change signs on the interval $\omega \in (0,\pi)$ and $\omega \in (-\pi, 0)$.  Thus, since the magnitude of $\gamma_3$ is determined by taking the $\arccos$ of (\ref{eq:simplifyCos}), we need only determine the sign by, for instance, approximating the parameterization for $\sin(\gamma_3)$ for $\omega \ll 1$.  In this direction, we observe that, to leading order in $\omega$,
\begin{equation}
\begin{aligned}
&\mathbf{t}_3  \cdot ( \mathbf{R}_4(\omega) \mathbf{t}_1 \times \mathbf{R}^T_2(\sigma \omega) \mathbf{t}_1)  \\
&\qquad \approx \begin{cases}
2(c_{\alpha} - c_{\beta})\big(c_{\frac{\alpha+\beta}{2}}\big)^2\omega  & \text{ if } \sigma = + \\
2(c_{\alpha} + c_{\beta})\big(s_{\frac{\alpha+\beta}{2}}\big)^2\omega & \text{ if } \sigma = -.
\end{cases}
\end{aligned}
\end{equation}
Note that $(c_{\frac{\alpha+\beta}{2}})^2 > 0$ since $0 < \alpha + \beta < 2\pi$, and $(s_{\frac{\alpha+\beta}{2}})^2 > 0$ since $\alpha = \pi - \beta$ for $\sigma = -$ is not relevant by (\ref{eq:specialAgain}).  Thus, as we only really require the sign from the formula, we conclude that 
\begin{equation}
\begin{aligned}\label{eq:gamma3Solve}
\gamma_3 &= \bar{\gamma}_3^{\sigma}(\omega)\\
&=  \text{sign}\Big((c_{\alpha} - \sigma c_{\beta}) \omega \Big) \arccos\Big(\tfrac{(\sigma 1 - c_{\alpha} c_{\beta})c_{\omega} + s_{\alpha} s_{\beta}}{(\sigma 1 - c_{\alpha} c_{\beta}) + s_{\alpha} s_{\beta} c_{\omega} } \Big)
\end{aligned}
\end{equation}
whenever $(\omega, \alpha, \beta, \sigma) \in [-\pi, \pi] \times (0,\pi)^2 \times \{ \pm \}$ do not satisfy the parameterizations in (\ref{eq:case11}), (\ref{eq:case12}) or (\ref{eq:specialAgain}).  At this point, we notice that the formula here is also valid for all $\omega \in [-\pi, \pi]$ and $\sigma \in \mathcal{A}$, as asserted in (\ref{kin1Sup}). That is, the singularity $\mathbf{P}_3 \mathbf{R}_4(\omega) \mathbf{t}_1 = 0$ disappears  under the simplification in (\ref{eq:simplifyCos}) when $\sigma = +$, $\omega = 0$ and $\alpha = \pi - \beta$ and when $\sigma  = -$, $\omega \in \{ \pm \pi \}$ and $\alpha = \beta$.  Thus, (\ref{eq:gamma3Solve}) for $\sigma \in \mathcal{A}$ and the special cases in (\ref{eq:specialAgain}) provide a complete parameterization for case 2.

To summarize, we have completely solved the necessary condition in (\ref{eq:secondNec}) by the parameterizations outlined.  To complete the proof, we focus first on the case that $\sigma \in \mathcal{A}$, which gives that $\gamma_3 = \bar{\gamma}_3^{\sigma}(\omega)$ for all $\omega \in [-\pi, \pi]$.  Since the necessary condition (\ref{eq:secondNec}) holds for this parameterization, we conclude that 
\begin{align}
\mathbf{R}_2(\sigma \omega) \mathbf{R}_3(\bar{\gamma}_3^{\sigma}(\omega))\mathbf{R}_4(\omega) \mathbf{t}_1 = \mathbf{t}_1.
\end{align}  
Consequently, there exists a $\gamma_1 \in [-\pi, \pi]$ such that $\mathbf{R}_1(\gamma_1) \mathbf{R}_2(\sigma \omega) \mathbf{R}_3(\bar{\gamma}_3^{\sigma}(\omega))\mathbf{R}_4(\omega) = \mathbf{I}$.  To determine it explicitly, we notice that we must have $|\gamma_1| = |\bar{\gamma}_3^{\sigma}(\omega)|$ by an argument similar to that of (\ref{eq:firstNec}) and (\ref{eq:derive1}).  In addition, since all the folding angles are engaged in this parameterization, Maekawa's theorem tells us that the folding angles (when each has magnitude $|\gamma_i| \in (0, \pi)$) should correspond to three mountains and a valley or three valleys and a mountain.  In other words, $\gamma_1 \gamma_2 \gamma_3 \gamma_4 \leq 0$ under this parameterization.  In combination, we obtain $\gamma_1 = -\sigma \bar{\gamma}_3^{\sigma}(\omega)$ \footnote{One could, of course, also deduce this result by direct calculation without resorting to Maekawa's theorem.}.   This completes the proof for the four-fold solutions.  The other solutions (i.e., those indicated by the special cases (\ref{eq:case11}), (\ref{eq:case12}) or (\ref{eq:specialAgain})) are simply describing the process of folding-in-half and folding-in-half again.  It is thus straightforward to confirm that (\ref{kin2Sup}) gives the complete parameterization for this case.  
\end{proof}

\subsection{On HMO as objective structures.}
An objective structure is the orbit of a point (or a collection of points) under a discrete group of isometries.  An \textit{isometry} is any map $g = (\mathbf{R}|\mathbf{c})$ such that $g(\mathbf{x}) = \mathbf{R}\mathbf{x} + \mathbf{c}$ for $\mathbf{R} \in SO(3)$ and $\mathbf{c} \in \mathbb{R}^3$ \footnote{Technically, one should also include reflections in the definition of an isometry, but the reflections are not relevant to this work.  Hence, their exclusion here.}.  A \textit{proper} rotation is any $\mathbf{R} \in SO(3)$ that is not $\mathbf{I}$, and it is characterized by a unique (up to a change in sign) axis given by $\mathbf{e} \in \mathbb{S}^2$ such that $\mathbf{R} \mathbf{e} = \mathbf{e}$.  We will always define the axis $\mathbf{e}$ such that a proper rotation $\mathbf{R}$ is a right-hand rotation about this axis by an angle $\theta \in (-\pi, \pi]\setminus \{0\}$.  We say that two proper rotations $\mathbf{R}_1$ and $\mathbf{R}_2$ have distinct axes $\mathbf{e}_1$ and $\mathbf{e}_2 \in \mathbb{S}^2$ whenever $\mathbf{e}_1 \neq \pm \mathbf{e}_2$.  We have the following classification theorem on the abelian groups:
\begin{theorem}\label{CharacterizeTheorem} The complete enumeration of the discrete and abelian groups of isometries is as follows.
\begin{enumerate}
\item[(i).] Those whose rotation component does not contain a proper rotation:
\begin{itemize}
\item ``Points" $-$ $\mathcal{G} = \{ id\}$;
\item ``Lines" $-$   $\mathcal{G} = \{t_1^p \colon p \in \mathbb{Z}\}$;
\item ``2D Lattices" $-$ $\mathcal{G} = \{t_1^p t_2^q \colon p,q \in \mathbb{Z} \}$; 
\item ``3D Lattices" $-$ $\mathcal{G} = \{t_1^p t_2^q t_3^r \colon p,q,r \in \mathbb{Z}\}$.
\end{itemize}
Here, $t_i = (\mathbf{I}| \mathbf{c}_i)$, $i =1,2,3,$ and $\text{span} \{ \mathbf{c}_1, \mathbf{c}_2, \mathbf{c}_3\} = \mathbb{R}^3$.
\item[(ii).] Those whose rotation component contains a proper rotation but does not contain two proper rotations of distinct axes:
\begin{itemize}
\item ``Rings" $-$ $\mathcal{G} = \{g_0^r \colon r \in \mathbb{Z} \} $; 
\item ``Cylinders" $-$ $\mathcal{G} = \{ g_1^p g_2^q \colon p,q \in \mathbb{Z} \}$.
\end{itemize}
Here, $g_i = (\mathbf{R}_{\theta_i}| (\mathbf{I} - \mathbf{R}_{\theta_i})\mathbf{z} + \tau_i \mathbf{e})$, $i = 0,1,2,$ for some axis of rotation $\mathbf{e} \in \mathbb{S}^2$ and some origin $\mathbf{z} \in \mathbb{R}^3$ with $\mathbf{z} \cdot \mathbf{e} = 0$.   In addition, the extension $\tau_i \in \mathbb{R}$ and twist $\theta_i \in (-\pi, \pi]$ satisfy
\begin{equation}
\begin{aligned}
&(\text{for rings:}) \quad \tau_0 = 0 \quad \text{ and } \quad  r^{\star} \theta_0 = 2\pi, \\
&(\text{for cylinder:}) \quad \begin{cases}
\tau_1^2 + \tau_2^2 > 0  \\
p^{\star} \tau_1 + q^{\star} \tau_2 = 0 \\ 
p^{\star} \theta_1 + q^{\star} \theta_2 = 2\pi, 
\end{cases}
\end{aligned}
\end{equation}
for some $p^{\star}, q^{\star}, r^{\star} \in \mathbb{Z}$.
\item[(iii).] Those whose rotation component contains at least two proper rotations of distinct axes:
\begin{itemize}
\item ``Tetrahedron" $-$ $\{ h_1^p h_2^q \colon p,q \in \{ 0,1\}\};$
\end{itemize}
Here,   $h_i = (\mathbf{R}_i |\mathbf{c}_i)$, $i =1,2$ satisfy
\begin{equation}
\begin{aligned}
&\mathbf{R}_i = 2 \mathbf{e}_i \otimes \mathbf{e}_i - \mathbf{I}, \quad i =1,2,\\
&\mathbf{c}_1 = \eta \mathbf{e}_2 + \xi \mathbf{e}_3 , \quad \mathbf{c}_2 = \mu \mathbf{e}_1 + \xi \mathbf{e}_3, 
\end{aligned}
\end{equation}
for an orthonormal basis $\{ \mathbf{e}_1, \mathbf{e}_2, \mathbf{e}_3\} \subset \mathbb{R}^3$ and an $(\eta, \xi, \mu) \in \mathbb{R}^3$.
\end{enumerate}
\end{theorem}
\noindent This is proved elsewhere \cite{dayal2010objective}.  

In this work, we focus on how the partially folded Miura parallelogram unit cell interacts with the so-called ``cylinder group" above.  In particular, we provide a complete classification of the compatible structures generated by such an interaction and thus, a complete characterization of HMO structures. 

\subsection{On the HMO design equations}
We assume the reference Miura parallelogram $\Omega$ is partially folded by a deformation $\mathbf{y} \equiv \mathbf{y}_{\omega}^{\sigma}$ parameterized by (\ref{eq:yDefSup}) with folding angles $\gamma_i  \equiv \gamma_i(\omega, \sigma)$ (for $i = 1,2,3,4$) that satisfy either (\ref{kin1Sup}) or the first fold-in-half parameterization in (\ref{kin2Sup}) for some $\omega \in [-\pi, \pi]$ and $\sigma \in \{ \pm \}$.  As such, we denote this partially folded Miura parallelogram with $\mathbf{y}_{\omega}^{\sigma}(\Omega)$ and thereby determine the corners of the parallelogram by $\bfy_i = \mathbf{y}_{\omega}^{\sigma}(\mathbf{x}_i)$ for $i=1, 2,3 ,4$.  With the corners, we suppress the dependence on $\omega$ and $\sigma$ to simplify the notation, but it is always implicit in what follows.  

 We consider two screw transformations $g_i = (\bfR_{\theta_i} | \tau_i \bfe  + (\bfI - \bfR_{\theta_i})\bfz)$ ($i = 1,2$) with parameters $\mathbf{R}_{\theta_i} \in SO(3)$, $\theta_{i} \in (-\pi,\pi]$, $\tau_{i} \in \mathbb{R}$, $\mathbf{e} \in \mathbb{S}^2$ and $\mathbf{z} \in \mathbb{R}^3$, $\mathbf{z} \cdot \mathbf{e} =0$ characterizing the rotation, rotation angle, translation, rotation axis and origin of the isometry, respectively.  These operate on a point $\mathbf{x} \in \mathbb{R}^3$ in the standard way $g_i(\mathbf{x}) = \mathbf{R}_{\theta_i}( \mathbf{x} - \mathbf{z}) + \tau_i \mathbf{e}_i + \mathbf{z}$.   
 
 To construct HMO structures, we apply these transformations to the corners of the Miura parallelogram, but there are restrictions.  As discussed in the main text, the parameters of these transformations must satisfy local compatibility and the non-degeneracy condition, i.e., 
 \begin{equation}
 \begin{aligned}\label{eq:localCompSup}
&g_1(\mathbf{y}_4) = \mathbf{y}_1,\quad g_1(\mathbf{y}_3) = \mathbf{y}_2, \\
&g_2(\mathbf{y}_1) = \mathbf{y}_2,\quad  g_2(\mathbf{y}_4) = \mathbf{y}_3.
\end{aligned}
\end{equation} 
Below, we derive all possible solutions to this system of equations.

To consolidate the notation in what follows, we define the side length vectors
\begin{equation}
\begin{aligned}\label{eq:vectorDefs}
\mathbf{u}_a =  \mathbf{y}_3 - \mathbf{y}_4, \quad \mathbf{u}_b = \mathbf{y}_2 - \mathbf{y}_1, \\
\mathbf{v}_a = \mathbf{y}_1 -\mathbf{y}_4, \quad \mathbf{v}_b = \mathbf{y}_2 - \mathbf{y}_3.
\end{aligned}
\end{equation}
We claim the following under the stated hypotheses on the parameters: If $\theta_1 \neq 0$, then (\ref{eq:localCompSup}) is  equivalent to 
\begin{equation}
\begin{aligned}\label{eq:rewriteLocal}
&\tau_1 = \mathbf{v}_a \cdot \mathbf{e}, \quad  \tau_2 = \mathbf{u}_a \cdot \mathbf{e}, \quad (\mathbf{u}_a - \mathbf{u}_b) \cdot \mathbf{e} = 0, \\
&\mathbf{R}_{\theta_1} \mathbf{P}_{\mathbf{e}} \mathbf{u}_a = \mathbf{P}_{\mathbf{e}} \mathbf{u}_b, \quad \mathbf{R}_{\theta_2} \mathbf{P}_{\mathbf{e}} \mathbf{v}_a = \mathbf{P}_{\mathbf{e}} \mathbf{v}_b, \\
&(\mathbf{I} - \mathbf{R}_{\theta_1}) \mathbf{z} = \mathbf{P}_{\mathbf{e}}(\mathbf{y}_2 - \mathbf{R}_{\theta_1} \mathbf{y}_3),
\end{aligned}
\end{equation}
where $\mathbf{P}_{\mathbf{e}}= \mathbf{I} - \mathbf{e} \otimes \mathbf{e}$.  Note, in this system there are nine equations for local compatibility, whereas in (\ref{eq:localCompSup}), there are twelve.  Thus, by this equivalence, we are exposing the redundancies in the original characterization of local compatibility.   For  the case $\theta_1 = 0$, (\ref{eq:rewriteLocal}) is necessary for (\ref{eq:localCompSup}) but not sufficient.  This special case is treated separately below.
\begin{proof}
((\ref{eq:localCompSup}) $\Rightarrow$ (\ref{eq:rewriteLocal})).  We begin with the easier direction.  The first equation $\tau_1 = \mathbf{v}_a \cdot \mathbf{e}$ is obtained by dotting $g_1(\mathbf{y}_4)= \mathbf{y}_1$ with $\mathbf{e}$ and using the stated properties of the parameters.  The second equation $\tau_2 = \mathbf{u}_a \cdot \mathbf{e}$ is obtained, in similar fashion, by dotting $g_2(\mathbf{y}_4) = \mathbf{y}_3$ with $\mathbf{e}$.  For the third, we first dot $g_2(\mathbf{y}_1) = \mathbf{y}_2$ with $\mathbf{e}$ to obtain $\tau_2 = \mathbf{u}_b \cdot \mathbf{e}$ (just like  the other cases), and we combine this with $\tau_2 = \mathbf{u}_a \cdot \mathbf{e}$ to get $(\mathbf{u}_a - \mathbf{u}_b) \cdot \mathbf{e} = 0$. Moving on to the second line, we obtain $\mathbf{R}_{\theta_1} \mathbf{P}_{\mathbf{e}} \mathbf{u}_a = \mathbf{P}_{\mathbf{e}} \mathbf{u}_b$ by first taking the difference between $g_1(\mathbf{y}_4) = \mathbf{y}_1$ and $g_1(\mathbf{y}_3) = \mathbf{y}_2$. We then premultiply this difference by $\mathbf{P}_{\mathbf{e}}$ and use the fact that $\mathbf{P}_{\mathbf{e}} \mathbf{R}_{\theta_1} = \mathbf{R}_{\theta_1} \mathbf{P}_{\mathbf{e}}$ to get the result.  We obtain $\mathbf{R}_{\theta_2} \mathbf{P}_{\mathbf{e}} \mathbf{v}_a = \mathbf{P}_{\mathbf{e}} \mathbf{v}_b$ by a similar manipulation with the two equations $g_2(\mathbf{y}_1) = \mathbf{y}_2$ and  $g_2(\mathbf{y}_4) = \mathbf{y}_3$.  Finally, for the third line, we obtain $(\mathbf{I} - \mathbf{R}_{\theta_i}) \mathbf{z} = \mathbf{P}_{\mathbf{e}}(\mathbf{y}_2 - \mathbf{R}_{\theta_1} \mathbf{y}_3)$ by premultiplying the equation $g_1(\mathbf{y}_3) = \mathbf{y}_2$ by $\mathbf{P}_{\mathbf{e}}$ and rearranging.  

((\ref{eq:localCompSup}) $\Leftarrow$ (\ref{eq:rewriteLocal}) when $\theta_1 \neq 0$).  For this direction, we first observe the following equivalences:
\begin{equation}
\begin{aligned}
&g_1(\mathbf{y}_3) - g_1(\mathbf{y}_4) = \mathbf{y}_2 - \mathbf{y}_1 \quad \Leftrightarrow \quad \mathbf{R}_{\theta_1} \mathbf{u}_a = \mathbf{u}_b  \\
&\qquad \Leftrightarrow \quad \mathbf{R}_{\theta_1} \mathbf{P}_{\mathbf{e}} \mathbf{u}_a + \big(\mathbf{e} \cdot (\mathbf{u}_a - \mathbf{u}_b) \big) \mathbf{e} = \mathbf{P}_{\mathbf{e}} \mathbf{u}_b. 
\end{aligned}
\end{equation}
Consequently, (\ref{eq:rewriteLocal}) implies $g_1(\mathbf{y}_3) - g_1(\mathbf{y}_4) = \mathbf{y}_2 - \mathbf{y}_1$.   In similar fashion, we observe that  
\begin{equation}
\begin{aligned}
&g_2(\mathbf{y}_1) - g_2(\mathbf{y}_4) = \mathbf{y}_2 - \mathbf{y}_3 \quad \Leftrightarrow \quad \mathbf{R}_{\theta_1} \mathbf{v}_a = \mathbf{v}_b  \\
&\qquad \Leftrightarrow \quad \mathbf{R}_{\theta_1} \mathbf{P}_{\mathbf{e}} \mathbf{v}_a + \big(\mathbf{e} \cdot (\mathbf{v}_a - \mathbf{v}_b) \big) \mathbf{e} = \mathbf{P}_{\mathbf{e}} \mathbf{v}_b. 
\end{aligned}
\end{equation}
Consequently, (\ref{eq:rewriteLocal}) implies $g_2(\mathbf{y}_1) - g_2(\mathbf{y}_4) = \mathbf{y}_2 - \mathbf{y}_3$ since $(\mathbf{u}_a - \mathbf{u}_b) \cdot \mathbf{e} = (\mathbf{v}_a - \mathbf{v}_b) \cdot \mathbf{e}$.   Next, we observe that 
\begin{equation}
\begin{aligned}\label{eq:equivalence3Sup}
&g_1(\mathbf{y}_3) = \mathbf{y}_2  \quad  \Leftrightarrow  \\
&\quad (\mathbf{I} - \mathbf{R}_{\theta_i}) \mathbf{z} + \big(\tau_1 - \mathbf{v}_b \cdot \mathbf{e} \big) \mathbf{e} = \mathbf{P}_{\mathbf{e}}( \mathbf{y}_2 - \mathbf{R}_{\theta_1} \mathbf{y}_3).
\end{aligned}
\end{equation}
In addition, (\ref{eq:rewriteLocal}) implies $\tau_1 = \mathbf{v}_b \cdot \mathbf{e}$ since it gives $(\mathbf{u}_a - \mathbf{u}_b) \cdot \mathbf{e} = 0$  and $\tau_1 = \mathbf{v}_a \cdot \mathbf{e}$ and since  $(\mathbf{v}_a - \mathbf{v}_b) \cdot \mathbf{e} =(\mathbf{u}_a - \mathbf{u}_b) \cdot \mathbf{e}$.   In combination with the equivalence (\ref{eq:equivalence3Sup}), we conclude that (\ref{eq:rewriteLocal}) implies $g_1(\mathbf{y}_3) = \mathbf{y}_2$. 

In summary, we have shown so far that (\ref{eq:rewriteLocal}) implies 
\begin{equation}
\begin{aligned}\label{eq:almostDone}
&g_1(\mathbf{y}_3) = \mathbf{y}_2 , \quad g_1(\mathbf{y}_4) = \mathbf{y}_1, \\
&g_2(\mathbf{y}_1) - \mathbf{y}_2 = g_2(\mathbf{y}_4) - \mathbf{y}_3.
\end{aligned}
\end{equation}
We can say more.  Observe that $g_i(\mathbf{x}- \mathbf{y}) = g_i(\mathbf{x}) - g_i(\mathbf{y}) + g_i(\mathbf{0})$ for any $\mathbf{x}, \mathbf{y} \in \mathbb{R}^3$ and $i = 1,2$.  In addition, the groups are abelian, i.e., $g_1g_2(\mathbf{x}) = g_2g_1(\mathbf{x})$.   Thus, from (\ref{eq:almostDone}), we evidently have the identity: 
\begin{equation}
\begin{aligned}\label{eq:coolIdent}
g_1\Big(g_2(\mathbf{y}_1) - \mathbf{y}_2 \Big) &= g_1 \Big(g_2(\mathbf{y}_4) - \mathbf{y}_3\Big) \\
&= g_1g_2(\mathbf{y}_4) - g_1(\mathbf{y}_3) + g_1(\mathbf{0}) \\
&= g_2g_1(\mathbf{y}_4) - \mathbf{y}_2 + g_1(\mathbf{0}) \\
&= g_2(\mathbf{y}_1) - \mathbf{y}_2 + g_1(\mathbf{0}).
\end{aligned}
\end{equation}
By subtracting $g_1(\mathbf{0})$ from both sides and  expanding out the $g_1(\cdot)$ terms, the above identity takes on the revealing form 
\begin{equation}
\begin{aligned}
&\mathbf{R}_{\theta_1}\Big( g_2(\mathbf{y}_1) - \mathbf{y}_2 \Big) = g_2(\mathbf{y}_1) - \mathbf{y}_2,
\end{aligned} 
\end{equation}
especially given that
\begin{align}
\mathbf{e} \cdot ( g_2(\mathbf{y}_1) - \mathbf{y}_2) = \tau_2 - \mathbf{e} \cdot \mathbf{v}_b = 0
\end{align}
from (\ref{eq:rewriteLocal}).  In particular, since $g_2(\mathbf{y}_1) - \mathbf{y}_2 \perp \mathbf{e}$ and since the identity (\ref{eq:coolIdent}) is evidently implied by (\ref{eq:almostDone}), we can infer that one of the two possibilities must be true: (a) $\theta_1 = 0$ or (b) $g_2(\mathbf{y}_1) = \mathbf{y}_2$.  As we have assumed $\theta_1\neq 0$ for this characterization, we have $g_2(\mathbf{y}_1) = \mathbf{y}_2$.  It therefore follows from (\ref{eq:almostDone}) that $g_2(\mathbf{y}_4) = \mathbf{y}_3$, and thus (\ref{eq:localCompSup}) holds as desired.
\end{proof}

We now turn to the case that $\theta_1 = 0$.  In this direction, consider the following system of equations:
\begin{equation}
\begin{aligned}\label{eq:theta1Eq0}
&\theta_1 = 0, \quad \mathbf{u}_a = \mathbf{u}_b, \quad \tau_2 = \mathbf{u}_a \cdot \mathbf{e}, \\
&\mathbf{v}_a = \mathbf{v}_b, \quad \mathbf{v}_b = \tau_1 \mathbf{e}, \\
&(\mathbf{I} - \mathbf{R}_{\theta_2}) \mathbf{z} = \mathbf{P}_{\mathbf{e}}(\mathbf{y}_2 - \mathbf{R}_{\theta_2} \mathbf{y}_1), \quad \theta_2 \neq 0. 
\end{aligned}
\end{equation}
We claim that 
\begin{equation}
\begin{aligned}\label{eq:implicationsTheta1}
 (\ref{eq:localCompSup}) \text{ with } \theta_1 = 0, \theta_2 \neq 0 \quad \Leftrightarrow \quad  (\ref{eq:theta1Eq0}).
\end{aligned}
\end{equation}
\begin{proof} 
($\Rightarrow.$) From the previous result, we have that (\ref{eq:localCompSup}) $\Rightarrow$ (\ref{eq:rewriteLocal}) in general.  Therefore, the implication holds for $\theta_1 = 0$.  Thus, $\tau_2 = \mathbf{u}_a \cdot \mathbf{e}$ is trivial. Further,  by substituting $\theta_1 = 0$ into (\ref{eq:rewriteLocal}), we easily deduce the $\mathbf{u}_a = \mathbf{u}_b$ and $\mathbf{v}_a = \mathbf{v}_b$.  By substituting the latter back into (\ref{eq:rewriteLocal}), we have $\mathbf{R}_{\theta_2} \mathbf{P}_{\mathbf{e}} \mathbf{v}_b = \mathbf{P}_{\mathbf{e}} \mathbf{v}_b$.  Notice if $\mathbf{P}_{\mathbf{e}} \mathbf{v}_a \neq 0$, then evidently $\theta_2 = 0$. We therefore have $\mathbf{P}_{\mathbf{e}}\mathbf{v}_b= \mathbf{0}$, as $\theta_2 \neq 0$ is assumed.  Since additionally $\tau_2 = \mathbf{e} \cdot \mathbf{v}_a$ in (\ref{eq:rewriteLocal}), we conclude $\mathbf{v}_b =  \tau_2 \mathbf{e}$.   Thus, for this direction, the only result in (\ref{eq:theta1Eq0}) that remains to be justified is $(\mathbf{I} - \mathbf{R}_{\theta_2}) \mathbf{z} = \mathbf{P}_{\mathbf{e}}(\mathbf{y}_2 - \mathbf{R}_{\theta_2} \mathbf{y}_1$.  This identity is obtained from (\ref{eq:localCompSup}) by projecting the identity $g_2(\mathbf{y}_1) = \mathbf{y}_2$ onto the plane with normal $\mathbf{e}$.  

($\Leftarrow$.) We observe that (\ref{eq:theta1Eq0}) is a solution to (\ref{eq:rewriteLocal}). (Note, $(\mathbf{I} - \mathbf{R}_{\theta_1}) \mathbf{z} = \mathbf{0}$, and also  $\mathbf{P}_{\mathbf{e}}(\mathbf{y}_2 - \mathbf{R}_{\theta_1} \mathbf{y}_3) = \mathbf{0}$ in this case.)  Thus, the previous analysis guarantees the three identities in (\ref{eq:almostDone}).  We claim also that $g_2(\mathbf{y}_1)=\mathbf{y}_2$.  Indeed, this follows by combining three identities from (\ref{eq:theta1Eq0}): $(\mathbf{I} - \mathbf{R}_{\theta_2}) \mathbf{z} = \mathbf{P}_{\mathbf{e}}(\mathbf{y}_2 - \mathbf{R}_{\theta_2} \mathbf{y}_1)$,  $\mathbf{u}_a = \mathbf{u}_b$ and  $\tau_2 = \mathbf{u}_a \cdot \mathbf{e}$.  Thus, (\ref{eq:localCompSup}) follows by combining (\ref{eq:almostDone}) with $g_2(\mathbf{y}_1) = \mathbf{y}_2$.
\end{proof}

Note, we do not include the equivalence for the case $\theta_1 = \theta_2 = 0$, as such parameterizations of local compatibility in (\ref{eq:localCompSup}) can never satisfy the discreteness condition.

\begin{figure}[!t]
	\includegraphics[width=\columnwidth]{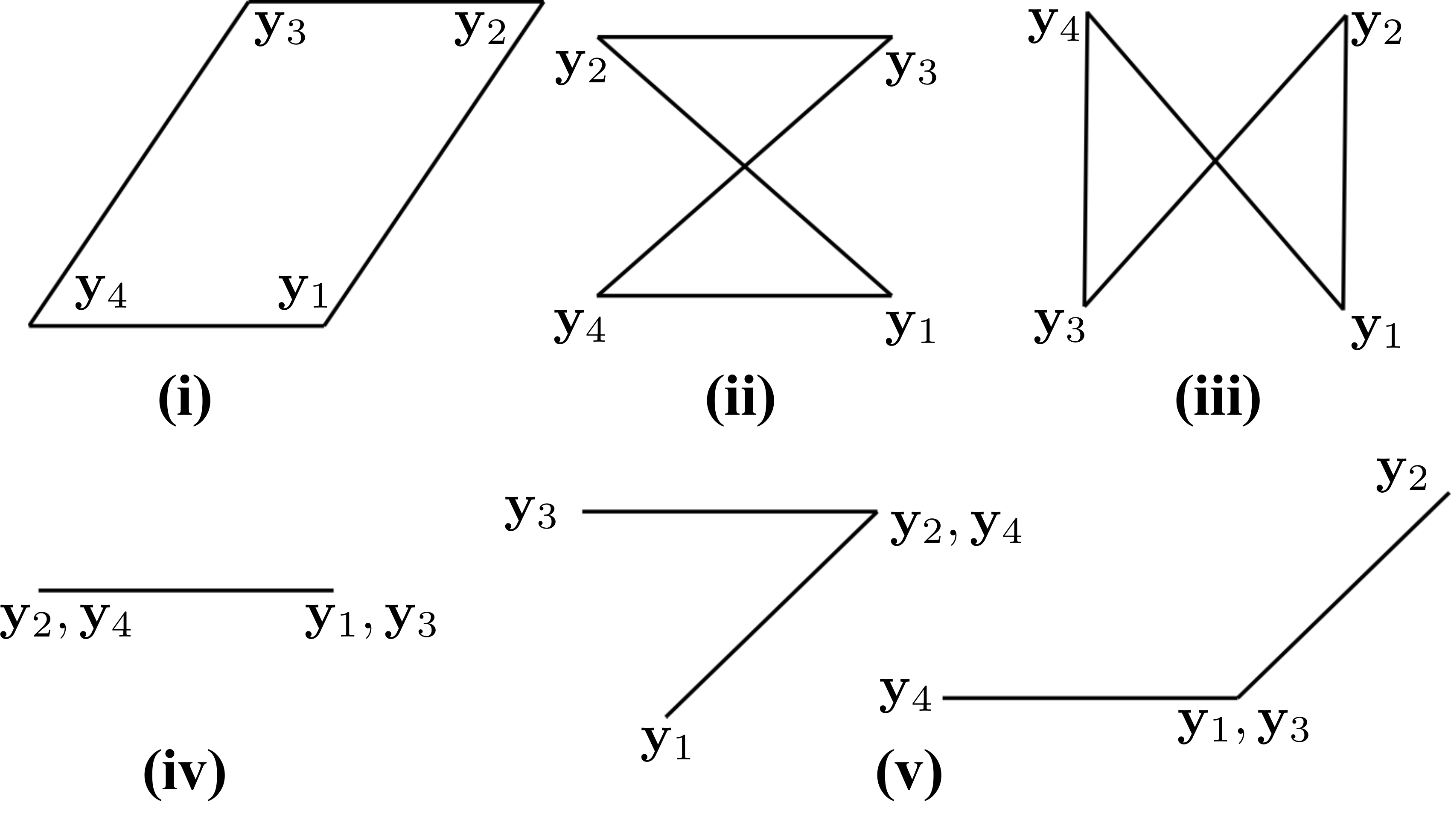}
	\caption{The boundary of a Miura parallelogram (with corner point displayed) when it is unfolded (i) and in various folded flat configurations (ii-v).  This observation is used in the Proposition \ref{vectorsProp}. }
	\label{fig:Flat}
\end{figure}

We aim now to fully parameterize the two possible solutions to local compatibility and non-degeneracy given by (\ref{eq:rewriteLocal}) and (\ref{eq:theta1Eq0}).  To do this, we make the following  observations relating the side length vectors $\mathbf{u}_{a,b}$ and $\mathbf{v}_{a,b}$ in (\ref{eq:vectorDefs}) to the fold angle parameter $\omega$.  
\begin{proposition}\label{vectorsProp}
For the Miura parallelogram $\mathbf{y}_{\omega}^{\sigma}(\Omega)$, the following statements are true:
\begin{enumerate}
\item[(i).] If $\omega = 0$, then $\mathbf{u}_a = \mathbf{u}_b$ and $\mathbf{v}_a = \mathbf{v}_b$.  
\item[(ii).] If $\omega \in \{ \pm \pi \}$ and $1 = |\mathbf{u}_a| > |\mathbf{v}_a| = l$, then $\mathbf{u}_a \nparallel \mathbf{u}_b$ and $\mathbf{v}_a = - \mathbf{v}_b$.
\item[(iii).] If $\omega \in \{ \pm \pi \}$ and $1 = |\mathbf{u}_a| < |\mathbf{v}_a| = l$, then $\mathbf{u}_a = - \mathbf{u}_b$ and $\mathbf{v}_a \nparallel \mathbf{v}_b$.
\item[(iv).] If $\omega \in \{ \pm \pi \}$, $1 = |\mathbf{u}_a| = |\mathbf{v}_a| = l$, and $\sigma \in \mathcal{A}$, then $\mathbf{u}_a = - \mathbf{u}_b = \mathbf{v}_a = - \mathbf{v}_b$. 
\item[(v).] If $\omega \in \{ \pm \pi \}$, $1 = |\mathbf{u}_a| = |\mathbf{v}_a| = l$, and $\sigma \notin \mathcal{A}$, then $\mathbf{u}_a \nparallel \mathbf{u}_b$ and $\mathbf{v}_a \nparallel \mathbf{v}_b$.
\item[(vi).] If $\omega \in (-\pi, \pi) \setminus \{ 0\}$, then $\mathbf{u}_a \nparallel \mathbf{u}_b$ and $\mathbf{v}_a \nparallel \mathbf{v}_b$. 
\end{enumerate}
\end{proposition}
\begin{proof}
The first five statements follow from a geometric argument.   We recognize that, when $\omega = 0$ or $\in \{ \pm \pi\}$, the Miura parallelogram $\mathbf{y}_{\omega}^{\sigma}(\Omega)$ is flat or folded-flat.  This means that the corner points $\mathbf{y}_i$, $i = 1,2,3,4,$ all lie in the same plane. Since the side-length vectors (\ref{eq:vectorDefs}) satisfy $|\mathbf{u}_a| = |\mathbf{u}_b| = 1$ and $|\mathbf{v}_a| = |\mathbf{v}_b| = l$, this plane condition is very restrictive.   One can convince oneself that the only way to put the corner points in a plane in a manner consistent with isometric origami is as displayed in Fig,\;\ref{fig:Flat}(i-v).  These correspond to the unfolded state, or various fully folded configurations with the properties as stated in the proposition.

Now, for the last statement (vi), we show that $\mathbf{u}_a \parallel \mathbf{u}_b$ or $\mathbf{v}_a \parallel \mathbf{v}_b$ implies $\omega \in \{ 0, \pm \pi\}$.  This then proves the claim.   Note, the statement is geometrically obvious for the fold-in-half cases $\sigma \notin \mathcal{A}$. Thus we focus on the generic case $\sigma \in \mathcal{A}$.  We first recall the parameter $\eta \in (0,\pi)$, which is the angle between $\mathbf{x}_2 - \mathbf{x}_1$ and $\mathbf{x}_4 - \mathbf{x}_1$.  We, therefore, let $\eta^{\sigma}(\omega) \in [0, \pi]$ be the  angle between $\mathbf{y}_2 - \mathbf{y}_1$ and $\mathbf{y}_4 - \mathbf{y}_1$ (note, this depends on $\omega$ since the corner points implicitly depend on $\omega$).  By a direct calculation, one can show that $\eta^{\sigma}(\omega)$, $\sigma \in \mathcal{A}$, is an even function (i.e., $\eta^{\sigma}(-\omega) = \eta^{\sigma}(\omega)$ for $\omega \in [-\pi, \pi]$) and, further, that it is a strictly decreasing function on the interval $[0, \pi]$.   We use these properties to verify the result.  

Let $\sigma \in \mathcal{A}$ and $\mathbf{u}_a \parallel \mathbf{u}_b$ and/or $\mathbf{v}_a \parallel \mathbf{v}_b$. Since $|\mathbf{u}_a| = |\mathbf{u}_b|$ and $|\mathbf{v}_a| = |\mathbf{v}_b|$, at least one of the following is true: $\mathbf{u}_a = \mathbf{u}_b$; $\mathbf{u}_a = - \mathbf{u}_b$; $\mathbf{v}_a = \mathbf{v}_b$; $\mathbf{v}_a = - \mathbf{v}_b$.  For the case, $\mathbf{u}_a = \mathbf{u}_b$, it follows that $\mathbf{v}_a = \mathbf{v}_b$ since $\mathbf{u}_a - \mathbf{u}_b = \mathbf{v}_a - \mathbf{v}_b$ (see (\ref{eq:vectorDefs})).  Consequently, the boundary of the Miura parallelogram (up to rigid motion) is as depicted in Fig.\;\ref{fig:Flat}(i). This means that $\eta^{\sigma}(\omega) = \eta$.  We also know that $\eta = \eta^{\sigma}(0)$.  Thus, due to the properties of $\eta^{\sigma}(\omega)$, it must be that $\omega = 0$.   Therefore $\mathbf{u}_a = \mathbf{u}_b$ implies $\omega = 0$.  Similarly, $\mathbf{v}_a = \mathbf{v}_b$ implies $\omega = 0$. Now suppose $\mathbf{u}_a = -\mathbf{u}_b$.  It follows  $|\mathbf{u}_a| \leq  |\mathbf{v}_a|$ and that the boundary of the Miura parallelogram (up to rigid motion) is as depicted in Fig.\;\ref{fig:Flat}(iii) or (iv).  We therefore have that $\eta^{\sigma}(\omega) = \eta^{\sigma}(\pi)$ in this case. Due to the properties of $\eta^{\sigma}(\omega)$, we conclude that $\omega \in \{ \pm \pi\}$.   Therefore $\mathbf{u}_a = - \mathbf{u}_b$ implies $\omega \in \{ \pm \pi\}$. The only case remaining is $\mathbf{v}_a = - \mathbf{v}_b$ and $\mathbf{u}_a \nparallel \mathbf{u}_b$.  This case gives that the boundary of the Miura parallelogram is (up to rigid motion) as depicted in Fig.\;\ref{fig:Flat}(ii).  Arguing as before, we again deduce that $\omega \in \{ \pm \pi\}$.  This exhausts all cases and completes the proof.
\end{proof}

\textbf{The degenerate cases.}  These are the cases where $\omega = 0$.  We have two parameterizations to consider.  

From Proposition \ref{vectorsProp},  $\omega= 0$ implies $\mathbf{u}_a  = \mathbf{u}_b$ and $\mathbf{v}_a = \mathbf{v}_b$.  Thus one way of solving for the group parameters  in this case is to set $\theta_1 = 0$.  Consequently, a complete parameterization of (\ref{eq:theta1Eq0}) (and by equivalence (\ref{eq:localCompSup}) with $\theta_1 = 0$) is thus
\begin{equation}
\begin{aligned}\label{eq:degenParamSup1}
&\theta_1 = 0, \quad \omega = 0, \quad \mathbf{e} = \pm \frac{\mathbf{x}_2 - \mathbf{x}_3}{|\mathbf{x}_2 - \mathbf{x}_3|},   \\
&\tau_1 = \pm |\mathbf{x}_2 - \mathbf{x}_3|, \quad  \tau_2 =  (\mathbf{x}_3 - \mathbf{x}_4) \cdot \mathbf{e}, \\
&\mathbf{z} = ( \mathbf{I} - \mathbf{R}_{\theta_2} + \mathbf{e} \otimes \mathbf{e})^{-1} \mathbf{P}_{\mathbf{e}}(\mathbf{x}_2 - \mathbf{R}_{\theta_2} \mathbf{x}_1), \\
&\qquad \text{for:} \quad \theta_2 \in (-\pi, \pi]\setminus \{ 0\}.
\end{aligned}
\end{equation}
Here, since $\omega = 0$, $\mathbf{y}_i = \mathbf{x}_i$ for all $i = 1,2,3,4$.  Also, to obtain the inversion for $\mathbf{z}$ (compare the $\mathbf{z}$ equations in (\ref{eq:theta1Eq0}) to (\ref{eq:degenParamSup1})), we used the fact that $\mathbf{z} \cdot \mathbf{e} = 0$ to introduce $\mathbf{e} \otimes \mathbf{e}$ on the left-hand side in (\ref{eq:theta1Eq0}).  The quantity $(\mathbf{I} - \mathbf{R}_{\theta_2} + \mathbf{e} \otimes \mathbf{e})$ is alway invertible for  $\theta_2 \in (-\pi, \pi]\setminus \{ 0\}$.  
Notice also that $\theta_2$ is a free parameter here and there is a degeneracy in the choice of sign of $\mathbf{e}$.  

A second degenerate case is obtained by solving the parameterization (\ref{eq:rewriteLocal}) for case $\omega = 0$ and $\theta_1 \neq 0$. We have $\mathbf{u}_a = \mathbf{u}_b$ and $\mathbf{v}_a = \mathbf{v}_b$ by Proposition \ref{vectorsProp} for $\omega =0$.  It follows from (\ref{eq:rewriteLocal}) that, since $\theta_1$ cannot be zero, we require $\mathbf{u}_a = \mathbf{u}_b = \tau_2 \mathbf{e}$ in this case (i.e., we simply reverse the $``1"$ and $``2"$ dependence in the parameterization (\ref{eq:degenParamSup1})).  The complete parameterization of this case is 
\begin{equation}
\begin{aligned}\label{eq:degenParamSup2}
&\theta_2 = 0, \quad \omega = 0, \quad \mathbf{e} = \pm \frac{\mathbf{x}_3 - \mathbf{x}_4}{|\mathbf{x}_3 - \mathbf{x}_4|},   \\
&\tau_2 = \pm |\mathbf{x}_3 - \mathbf{x}_4|, \quad  \tau_1 =  (\mathbf{x}_2 - \mathbf{x}_3) \cdot \mathbf{e}, \\
&\mathbf{z} = ( \mathbf{I} - \mathbf{R}_{\theta_1} + \mathbf{e} \otimes \mathbf{e})^{-1} \mathbf{P}_{\mathbf{e}}(\mathbf{x}_2 - \mathbf{R}_{\theta_1} \mathbf{x}_3), \\
&\qquad \text{for:} \quad \theta_1 \in (-\pi, \pi]\setminus \{ 0\}.
\end{aligned}
\end{equation} 
Here, $\theta_1$ is a free parameter and there is again a degeneracy in the choice of sign of $\mathbf{e}$.  

\textbf{The generic cases.}  These are cases for which $\omega \in (-\pi, \pi) \setminus \{ 0\}$.  By Proposition \ref{vectorsProp}, the side length vectors satisfy $\mathbf{u}_a \nparallel \mathbf{u}_b$ and $\mathbf{v}_a \nparallel \mathbf{v}_b$ in these cases.   Thus, the parameterization of local compatibility cannot by as in (\ref{eq:theta1Eq0}).  Instead, it must solve (\ref{eq:rewriteLocal}) (and by equivalence (\ref{eq:localCompSup}) with $\theta_1 \neq 0$).   Notice that, in this parameterization, $\mathbf{P}_{\mathbf{e}} \mathbf{v}_a \neq \mathbf{P}_{\mathbf{e}} \mathbf{v}_b$ and   $\mathbf{P}_{\mathbf{e}} \mathbf{v}_a \neq \mathbf{P}_{\mathbf{e}} \mathbf{v}_b$ since $\mathbf{u}_a \neq \mathbf{u}_b$ and $\mathbf{v}_a \neq \mathbf{v}_b$. 

Thus, to introduce the parameterization, we first choose a convenient coordinate system.  Since $\mathbf{u}_a \nparallel \mathbf{u}_b$, we choose 
\begin{equation}
\begin{aligned}
\bff_1 = \frac{\bfu_a + \bfu_b}{|\bfu_a + \bfu_b|}, \quad
\bff_2 = \frac{\bfu_a \times \bfu_b}{|\bfu_a \times \bfu_b|}, \quad
\bff_3=\frac{\bfu_a - \bfu_b}{|\bfu_a - \bfu_b|},
\end{aligned}
\end{equation}
and the complete parameterization for (\ref{eq:rewriteLocal}) is 
\begin{equation}
\begin{aligned}\label{eq:mainParam}
&\mathbf{e} \equiv \mathbf{e}^{\sigma}(\omega, \varphi) = c_{\varphi} \mathbf{f}_1  + s_{\varphi} \mathbf{f}_2, \\
&\theta_1 \equiv \theta_1^{\sigma}(\omega, \varphi) = \text{sign}( \mathbf{e} \cdot (\mathbf{u}_a \times \mathbf{u}_b)) \arccos\Big( \frac{\mathbf{u}_a \cdot \mathbf{P}_{\mathbf{e}} \mathbf{u}_b}{|\mathbf{P}_{\mathbf{e}} \mathbf{u}_a|^2} \Big), \\
&\theta_2 \equiv \theta_2^{\sigma}(\omega, \varphi) = \text{sign}( \mathbf{e} \cdot (\mathbf{v}_a \times \mathbf{v}_b)) \arccos\Big( \frac{\mathbf{v}_a \cdot \mathbf{P}_{\mathbf{e}} \mathbf{v}_b}{|\mathbf{P}_{\mathbf{e}} \mathbf{v}_a|^2} \Big), \\
&\tau_1 \equiv \tau_1^{\sigma}(\omega, \varphi) = \mathbf{e} \cdot \mathbf{v}_a, \quad \tau_2 \equiv \tau_2^{\sigma}(\omega, \varphi) = \mathbf{e} \cdot \mathbf{u}_a,  \\
&\mathbf{z} \equiv \mathbf{z}^{\sigma}(\omega, \varphi)  = (\mathbf{I} - \mathbf{R}_{\theta_1} + \mathbf{e} \otimes \mathbf{e})^{-1} \mathbf{P}_{\mathbf{e}} (\mathbf{y}_2 - \mathbf{R}_{\theta_1} \mathbf{y}_3),\\
&\qquad \text{for:} \quad \omega \in (-\pi, \pi)\setminus \{ 0\}, \quad \varphi \in (-\pi, \pi].
\end{aligned}
\end{equation}
(Note, there is a slight caveat to the parameterization here. If $\mathbf{e} \cdot (\mathbf{u}_a \times \mathbf{u}_b) = 0$, then we set $\theta_1 = \pi$. Alternatively, if $\mathbf{e} \cdot (\mathbf{v}_a \times \mathbf{v}_b) = 0$, we set $\theta_2 = \pi$.)

\textbf{The folded flat cases.}   We now set $\omega \in \{ \pm \pi \}$. Following Proposition \ref{vectorsProp}, this parameterization is characterized by cases (ii-v).

(\textit{Case (ii)}).  In this case, we need to satisfy the parameterization in (\ref{eq:rewriteLocal}) under the assumption that $\mathbf{u}_a \nparallel \mathbf{u}_b$ and $\mathbf{v}_a = -\mathbf{v}_b$.  Since $\mathbf{u}_a - \mathbf{u}_b = \mathbf{v}_a - \mathbf{v}_b$, it follows that $\tau_1 = \mathbf{v}_a \cdot \mathbf{e} = \mathbf{v}_b \cdot \mathbf{e} = 0$ and $\theta_2 = \pi$.  The complete parameterization is thus 
\begin{equation}
\begin{aligned}\label{eq:mainParam1}
&\mathbf{e} \equiv \mathbf{e}^{\sigma}( \varphi) = c_{\varphi} \mathbf{f}_1  + s_{\varphi} \mathbf{f}_2, \\
&\theta_1 \equiv \theta_1^{\sigma}(\varphi) = \text{sign}( \mathbf{e} \cdot (\mathbf{u}_a \times \mathbf{u}_b)) \arccos\Big( \frac{\mathbf{v}_a \cdot \mathbf{P}_{\mathbf{e}} \mathbf{v}_b}{|\mathbf{P}_{\mathbf{e}} \mathbf{v}_a|^2} \Big), \\
&\theta_2 = \pi, \quad \tau_1 = 0, \quad \tau_2 \equiv \tau_2^{\sigma}( \varphi) = \mathbf{e} \cdot \mathbf{u}_a,   \\
&\mathbf{z} \equiv \mathbf{z}^{\sigma}(\varphi)  = (\mathbf{I} - \mathbf{R}_{\theta_1} + \mathbf{e} \otimes \mathbf{e})^{-1} \mathbf{P}_{\mathbf{e}} (\mathbf{y}_2 - \mathbf{R}_{\theta_1} \mathbf{y}_3),\\
&\qquad \text{for:} \quad \varphi \in (-\pi, \pi] \setminus \{ 0\}
\end{aligned}
\end{equation}
(if $\varphi = 0$, then the parameterization is also as above except that $\theta_1 = \pi$).  

(\textit{Case (iii)}.) In this case, we need to satisfy the parameterization in (\ref{eq:rewriteLocal}) under the assumption that $\mathbf{u}_a = - \mathbf{u}_b$ and $\mathbf{v}_a \nparallel \mathbf{v}_b$.  It follows that $\tau_2 = \mathbf{u}_a \cdot \mathbf{e} = \mathbf{u}_b \cdot \mathbf{e} = 0$, and $\theta_1 = \pi$.  For the axis $\mathbf{e}$, we introduce the coordinate system 
\begin{equation}
\begin{aligned}
\bfh_1 = \frac{\bfv_a + \bfv_b}{|\bfv_a + \bfv_b|}, \quad
\bfh_2 = \frac{\bfv_a \times \bfv_b}{|\bfv_a \times \bfv_b|}, \quad
\bfh_3=\frac{\bfv_a - \bfv_b}{|\bfv_a - \bfv_b|}
\end{aligned}
\end{equation}  
since $\mathbf{v}_a \nparallel \mathbf{v}_b$. The complete parameterization in this case is thus
\begin{equation}
\begin{aligned}\label{eq:mainParam2}
&\mathbf{e} \equiv \mathbf{e}^{\sigma}(\varphi) = c_{\varphi} \mathbf{h}_1  + s_{\varphi} \mathbf{h}_2, \\
&\theta_2 \equiv \theta_2^{\sigma}(\varphi) = \text{sign}( \mathbf{e} \cdot (\mathbf{v}_a \times \mathbf{v}_b)) \arccos\Big( \frac{\mathbf{v}_a \cdot \mathbf{P}_{\mathbf{e}} \mathbf{v}_b}{|\mathbf{P}_{\mathbf{e}} \mathbf{v}_a|^2} \Big), \\
&\theta_1= \pi, \quad \tau_1 \equiv \tau_1^{\sigma}(\varphi) = \mathbf{e} \cdot \mathbf{v}_a, \quad \tau_2 = 0,  \\
&\mathbf{z} \equiv \mathbf{z}^{\sigma}(\varphi)  = (\mathbf{I} - \mathbf{R}_{\theta_1} + \mathbf{e} \otimes \mathbf{e})^{-1} \mathbf{P}_{\mathbf{e}} (\mathbf{y}_2 - \mathbf{R}_{\theta_1} \mathbf{y}_3),\\
&\qquad \text{for:} \quad \varphi \in (-\pi, \pi] \setminus \{ 0\}
\end{aligned}
\end{equation}
(if $\varphi = 0$, then the parameterization is also as above except that $\theta_2 = \pi$).  

(\textit{Case (iv).)} This case can only generate a ring structure which oscillates back and forth between the points $\mathbf{y}_1$ and $\mathbf{y}_2$.  It is therefore not relevant. 

(\textit{Case (v).}) In this case, the complete parameterization of (\ref{eq:rewriteLocal}) is as in (\ref{eq:mainParam}) with $\omega \in \{ \pm  \pi\}$.

\textbf{The discreteness conditions.}  For the first degenerate parameterization in (\ref{eq:degenParamSup1}), the discreteness conditions reduce to 
\begin{equation}
\begin{aligned}
 \theta_2 = \frac{2\pi}{q^{\star}} ,  \quad  \frac{\cos(\eta)}{l} = \frac{p^{\star}}{q^{\star}}
\end{aligned}
\end{equation}
for some $p^{\star} \in \mathbb{Z}$ and $q^{\star} \in \mathbb{Z} \setminus \{ 0\}$.  Here, we recall the definitions of the parameters $l$ and $\eta$ (see Fig.\;2 in the main text).  The second degenerate parameterization (\ref{eq:degenParamSup2}) is similar.  The discreteness conditions reduce to 
\begin{equation}
\begin{aligned}
 \theta_1 = \frac{2\pi}{p^{\star}} ,  \quad   \cos(\eta)l  = \frac{q^{\star}}{p^{\star}}
\end{aligned}
\end{equation}
for some $p^{\star} \in \mathbb{Z}\setminus\{ 0\}$ and $q^{\star} \in \mathbb{Z}$ in this case.  Finally, in the case of the parameterization in (\ref{eq:mainParam}), we first take the axis angle $\varphi \in (-\pi/2,\pi/2]$ without loss of generality (since $(\mathbf{e}, \theta_{1,2}, \tau_{1,2}) \mapsto - (\mathbf{e}, \theta_{1,2}, \tau_{1,2})$  generates the same structure).  Then the first discreteness condition $p^{\star} \tau_1^{\sigma}(\omega, \varphi) + q^{\star} \tau_2^{\sigma}(\omega, \varphi) = 0$ is equivalent to  
\begin{equation}
\begin{aligned}\label{eq:varphiStar}
\varphi \equiv \varphi_{\star}^{\sigma}(\omega) = \arctan\Big(\frac{-\mathbf{f}_1\cdot ( p^{\star} \mathbf{v}_a + q^{\star} \mathbf{u}_a)}{\mathbf{f}_2 \cdot ( p^{\star} \mathbf{v}_a + q^{\star} \mathbf{u}_a) }\Big)
\end{aligned}
\end{equation}
(with $\varphi \equiv \varphi_{\star}^{\sigma}(\omega)$ whenever $\omega \in (-\pi, \pi) \setminus \{ 0\}$ is such that $\mathbf{f}_2 \cdot (p^{\star} \mathbf{v}_a + q^{\star} \mathbf{u}_a) = 0$). Consequently, the second condition
\begin{equation}
\begin{aligned}\label{eq:designSup}
p^{\star} \theta_1^{\sigma}(\omega, \varphi_{\star}^{\sigma}(\omega)) + q^{\star} \theta_2^{\sigma}(\omega, \varphi_{\star}^{\sigma}(\omega)) = 2\pi
\end{aligned}
\end{equation} 
can be easily  evaluated by cycling through $\omega \in (-\pi, \pi) \setminus \{ 0\}$ numerically.  

The discreteness conditions for the  parameterization in (\ref{eq:mainParam1}) are solved analgously by taking $\varphi = \varphi_{\star}^{\sigma}(\pi)$ above, and finding a $(p^{\star}, q^{\star}) \in \mathbb{Z}^2$ such that $p^{\star} \theta_1(\varphi_{\star}(\pi)) + q^{\star} \pi = 2\pi$. Also the discreteness conditions for the  parameterization in (\ref{eq:mainParam2}) are solved analogously by replacing $\mathbf{f}_{1,2}$ with $\mathbf{h}_{1,2}$ in (\ref{eq:varphiStar}), taking $\varphi = \varphi_{\star}(\pi)$ after this replacement, and finding a $(p^{\star}, q^{\star}) \in \mathbb{Z}^2$ such that $p^{\star} \pi + q^{\star} \theta_2^{\sigma}(\varphi_{\star}(\pi)) = 2\pi$.  Notice that there are no free parameters for the last condition.  So these flattened configurations are highly non-generic.

\subsection{On the phase diagram for HMO}
In the main text, we highlighted for illustrative purposes two slices of the phase-space for HMO structures\textemdash specifically, the cases $(p^{\star}, q^{\star}) = (3,7)$ and $=(8,0)$ at $\lambda = 0.5$ for the crease parameter\textemdash where we cycled through the reference parameters $(l, \eta) \in (0,2) \times (0, \pi)$ and $\sigma \in \{ \pm \}$ to determine all HMO solutions.   We now provide a more comprehensive view of the phase diagrams for these discretenesses below in Fig.\;\ref{fig:ForPhaseSup}.  This further reinforces the notion that multistable regions are ubiquitous to HMO structures.

\begin{figure*}
	\includegraphics[width=\textwidth]{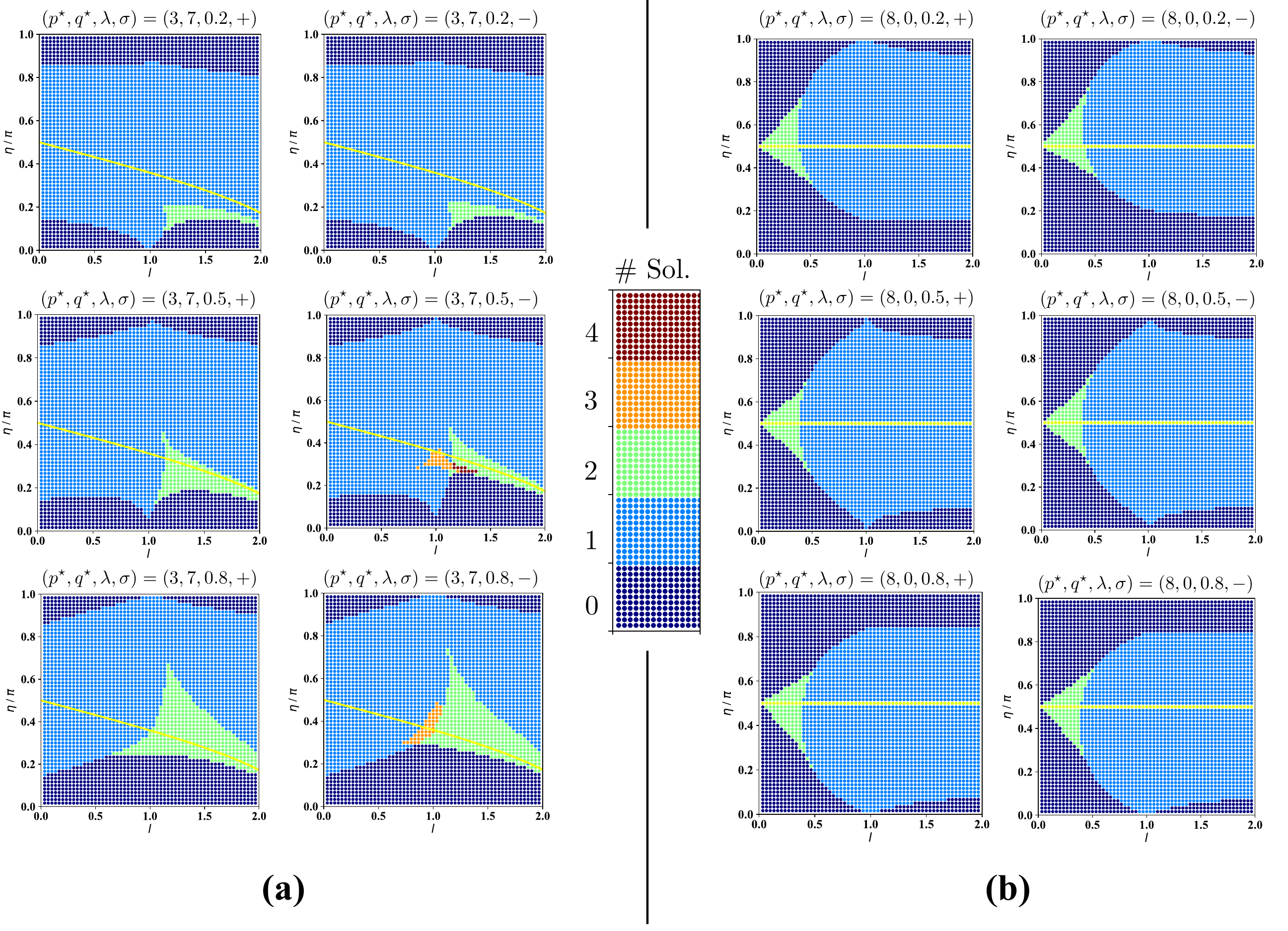}
	\caption{The phase space for HMO structures for $(p^{\star} ,q^{\star})  = (3,7)$ in (a) and $(p^{\star},q^{\star}) = (8,0)$ in (b).  The slices of the phase space are given at crease parameter $\lambda = 0.2,0.5,0.8$, resepctively.  There are two slices per discreteness, since there are two possible mountain-valley assignments $\sigma \in \{ \pm\}$.  The number of HMO solutions for a fixed mountain-valley assignment is displayed according to the indicated coloring scheme: 0 (purple), 1 (light blue), 2 (green), 3 (orange) and 4 (red).  The yellow curve denotes the degenerate line for which the solutions correspond to $\omega = 0$ and one of the parameterizations in (\ref{eq:degenParamSup1}-\ref{eq:degenParamSup2}). This curve is independent of $\lambda$. }
	\label{fig:ForPhaseSup}
\end{figure*}

\subsection{The Kresling pattern}
\begin{figure}[!t]
	\includegraphics[width=\columnwidth]{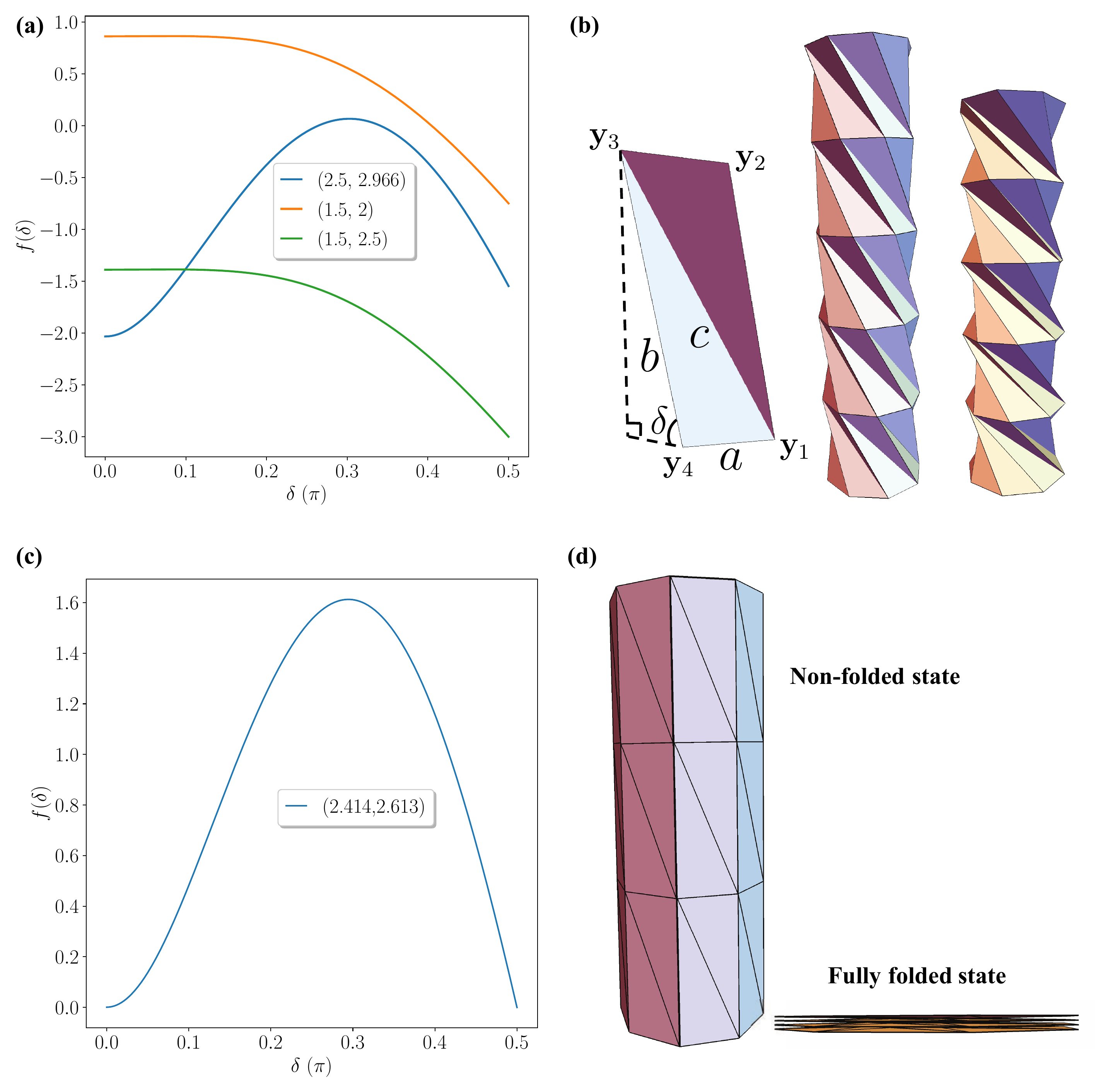}
	\caption{(a) The equation $f(\delta)=0$ has 2, 1 and 0 root respectively when $(\frac{b}{a}, \frac{c}{a})= (2.5, 2.966), (1.5, 2), (1.5, 2.5)$ and $n=8$.  (b) The geometry of unit cell of Kresling pattern and an example of bistable Kresling pattern with $(\frac{b}{a}, \frac{c}{a})= (2.5, 2.966)$. (c) The equation $f(\delta)=0$ has two roots $\delta=0, \pi/2$ with $(\frac{b}{a}, \frac{c}{a})= (2.414, 2.613)$ (the special Kresling pattern) and $n=8$. (d) The non-folded state and fully folded state of the special Kresling pattern.}
	\label{fig:kresling}
\end{figure}

A common tubular structure made of an origami unit cell is the Kresling pattern.  These are the limiting cases of HMO corresponding to  the crease pattern parameter $\lambda \rightarrow 0$ or $\rightarrow 1$ (described in (i-iii) in the main text).  For the case of a ring-type Kresling pattern, Cai et.\;al.\;\cite{jianguo2015bistable} derived the geometrical relation for this pattern as:
\beqs
(\frac{c}{a})^2=&&[\frac{1}{\sin(\frac{\pi}{n})}\sin(\frac{\pi}{n}+\arcsin(\frac{b}{a}\cos\delta \sin(\frac{\pi}{n})))]^2  \nonumber\\
	&&+(\frac{b}{a}\sin\delta)^2,
\label{eq:kresling}
\eeqs
where $a=|\bfy_1 - \bfy_4|$, $b= |\bfy_1 - \bfy_2|$, $c=|\bfy_1 - \bfy_3|$, $\delta$ is the angle between $\bfy_3 - \bfy_4$ and the horizontal polygon, and $n$ is the number of elements in a horizontal role (see Fig. \ref{fig:kresling}(b)). 
The function 
\beqs
f(\delta)=&&[\frac{1}{\sin(\frac{\pi}{n})}\sin(\frac{\pi}{n}+\arcsin(\frac{b}{a}\cos\delta \sin(\frac{\pi}{n})))]^2  \nonumber\\
&&+(\frac{b}{a}\sin\delta)^2 -(\frac{c}{a})^2
\eeqs
has 0, 1, or 2 roots for $\delta \in [0, \pi/2]$  depending on parameters $b/a,\ c/a,\ n$ (see, for instance, the examples in Fig. \ref{fig:kresling}(a)). This implies that the Kresling pattern is either inaccessible, monostable, or bistable.  For the special Kresling pattern that gives the maximal transforming extension with $n = 8$ (Fig.\;7(d) in the main text), the equation (\ref{eq:kresling}) holds for both $\delta=0$ and $\delta=\pi/2$, i.e., this particular Kresling pattern is stable when the unit cell is fully unfolded or fully folded. In substituting $\delta=0, \pi/2$ and then subtracting the equations, we have 
\beq
[\frac{1}{\sin(\frac{\pi}{n})}\sin(\frac{\pi}{n} + \arcsin(\frac{b}{a}\sin(\frac{\pi}{n})))]^2 = (\frac{b}{a})^2 + 1,
\eeq
which has a solution for the geometric parameters 
$(\frac{b}{a}, \frac{c}{a})=(2.414, 2.613)$ when $n=8$. This corresponds exactly to the reference parameters $(l, \eta) = (|\bfx_1 - \bfx_4|, \angle \bfx_4 \bfx_1 \bfx_2) = (0.414, \pi/2)$ since $b =|\mathbf{y}_2 - \mathbf{y}_1|=1$ has been assumed throughout.
\subsection{On compatibility for the coexistence of two phases}\label{ssec:TwoPhase}
Though we already have the design equations for the HMO, it is not obvious that different phases can fit together and remain cylindrical.   Here, we derive the compatibility conditions for the co-existence of two phases.  

Local compatibility for the two phases (i.e., (\ref{eq:localCompSup})) is certainly necessary.  Thus, we consider two locally compatible origami structures generated by the same underlying tessellation: $\check{\mathcal{G}} \mathbf{y}_{\check{\omega}}^{\check{\sigma}}(\Omega) = \{ \check{g}_1^p \check{g}_2^q( \mathbf{y}_{\check{\omega}}^{\check{\sigma}}(\Omega)) \colon (p,q) \in \mathbb{Z}^2 \}$ and $\hat{\mathcal{G}} \mathbf{y}_{\hat{\omega}}^{\hat{\sigma}}(\Omega) = \{\hat{g}_1^p \hat{g}_2^q( \mathbf{y}_{\hat{\omega}}^{\hat{\sigma}}(\Omega)) \colon (p,q) \in \mathbb{Z}^2\}$, where $\check{g}_i$ (respectively, $\hat{g}_i$) have group parameters as in (\ref{eq:mainParam}) with $(\omega, \varphi, \sigma) = (\check{\omega}, \check{\varphi}, \check{\sigma})$ (respectively, $(\omega, \varphi, \sigma) = (\hat{\omega}, \hat{\varphi}, \hat{\sigma})$) on the domain $(-\pi,\pi) \setminus \{ 0\} \times (-\pi/2,\pi/2] \times \{ \pm \}$.  We claim that the necessary and sufficient conditions for a closed cylindrical origami of these (potentially) two phases are:
\begin{equation}
\begin{aligned}\label{eq:twoPhaseSup}
&\tau_{1}^{\check{\sigma}}(\check{\omega}, \check{\varphi}) = \tau_{1}^{\hat{\sigma}}(\hat{\omega}, \hat{\varphi}),  \\
&\theta_{1}^{\check{\sigma}}(\check{\omega}, \check{\varphi}) = \theta_{1}^{\hat{\sigma}}(\hat{\omega}, \hat{\varphi}),  \\
&p^{\star} \tau_1^{\check{\sigma}}(\check{\omega}, \check{\varphi}) +  \tilde{q} \tau_2^{\check{\sigma}}(\check{\omega}, \check{\varphi}) + (q^{\star}-\tilde{q})  \tau_{2}^{\hat{\sigma}}(\hat{\omega}, \hat{\varphi}) = 0, \\
&p^{\star} \theta_1^{\check{\sigma}}(\check{\omega}, \check{\varphi}) + \tilde{q} \theta_2^{\check{\sigma}}(\check{\omega}, \check{\varphi}) + (q^{\star}-\tilde{q})  \theta_{2}^{\hat{\sigma}}(\hat{\omega}, \hat{\varphi}) = 2\pi,
\end{aligned}
\end{equation}
for some $p^{\star},\tilde{q},q^{\star} \in \mathbb{Z}$ with $|\tilde{q}| \leq |q^{\star}|$ and $q^{\star} \tilde{q} \geq 0$, or we can also exchange the roles of $(\cdot)_2$ and $(\cdot)_1$ above to generate the cylindrical origami.  We focus on the system in (\ref{eq:twoPhaseSup}) without loss of generality.

We justify this claim through a series of propositions.   The first proposition gives that, if the first two identities hold in the (\ref{eq:twoPhaseSup}), then the radius of the cylinders generated by the two phases are the same.  To make this precise, note that the radiuses of the cylinders are 
\begin{equation}
\begin{aligned}
\check{r} = |\mathbf{P}_{\check{\mathbf{e}}} (\check{\mathbf{y}}_3 - \check{\mathbf{z}})| , \quad \hat{r} =  |\mathbf{P}_{\hat{\mathbf{e}}} (\hat{\mathbf{y}}_3 - \hat{\mathbf{z}})| 
\end{aligned}
\end{equation}
where $\hat{\mathbf{e}} \equiv \mathbf{e}^{\hat{\sigma}}(\hat{\omega},\hat{\varphi})$, etc., from (\ref{eq:mainParam}).  We establish the following result:
\begin{proposition}\label{radiusProp}
If $\check{\tau}_1 \equiv \tau_{1}^{\check{\sigma}}(\check{\omega}, \check{\varphi}) = \tau_{1}^{\hat{\sigma}}(\hat{\omega}, \hat{\varphi}) \equiv \hat{\tau}_1$ and $\check{\theta}_1 \equiv \theta_{1}^{\check{\sigma}}(\check{\omega}, \check{\varphi}) = \theta_{1}^{\hat{\sigma}}(\hat{\omega}, \hat{\varphi}) \equiv \hat{\tau}_1$, then $\check{r} = \hat{r}$.
\end{proposition}
\begin{proof}
	We calculate $\check{r}$ first. By substituting the formula of $\check{\mathbf{z}} \equiv \mathbf{z}^{\check{\sigma}}(\check{\omega},\check{\varphi})$ into the radius equation,  we obtain
	\beqs\label{eq:longCalcSup}
	\check{r} =&&|\mathbf{P}_{\check{\mathbf{e}}}(\check{\bfy}_{3} - (\bfI - \check{\bfR}_{\check{\theta}_{1}} +\check{\mathbf{e}} \otimes \check{\mathbf{e}})^{-1}\mathbf{P}_{\check{\mathbf{e}}}(\check{\bfy}_{2} - \check{\bfR}_{\check{\theta}_{1}}\check{\bfy}_{3}))| \nonumber \\
	=&&|(\bfI - \check{\bfR}_{\check{\theta}_{1}}+ \check{\mathbf{e}} \otimes \check{\mathbf{e}} )^{-1}\mathbf{P}_{\check{\mathbf{e}}}((\bfI-\check{\bfR}_{\check{\theta}_{1}})\check{\bfy}_{3} - (\check{\bfy}_{2} - \check{\bfR}_{\check{\theta}_{1}} \check{\bfy}_{3})| \nonumber \\
	=&&|(\bfI - \check{\bfR}_{\check{\theta}_{1}} + \check{\mathbf{e}} \otimes \check{\mathbf{e}})^{-1}(\check{\bfy}_{2} -\check{\bfy}_{3} - \check{\tau}_{1} \check{\bfe})| \nonumber \\
	=&& \frac{\sqrt{|\bfx_3 - \bfx_2|^2 - (\check{\tau}_{1})^2}}{|2 \sin (\check{\theta}_{1}/2)|}.
	\eeqs
	Here, the first equality is by definition, the second uses commutativity of the transformations, and the third follows from the substitution $\check{\tau}_{1} = \check{\bfe} \cdot (\check{\bfy}_{2}-\check{\bfy}_{3})$. Finally, since $\check{\bfy}_{2}-\check{\bfy}_{3} - \check{\tau}_{1} \check{\bfe}$ is on the plane perpendicular to $\check{\bfe}$, 
	we derive that
\begin{equation}
\begin{aligned}
&(\bfI - \check{\bfR}_{\check{\theta}_{1}} + \check{\mathbf{e}} \otimes \check{\mathbf{e}})^{-1}(\check{\bfy}_{2} -\check{\bfy}_{3} - \check{\tau}_{1} \check{\bfe}) \\
&\quad  =(2 \sin(\check{\theta}_1/2))^{-1} \check{\mathbf{R}}_{\tfrac{1}{2}(\pi-\check{\theta}_1)} (\check{\bfy}_{2} -\check{\bfy}_{3} - \check{\tau}_{1} \check{\bfe}),
\end{aligned}
\end{equation}
by an explicit calculation. The final identity in (\ref{eq:longCalcSup}) then easily follows.  Analogously, the radius of the other phase is
	\beq
	\hat{r} =\frac{\sqrt{|\bfx_3 - \bfx_2|^2 - (\hat{\tau}_{1})^2}}{|2 \sin (\hat{\theta}_{1}/2)|}.
	\eeq
	Thus, $\check{\theta}_{1}=\hat{\theta}_{1}$ and $\check{\tau}_{1}=\hat{\tau}_{1}$ gives $\check{r} = \hat{r}$.
\end{proof}
\begin{remark}\label{RemarkSup}
If $\check{r} = \hat{r}$, then actually $|\mathbf{P}_{\check{\mathbf{e}}} ( \check{g}_1^p \check{g}_2^q(\check{\mathbf{y}}_3)  - \check{\mathbf{z}})| = |\mathbf{P}_{\hat{\mathbf{e}}}(\hat{g}_1^{p} \hat{g}_2^q(\hat{\mathbf{y}}_3) - \hat{\mathbf{z}})| > 0$ for all $p,q \in \mathbb{Z}$.  
\end{remark}
\begin{proof}
It is easy to see that $\check{r}=|\mathbf{P}_{\check{\mathbf{e}}} ( \check{g}_1^p \check{g}_2^q(\check{\mathbf{y}}_3)  - \check{\mathbf{z}})|$ and $\hat{r} = |\mathbf{P}_{\hat{\mathbf{e}}}(\hat{g}_1^{p} \hat{g}_2^q(\hat{\mathbf{y}}_3) - \hat{\mathbf{z}})|$ for all $p,q \in \mathbb{Z}$.  So the equality follows.  The inequality also holds.  For otherwise, given (\ref{eq:localCompSup}), this would imply that all the corner points of a partially folded Miura parallelogram lie on the line $\{\check{\mathbf{z}} + \lambda \check{\mathbf{e}} \colon \lambda \in \mathbb{R}\}$ or $\{\hat{\mathbf{z}} + \lambda\hat{\mathbf{e}} \colon \lambda \in \mathbb{R} \}$, respectively.  This is impossible given Proposition \ref{prop:nonparallel}.
\end{proof}

\begin{proposition}\label{localCompatProp}
	Fix some $\check{\mathbf{x}} = \check{g}_2^q(\check{\mathbf{y}}_3)$ and $\hat{\mathbf{x}} = \hat{g}_2^{q}(\hat{\mathbf{y}}_3)$ for $q \in \mathbb{Z}$.  There exists a rotation $\bfR \in \text{SO}(3)$ and translation $\bft \in \mathbb{R}^3$ such that 
	\beq\label{eq:RotationSup}
	\bfR \check{g}_1^p(\check{\bfx}) + \bft = \hat{g}_1^p(\hat{\bfx}) \quad \text{for all } p \in \mathbb{Z}
	\eeq
	if and only if $\check{\tau}_1 \equiv \tau_{1}^{\check{\sigma}}(\check{\omega}, \check{\varphi}) = \tau_{1}^{\hat{\sigma}}(\hat{\omega}, \hat{\varphi}) \equiv \hat{\tau}_1$ and $\check{\theta}_1 \equiv \theta_{1}^{\check{\sigma}}(\check{\omega}, \check{\varphi}) = \theta_{1}^{\hat{\sigma}}(\hat{\omega}, \hat{\varphi}) \equiv \hat{\tau}_1$.
	\end{proposition}
\begin{proof}
We first note that, by a general result elucidated in \cite{feng2018compatibility}, the only types of compatible interfaces in helical structures that do not involve a fixed radius of the two phases are \textit{vertical interfaces}.  A vertical interface implies that $\mathbf{v}_a \parallel \mathbf{v}_b \parallel \mathbf{e}$ or $\mathbf{u}_a \parallel \mathbf{u}_b \parallel \mathbf{e}$.  Due to Proposition \ref{prop:degenerate1}, this is not possible for the locally compatible parameterizations we are considering in this framework since we assume $\omega \in (-\pi, \pi) \setminus\{ 0\}$, i.e., a partially folded configuration.  We therefore have the necessary condition that $\hat{r} = \check{r}$.  

Let us assume that the parameters of the two phases are such that $\check{r} = \hat{r}$.  By the multiplication rule of generators, we have 
	\begin{equation}
	\begin{aligned}
	&\check{g}_1^p(\check{\bfx})=\check{\bfR}_{p \check{\theta}_1} (\check{\bfx}-\check{\bfz}) + p \check{\tau}_1 \check{\bfe} + \check{\bfz}, \\
	&\hat{g}_1^p(\hat{\bfx})=\hat{\bfR}_{p \hat{\theta}_1} (\hat{\bfx}-\hat{\mathbf{z}}) + p \hat{\tau}_1 \hat{\bfe} + \hat{\mathbf{z}}.
	\end{aligned}
	\end{equation}
	Let $\bfR_{\check{\mathbf{e}} \times \hat{\mathbf{e}}}$ be a rotation about $\check{\bfe} \times \hat{\bfe}$  satisfying $\bfR_{\check{\bfe}\times\hat{\bfe}} \check{\bfe} = \hat{\bfe}$. 
	Since $\check{r} = |\mathbf{P}_{\check{\mathbf{e}}}(\check{\bfx}-\check{\bfz})|=|\mathbf{P}_{\hat{\mathbf{e}}}(\hat{\bfx} - \hat{\bfz})|= \hat{r}$ (Remark \ref{RemarkSup}), 
	there exits $\bfR_{\hat{\bfe}} \in \text{SO}(3)$, $\bfR_{\hat{\bfe}} \hat{\bfe} = \hat{\bfe}$ and $\rho \in \mathbb{R}$ such that 
	\beq\label{eq:cylinderSup}
	\hat{\bfx} - \hat{\bfz} = \bfR_{\hat{\bfe}} \bfR_{\check{\bfe} \times \hat{\bfe}} (\check{\bfx}-\check{\bfz}) + \rho \hat{\bfe}.
	\eeq
	Let $\bar{\bfR}= \bfR_{\hat{\mathbf{e}}} \bfR_{\check{\bfe}\times\hat{\bfe}} \in SO(3)$ and $\bar{\bft}= \hat{\bfz} - \bfR_{\hat{\bfe}} \bfR_{\check{\bfe}\times\hat{\bfe}} \check{\bfz} + \rho \hat{\bfe}$. Using the identity $\hat{\bfR}_{p\check{\theta}_1} \bfR_{\hat{\bfe}} \bfR_{\check{\bfe} \times \hat{\bfe}} = \bfR_{\hat{\bfe}} \bfR_{\check{\bfe} \times \hat{\bfe}}\check{\bfR}_{p\check{\theta}_1}$ and (\ref{eq:cylinderSup}), we obtain
	\begin{equation}
	\begin{aligned}\label{eq:longCalcCompat}
	\bar{\bfR} \check{g}_1^p(\check{\bfx}) + \bar{\bft}&=\bfR_{\hat{\bfe}} \bfR_{\check{\bfe}\times\hat{\bfe}} \Big(\check{\bfR}_{p\check{\theta}_1}(\check{\bfx}-\check{\bfz})+ p\check{\tau}_1 \check{\bfe} + \check{\bfz}\Big) + \bar{\bft}  \\
	&=\hat{\bfR}_{p\check{\theta}_1} \bfR_{\hat{\bfe}} \bfR_{\check{\bfe} \times \hat{\bfe}}(\check{\bfx}-\check{\bfz}) + p \hat{\tau}_1 \hat{\bfe} + \hat{\bfz} + \rho \hat{\bfe} \\
	&=\hat{\bfR}_{p\check{\theta}_1} (\hat{\bfx} - \hat{\bfz}) + p\check{\tau}_1\hat{\bfe} + \hat{\bfz}. 
	\end{aligned}
	\end{equation}
	Since $\hat{r} = \check{r}$ is necessary for the parameterization herein, the statement (\ref{eq:RotationSup}) is evidently equivalent to 
	\begin{equation}
	\begin{aligned}
	\mathbf{Q}\Big(\bar{\bfR} \check{g}_1^p(\bfx) + \bar{\bft}\Big) + \mathbf{Q} \mathbf{c} = \hat{g}_1^{p}(\mathbf{x})
	\end{aligned}
	\end{equation}
	for some $\mathbf{Q} \in SO(3)$ and $\mathbf{c} \in \mathbb{R}^3$ and for all $p\in \mathbb{Z}$.   By taking the norm of both sides, we deduce that $\mathbf{c}$ satisfies 
	\begin{equation}
	\begin{aligned}
	\mathbf{c} = \Big( \hat{\mathbf{R}}_{p \hat{\theta}_1} - \hat{\mathbf{R}}_{p\check{\theta}_1} \Big)(\hat{\mathbf{x}} - \hat{\mathbf{z}})  + p (\hat{\tau}_1 - \check{\tau}_1) \hat{\mathbf{e}}.
	\end{aligned}
	\end{equation}
	But also, $\mathbf{c}$ has to be independent of $p \in \mathbb{Z}$.  Notice that $\mathbf{c} \cdot \hat{\mathbf{e}} = p(\hat{\tau}_1 - \check{\tau}_1)$.   Thus for the $p$-independence, we conclude that $\hat{\tau}_1 = \check{\tau}_1$.  Now notice that $p = 0$ gives $\mathbf{c} = \mathbf{0}$.   Thus again for $p-$independence, we must have $\mathbf{R}_{p\hat{\theta}_1} (\hat{\mathbf{x}} - \hat{\mathbf{z}})  = \hat{\mathbf{R}}_{p \check{\theta}_1} (\hat{\mathbf{x}} - \hat{\mathbf{z}})$ for all $p \in \mathbb{Z}$.  Since $|\mathbf{P}_{\hat{\mathbf{e}}}(\hat{\mathbf{x}} - \hat{\mathbf{z}})| > 0$ (Remark \ref{RemarkSup}), the latter is solved if and only if $\check{\theta}_1 = \hat{\theta}_1$.   Consequently,  necessary conditions for (\ref{eq:cylinderSup}) are $\check{r} = \hat{r}$, $\check{\tau}_1 = \hat{\tau}_1$ and $\check{\theta}_1 = \hat{\theta}_1$.  
	
By Proposition \ref{radiusProp}, $\check{\tau}_1 = \hat{\tau}_1$ and $\check{\theta}_1 = \hat{\theta}_1$ gives $\check{r} = \hat{r}$.  This means that the formula in (\ref{eq:longCalcCompat}) is valid with the substitution $\check{\tau}_1 = \hat{\tau}_1$ and $\check{\theta}_1 = \hat{\theta}_1$.   In making this substitution, we conclude 
\begin{align}
\bar{\bfR} \check{g}_1^p(\check{\bfx}) + \bar{\bft} = \hat{g}_1^p(\hat{\mathbf{x}})
\end{align}
for all $p \in \mathbb{Z}$.  Hence, $\bar{\mathbf{R}}$ and $\bar{\mathbf{t}}$ are the rotation and translation, respectively, that yields (\ref{eq:cylinderSup}).  
\end{proof}

\begin{remark}\label{exchangeRemark}
Proposition \ref{radiusProp} and \ref{localCompatProp} are also true if we exchange the roles of $(\cdot)_1$ and $(\cdot)_2$.
\end{remark}

\begin{figure}[!t]
	\includegraphics[width=\columnwidth]{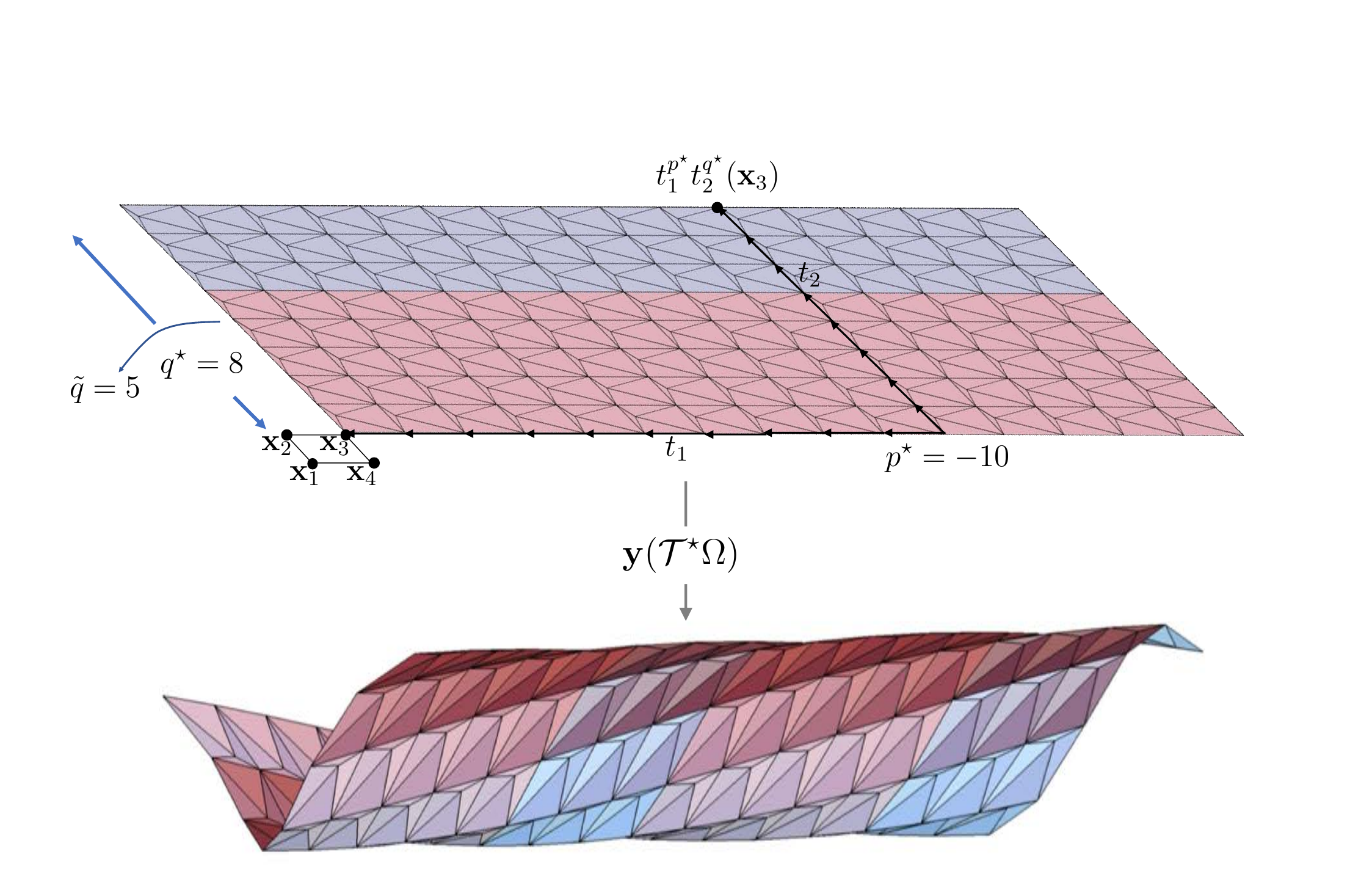}
	\caption{Co-existence of two phases along helical interfaces: The reference tessellated strip has a width of $q^{\star} = 8$ unit cells.  The width of $\tilde{q} = 5$ cells deforms by the $\check{(\cdot)}$ phase, and the width of $q^{\star} - \tilde{q} = 3$ cells deforms by the $\hat{(\cdot)}$ phase, as indicated by the differences in color.  The deformation $\mathbf{y} \colon \mathcal{T}^{\star} \Omega \rightarrow \mathbb{R}^3$ that achieves the cylindrical tube (bottom) maps the point $\mathbf{x}_3$ to the point $t_1^{p^{\star}} t_2^{q^{\star}}(\mathbf{x}_3)$, and this closes the cylinder.  The notation here is as established in Section\;\ref{ssec:TwoPhase}.}
	\label{fig:TwoPhaseFig}
\end{figure}

We are now able to justify the formula in (\ref{eq:twoPhaseSup}).  First note, the case of a horizontal interface is $\tilde{q} = q^{\star} = 0$, and this is treated completely in the main text of the paper.  Therefore, we focus on the justification in the case of a \textit{helical interface}.  That is, we assume $q^{\star} > \tilde{q} > 0$.  By Proposition \ref{localCompatProp}, we have 
\begin{align}\label{eq:interfaceIdentSup}
\mathbf{R}_{\tilde{q}} \check{g}_1^p \check{g}_2^{\tilde{q}}(\check{\mathbf{y}}_3) + \mathbf{t}_{\tilde{q}} = \hat{g}_1^p \hat{g}_2^{\tilde{q}}(\hat{\mathbf{y}}_3)
\end{align}
for all $p \in \mathbb{Z}$ if and only if $\check{\tau}_1 = \hat{\tau}_1$ and $\check{\theta}_1 = \hat{\theta}_1$.  We therefore introduce the tessellated strip (for instance, see Fig.\;\ref{fig:TwoPhaseFig}) $\mathcal{T}^{\star} \Omega = \big\{ t_1^p t_2^q(\Omega) \colon p \in \mathbb{Z}, q \in \{ 1, \ldots, q^{\star}\}\big\}$ for $t_1= (\mathbf{I}|\mathbf{x}_1 - \mathbf{x}_4)$ and $t_2 = (\mathbf{I}|\mathbf{x}_2 - \mathbf{x}_1)$.  As a consequence of the identity (\ref{eq:interfaceIdentSup}),  the map $\mathbf{y} \colon \mathcal{T}^{\star}\Omega \rightarrow \mathbb{R}^3$ defined by 
\begin{equation}
\begin{aligned}
&\mathbf{y}(t_1^p t_2^q(\mathbf{x}))\\
&\;\;  =   \begin{cases}
\check{g}_1^p \check{g}_2^q (\mathbf{y}_{\check{\omega}}^{\check{\sigma}}(\mathbf{x}))&  p \in \mathbb{Z}, q \in \{1,\ldots, \tilde{q}\}  \\
\bar{\mathbf{R}}^T\Big( \hat{g}_1^p\hat{g}_2^q(\mathbf{y}_{\hat{\omega}}^{\hat{\sigma}}(\mathbf{x})) - \bar{\mathbf{t}} \Big) & p \in \mathbb{Z}, q \in \{\tilde{q}+1, \ldots, q^{\star}\}
\end{cases}
\end{aligned}
\end{equation}
is a continuous isometric origami deformation of the tessellated strip $\mathcal{T}^{\star}\Omega$ if and only if $\check{\tau}_1 = \hat{\tau}_1$ and $\check{\theta}_1 = \hat{\theta}_1$.   

It remains to enforce the condition that the cylinder closes.  In this direction, we note that a vertex $\mathbf{x}_3$ is on the bottom boundary of tessellation and is deformed via $\check{\mathbf{y}}_3 = g_1^0 g_2^1(\check{\mathbf{y}}_4)$.  Similarly, the vertices $t_1^p t_2^{q^{\star}}(\mathbf{x}_3)$ are on the top boundary of the tessellation and are deformed by $\bar{\mathbf{R}}^T\big(\hat{g}_1^p \hat{g}_2^{q^{\star}}(\hat{\mathbf{y}}_3) - \bar{\mathbf{t}}\big)$ for $p \in \mathbb{Z}$ (see Fig.\;\ref{fig:TwoPhaseFig}).   The closure condition is equivalent to specifying that one vertex on the top boundary is equal to $\check{\mathbf{y}}_3$ since the rest of the cylinder will close due to the underlying symmetry.  Consequently, continuity and the closure condition are equivalent to
\begin{equation}
\begin{aligned}
\begin{cases}\label{eq:priorTwoPhase}
\check{\mathbf{y}}_3 =\bar{\mathbf{R}}^T\big(\hat{g}_1^{p^{\star}}\hat{g}_2^{q^{\star}}(\hat{\mathbf{y}}_3) - \bar{\mathbf{t}}\big) \;\; \text{ for some } p^{\star} \in \mathbb{Z}, \\
\tau_{1}^{\check{\sigma}}(\check{\omega}, \check{\varphi}) = \tau_1^{\hat{\sigma}}(\hat{\omega},\hat{\varphi}), \\
\theta_1^{\check{\sigma}}(\check{\omega}, \check{\varphi}) = \theta_1^{\hat{\sigma}}(\hat{\omega}, \hat{\varphi}).
\end{cases}
\end{aligned}
\end{equation}
in this setting.   

This reduces, in a suitable way, to the formula in (\ref{eq:twoPhaseSup}). Indeed, we first observe that the rotation $\bar{\mathbf{R}}$ and translation $\bar{\mathbf{t}}$ satisfy 
\begin{equation}
\begin{aligned}
&\bar{\mathbf{t}} = \hat{\mathbf{z}} - \bar{\mathbf{R}} \check{\mathbf{z}} + \rho \hat{\mathbf{e}} \\
&\bar{\mathbf{R}}\big(\check{g}_2^{\tilde{q}}(\check{\mathbf{y}}_3) - \check{\mathbf{z}}\big) =  \hat{g}_2^{\tilde{q}}(\hat{\mathbf{y}}_3) - \hat{\mathbf{z}} - \rho \hat{\mathbf{e}} 
\end{aligned}
\end{equation}
for some $\rho \in \mathbb{R}$ in light of Proposition \ref{localCompatProp} (and, in particular, the equation (\ref{eq:cylinderSup}) in this proposition).  Thus,  
\begin{equation}
\begin{aligned}
&\hat{g}_1^{p^{\star}}\hat{g}_2^{q^{\star}}(\hat{\mathbf{y}}_3) - \bar{\mathbf{t}}\\
&\qquad=\hat{g}_1^{p^{\star}}\hat{g}_2^{q^{\star}- \tilde{q}}(\hat{g}_2^{\tilde{q}}(\hat{\mathbf{y}}_3)) - \bar{\mathbf{t}} \\
&\qquad=  \hat{\mathbf{R}}_{p^{\star}\hat{\theta}_1 + (q^{\star} - \tilde{q}) \hat{\theta}_2} (\hat{g}_2^{\tilde{q}}(\hat{\mathbf{y}}_3) - \hat{\mathbf{z}} - \rho \hat{\mathbf{e}}) \\
&\qquad \qquad  + ( p^{\star}\hat{\tau}_1 + (q^{\star} - \tilde{q}) \hat{\tau}_2) \hat{\mathbf{e}}  + \bar{\mathbf{R}} \check{\mathbf{z}} \\
&\qquad = \hat{\mathbf{R}}_{p^{\star}\hat{\theta}_1 + (q^{\star} - \tilde{q}) \hat{\theta}_2} \bar{\mathbf{R}} \big(\check{g}_2^{\tilde{q}}(\check{\mathbf{y}}_3) - \check{\mathbf{z}}\big)  \\
&\qquad \qquad + ( p^{\star}\hat{\tau}_1 + (q^{\star} - \tilde{q}) \hat{\tau}_2) \hat{\mathbf{e}}  + \bar{\mathbf{R}} \check{\mathbf{z}}.
\end{aligned}
\end{equation}
We then combine this observation with the fact that  $\bar{\mathbf{R}}^T  \hat{\mathbf{R}}_{p^{\star}\hat{\theta}_1 + (q^{\star} - \tilde{q}) \hat{\theta}_2} \bar{\mathbf{R}}= \check{\mathbf{R}}_{p^{\star}\hat{\theta}_1 + (q^{\star} - \tilde{q}) \hat{\theta}_2}$ and $\bar{\mathbf{R}}^T \hat{\mathbf{e}} = \check{\mathbf{e}}$ to obtain
\begin{equation}
\begin{aligned}
&\bar{\mathbf{R}}^T\Big(\hat{g}_1^{p^{\star}}\hat{g}_2^{q^{\star}}(\hat{\mathbf{y}}_3) - \bar{\mathbf{t}}\Big) - \hat{\mathbf{z}}  \\
&\qquad = \check{\mathbf{R}}_{p^{\star}\hat{\theta}_1 + (q^{\star} - \tilde{q}) \hat{\theta}_2}  \big(\check{g}_2^{\tilde{q}}(\check{\mathbf{y}}_3) - \check{\mathbf{z}}\big) \\
&\qquad \qquad + ( p^{\star}\hat{\tau}_1 + (q^{\star} - \tilde{q}) \hat{\tau}_2) \check{\mathbf{e}}  \\
&\qquad = \check{\mathbf{R}}_{p^{\star}\hat{\theta}_1 + (q^{\star} - \tilde{q}) \hat{\theta}_2 + \tilde{q}\check{\theta}_2}  \big(\check{\mathbf{y}}_3 - \check{\mathbf{z}}\big) \\
&\qquad \qquad + ( p^{\star}\hat{\tau}_1 + (q^{\star} - \tilde{q}) \hat{\tau}_2 + \tilde{q} \check{\tau}_2) \check{\mathbf{e}}.
\end{aligned}
\end{equation}
Consequently, the formula in (\ref{eq:priorTwoPhase}) is equivalent to $\tau_{1}^{\check{\sigma}}(\check{\omega}, \check{\varphi}) = \tau_1^{\hat{\sigma}}(\hat{\omega},\hat{\varphi})$, $\theta_1^{\check{\sigma}}(\check{\omega}, \check{\varphi}) = \theta_1^{\hat{\sigma}}(\hat{\omega}, \hat{\varphi})$ and 
\begin{equation}
\begin{aligned}\label{eq:finalfinalTwoPhase}
&\check{\mathbf{y}}_3 - \check{\mathbf{z}} = \check{\mathbf{R}}_{p^{\star}\hat{\theta}_1 + (q^{\star} - \tilde{q}) \hat{\theta}_2 + \tilde{q}\check{\theta}_2}  \big(\check{\mathbf{y}}_3 - \check{\mathbf{z}}\big) \\
&\qquad \qquad + ( p^{\star}\hat{\tau}_1 + (q^{\star} - \tilde{q}) \hat{\tau}_2 + \tilde{q} \check{\tau}_2) \check{\mathbf{e}}.
\end{aligned}
\end{equation}
By dotting this with $\check{\mathbf{e}}$, we obviously require  $p^{\star}\hat{\tau}_1 + (q^{\star} - \tilde{q}) \hat{\tau}_2 + \tilde{q} \check{\tau}_2= 0$.  As $|\mathbf{P}_{\check{\mathbf{e}}}(\check{\mathbf{y}}_3 - \check{\mathbf{z}})| > 0$, we evidently also require that $\check{\mathbf{R}}_{p^{\star}\hat{\theta}_1 + (q^{\star} - \tilde{q}) \hat{\theta}_2 + \tilde{q}\check{\theta}_2}= \mathbf{I}$.  In order to guarantee that the structure wraps around exactly once before closing (i.e., it is invertible), the latter demands that $p^{\star}\hat{\theta}_1 + (q^{\star} - \tilde{q}) \hat{\theta}_2 + \tilde{q}\check{\theta}_2 \in \{\pm 2\pi\}$.   These two condition are also sufficient to solve (\ref{eq:finalfinalTwoPhase}); thus yielding the indentity (\ref{eq:priorTwoPhase}).  

To relate this characterization exactly to (\ref{eq:twoPhaseSup}), we recall the assumption $q^{\star} > \tilde{q} > 0$ at the start of this analysis.  Thus, if we find a $p^{\star} \in \mathbb{Z}$ such that $p^{\star}\hat{\theta}_1 + (q^{\star} - \tilde{q}) \hat{\theta}_2 + \tilde{q}\check{\theta}_2 = -2\pi$, we can always make the transformation $(p^{\star}, \tilde{q}, q^{\star}) \mapsto -(p^{\star}, \tilde{q}, q^{\star})$ to yield exactly (\ref{eq:twoPhaseSup}), which thus justifies the formula.   Finally, it is clear by symmetry that we can exchange the roles of $(\cdot)_1$ and $(\cdot)_2$ in all the of the result above (see Remark \ref{exchangeRemark}), and thus justify another compatibility condition akin to (\ref{eq:twoPhaseSup}) with the roles of $(\cdot)_1$ and $(\cdot)_2$ exchanged.

\bibliographystyle{ieeetr}
\bibliography{origami}